\documentclass{article}
\usepackage{amssymb}
\usepackage{amsbsy}
\usepackage{color}
\usepackage{amsmath}

\usepackage{float}
\usepackage{caption}

\usepackage{comment}

\usepackage{lscape}

\usepackage{multirow}

\usepackage{enumerate}
\usepackage{graphicx}
\newcommand{\be}{\begin{equation}}
\newcommand{\ee}{\end{equation}}

\def\bea{\begin{eqnarray}}
\def\eea{\end{eqnarray}}
\def\bean{\begin{eqnarray*}}
\def\eean{\end{eqnarray*}}

\newcommand{\barr}{\begin{array}}
\newcommand{\earr}{\end{array}}

\newcommand{\bed}{\begin{displaymath}}
\newcommand{\eed}{\end{displaymath}}
\newcommand{\bal}{\begin{array}{ll}}
\newcommand{\eal}{\end{array}}

\def\bvec#1{\raise1.5ex\hbox{$\rightarrow$}\mkern-16.5mu #1}

\def\m#1{\mathcal#1}

\newcommand{\bs}{\boldsymbol}
\begin{document}

\title{\hfill ~\\[-30mm]
       \hfill\mbox{\small }\\[30mm]
       \textbf{{{\huge $\mathbf{\theta_{13}^{}}\,$} and The Flavor Ring}}} 
\date{}
\author{\\ Jennifer Kile,\footnote{E-mail: {\tt jenkile@phys.ufl.edu}}~~M. Jay P\'erez,\footnote{E-mail: {\tt mjperez@phys.ufl.edu}}~~Pierre Ramond,\footnote{E-mail: {\tt ramond@phys.ufl.edu}}~~Jue Zhang,\footnote{E-mail: {\tt juezhang@phys.ufl.edu}}\\ \\
  \emph{\small{}Institute for Fundamental Theory, Department of Physics,}\\
  \emph{\small University of Florida, Gainesville, FL 32611, USA}}

\maketitle

\begin{abstract}
\noindent  We present the results of a numerical search for the Dirac Yukawa matrices of the Standard Model, consistent with the quark and lepton masses and their mixing angles. We assume a diagonal up-quark matrix, natural in $\bs{\m Z_7 \rtimes \m Z_3}$, Bimaximal or Tri-bimaximal seesaw mixing, and $SU(5)$ unification to relate the down-quark and charged lepton Dirac Yukawa matrices using Georgi-Jarlskog mechanisms. The measured value of $\theta_{13}$ requires an asymmetric down-quark Yukawa matrix. Satisfying the measured values of both $\theta_{13}$ and the electron mass restricts the number of solutions, underlying the importance of the recent measurement of the reactor angle.

\end{abstract}

\thispagestyle{empty}
\vfill
\newpage
\setcounter{page}{1}

\section{Introduction}

The measurement of the third neutrino mixing angle $\theta_{13}$ \cite{dayabay,reno,dchooz} affords us an interesting opportunity to stringently compare the predictions of flavor models to observation, and opens the door for substantial CP violation in the neutrino sector.\footnote{See \cite{neutrinos} for reviews on neutrinos and flavor model building.} In this work, we consider the ramifications of this result, along with those of other flavor measurements in the quark, charged-lepton, and neutrino sectors within the context of Grand Unified Theories (GUTs).

GUT patterns and relationships between quark and lepton Yukawa couplings, together with constraints from flavor observables, can be quite a powerful combination:  up- and down-quark Yukawa matrices are related by the Cabibbo-Kobayashi-Maskawa (CKM) matrix;   GUTs then suggest close relations between the down-quark and charged-lepton mass matrices; the Maki-Nakagawa-Sakata-Pontecorvo (MNSP) matrix then relates the charged-lepton Yukawa matrix to the neutrino mass matrix, itself a function of the high-scale Majorana mass matrix and the neutral Yukawa matrix; finally the latter can be related back to the up-quark Yukawa coupling via GUT-inspired relations. We call this set of relations the ``Flavor Ring''.

In a previous work \cite{Kile:2013gla}, we investigated the relation between the up-quark and seesaw sectors of the Flavor Ring within the framework of a $(\bs{\m Z_7 \rtimes \m Z_3})$ flavor symmetry \cite{z7z3singlet}.  In this paper, we close the ring by presenting the results of a numerical study  of  the interplay between the CKM parameters, charged lepton masses, and neutrino mixing angles.  Given a down-quark Yukawa matrix, possible charged-lepton Yukawa matrices can be deduced through $SU(5)$ GUT relations and the use of Georgi-Jarlskog (GJ) insertions \cite{Georgi:1979df}.  


In our numerical search, we first assume a diagonal up-quark Yukawa matrix, motivated by the $(\bs{\m Z_7 \rtimes \m Z_3})$ flavor symmetry.  
The down-quark Yukawa matrix is then determined by the CKM matrix, GUT-scale  masses, and a right-handed unitary mixing matrix, whose form we vary. Through $SU(5)$ relations we then obtain the charged-lepton Yukawa matrix, including up to eight GJ factors. We diagonalize this matrix to obtain lepton masses and its left-handed unitary mixing matrix. To make contact with the MNSP matrix, we assume the seesaw neutrino mixing matrix (neglecting phases) to be of the Tri-bimaximal (TBM) or Bimaximal (BM) forms which share the interesting feature of $(\mu-\tau)$ symmetry and vanishing seesaw one-three mixing, implying that CP violation in the lepton sector must originate in the quark sector. 

The search inputs are the rotation angles in the right-handed down-quark unitary mixing matrix, and all possible ways of assigning GJ factors; outputs are the MNSP angles, lepton mass ratios, and the MNSP Jarlskog invariant. We retain as solutions those outputs which are compatible with experimental constraints.


Our results display some interesting features:  first, there are no solutions for which the down-quark Yukawa matrix is symmetric; secondly,  satisfying both the electron-to-muon mass ratio $m_e/m_{\mu}$ and $\theta_{13}\sim 9^\circ$ severely constrains the down-quark and charged lepton Yukawa matrices, and is key to their asymmetry.  If $\theta_{13}$ had been measured to be  $\sim 3^\circ$ as predicted in many models, either the original Georgi-Jarlskog scenario or a symmetric down-quark Yukawa would have sufficed to reproduce all observed data.  This highlights the importance of the measurements of $\theta_{13}$.

The question of reproducing suitable neutrino parameters consistent with the value of $\theta_{13}$ has been studied previously by several authors. Some do not assume GUT relations between the quark and lepton sectors, but instead aim to phenomenologically study the deviations of the neutrino mixing parameters from popular models (Tri-bimaximal, etc.) within the lepton sector \cite{searchnoGUTs}. Others, however, do invoke the GUT framework, and explore ideas such as Cabibbo Haze \cite{Cabibbohaze} and Quark-Lepton Complementarity \cite{QLC}. Recently, suggested by the numerical relation $\theta_{13} \approx \lambda/\sqrt{2}$, where $\lambda$ is the Cabibbo angle, others have performed a more detailed study within the GUT framework \cite{lamsqrt2_2}. They focus on the upper $(2 \times 2)$ blocks of the down-quark and charged-lepton Yukawa matrices, and try to relate them by exploring many possible Clebsch-Gordan coefficients of the GUT groups. In this work, however, we restrict the Clebsch-Gordan coefficients to those of the Georgi-Jarlskog mechanism, and search for their allowed insertions into a full $(3 \times 3)$ down-quark Yukawa matrix.


The rest of this paper is organized as follows.  In Section \ref{sec:flavor_ring}, we describe the Flavor Ring in detail and give the relevant relations between the down-quark and charged-lepton Yukawa matrices and the neutrino mixing angles, as well as the basics of GJ insertions. 
The main numerical search results can be found in Section \ref{sec:search}, where we list the search assumptions, explain the search procedure and discuss its results by presenting some illustrative examples.  Section \ref{sec:analysis} presents analytic studies aimed at clarifying and understanding the search results. The effects of varying the down-quark mass ratios on the search results are discussed in Section \ref{sec:massratio}. We also give an alternative way of performing the search in Section \ref{sec:symmetric}. Section \ref{sec:conclusion} gives our conclusions.  The Appendices display detailed numerical results and a discussion of their subtleties.
\section{The Flavor Ring}
\label{sec:flavor_ring}

Flavor models aim to provide a satisfactory description of the masses and mixing between the three families of the Standard Model. For the up- and down-type quarks and charged leptons, these masses and mixings are generated by the Dirac Yukawa matrices $Y^{(2/3)}$, $Y^{(-1/3)}$, and $Y^{(-1)}$. Although the origin of neutrino masses is still an open question, the seesaw mechanism \cite{seesaw,lseesaw}, which  provides a satisfactory explanation for  small neutrino masses, adds to this list two more matrices: the neutral Dirac Yukawa matrix $Y^{(0)}$ and  $\m M$, the Majorana mass matrix of the right-handed neutrinos. 

Our aim is to provide a logical framework, using ``top-down'' ideas from the seesaw mechanism and GUTs, by which these matrices may be related to one another, forming a ``Flavor Ring'' shown in Fig. 1.

\begin{figure}[h!] 
\centering
\scalebox{0.7}{\includegraphics*[170,420][450,720]{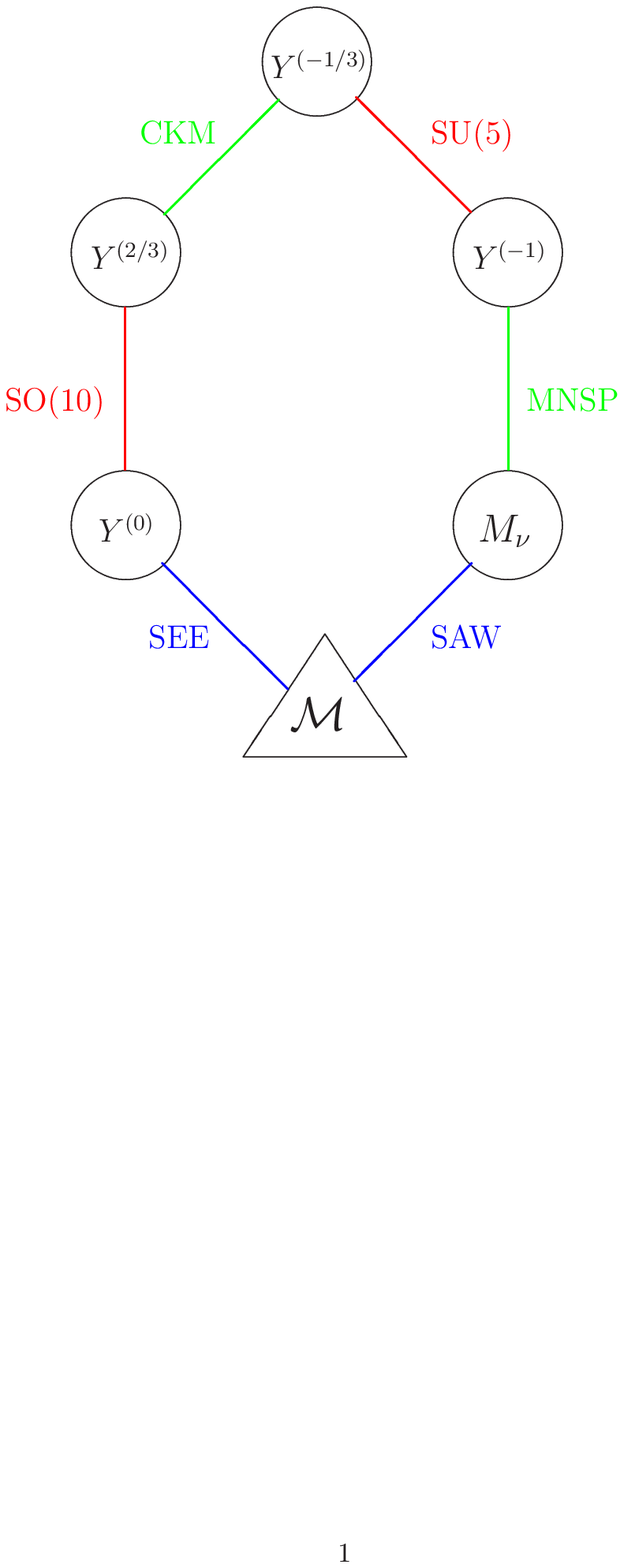}} 
\caption{The ``Flavor Ring''.}
\label{fig:flavor-ring}
\end{figure}

\subsection{Previous Work: the Seesaw Sector}

In a previous publication \cite{Kile:2013gla} we addressed the seesaw sector of the Flavor Ring. We assumed that the form of the Majorana matrix $\m M$ was determined by the family symmetry $(\bs{\m Z_7 \rtimes \m Z_3})$, the smallest non-Abelian discrete subgroup of $SU(3)$ that is not a subgroup of $SO(3)$ or $SU(2)$; it contains  $21$ elements and only $\bf 3$, $\bf\bar 3$, $\bf 1'$, $\bf \bar 1'$ and $\bf 1$ representations.

Using $SO(10)$ as a guide, where all $SU(5)$ matter fields, $\bf 10 (\chi)$, $\bf \bar{5}(\psi)$ and $\bf{1}$($N$) are contained in the sixteen-dimensional spinor representation, the three chiral families transform as one $(\bs{\m Z_7 \rtimes \m Z_3})$ triplet. 

The product of  two triplet representations is given by, 

$$ \bf 3 \otimes \bf 3 = (\bf 3 + \bf \bar{3})_+ + \bf \bar{3}_-  , 
$$
where the symmetric $\bf 3$ is diagonal, and the $\bf\bar 3_\pm$ are off-diagonal symmetric and antisymmetric, respectively. 

Our choice was to take the Higgs fields as anti-triplets, thereby assuring a diagonal $Y^{(2/3)}_{}$.  

In $SO(10)$, it is natural to relate the Dirac Yukawa matrices,

\begin{equation} \label{eq:so10relation}
 Y_{}^{(2/3)} \sim Y_{}^{(0)},
\end{equation}

where $Y^{(0)}$ appears in the numerator of the seesaw mass formula, 

\begin{equation} \label{eq:seesaw}
M^{}_\nu =  Y_{}^{(0)} \m M_{}^{-1} Y_{}^{(0) T}~=~ \m U^{}_{\rm \,seesaw}~ D^{}_\nu~ \m U^T_{\rm \,seesaw},
\end{equation}
where  $D_\nu={\rm diag}(m_1,m_2,m_3)$ is the diagonal light neutrino mass matrix, and $\m U_{\rm \,seesaw}$ is the seesaw neutrino mixing matrix. Eq.(\ref{eq:so10relation}) may then seem puzzling, since the large hierarchy of the up-quark sector is not replicated amongst the light neutrinos, implying that the Majorana matrix undoes the up-quark hierarchy.

Using one linear combination of two $(\bs{\m Z_7 \rtimes \m Z_3})$-invariant dimension-five couplings (or a single dimension-six coupling), we constructed a predictive and compelling Majorana matrix with the required large correlated hierarchy; it predicts normal hierarchy and the values of neutrino masses, leads to Tri-bimaximal and Golden Ratio seesaw mixing matrices, and its  correlated hierarchy is  entirely generated by the vacuum values of familon fields.   

Although inspired by our previous work, the conclusions of this paper do not depend on any family symmetry. The starting points of the numerical search presented in this paper are then a diagonal $Y^{(2/3)}$, a normal or inverted neutrino mass hierarchy, and either a Tri-bimaximal or Bimaximal $\m U^{}_{\rm seesaw}$.\footnote{The results for Golden Ratio mixing will be similar to TBM aside from the value of $\theta_{12}$. We consider Bimaximal mixing as well, since this pattern is distinct from TBM or Golden Ratio mixing.}  

\subsection{Closing the Flavor Ring}

There remains to determine the missing links required to close the flavor ring. 

One is the observable leptonic mixing matrix $\m U_{\rm MNSP}$, which relates  $M_\nu$, (and thus $Y^{(0)}$ and $\m M$), and the charged-lepton matrix $Y^{(-1)}$ through $\m U_{\rm \,seesaw}$,

\begin{equation} \label{eq:MNSP}
\m U^{}_{\rm MNSP}~=~\m U^\dagger_{-1}\,\m U^{}_{\rm seesaw} ,
\end{equation}
where $\m U^{}_{-1}$ is determined by the charged lepton Yukawa matrix $Y^{(-1)}_{}$. 

The second is the observed CKM mixing of the quarks which connects $Y^{(2/3)}$ to $Y^{(-1/3)}$, and since  $Y^{(2/3)}$ is diagonal,

\begin{equation}
\m U_{\rm CKM}^{} ~=~\m U^{}_{-1/3} ,
\end{equation}
where $\m U_{-1/3}$ is the left-handed unitary matrix which diagonalizes $Y^{(-1/3)}$. This is enough to determine a symmetric $Y^{(-1/3)}$, which, as we shall see, does not reproduce the Daya Bay/RENO measurements. 


The last step is to relate the down-quark and charged-lepton sectors. Such a relationship is natural in minimal $SU(5)$, where the Yukawa coupling $ \psi \chi H_d$ to a $\bf \bar{5}$ Higgs predicts, 

\begin{equation}
Y_{}^{(-1)} = Y_{}^{(-1/3) T},
\end{equation}
providing the final link. Although it successfully predicts $m_b = m_\tau$ at unification, it also yields $m_e = m_d$ and $m_\mu = m_s$, while renormalization group-running favors 

\begin{equation}\label{gj}
m_e = m_d / 3 , \qquad m_\mu = 3 m_s .
\end{equation}
An elegant solution to this problem, the Georgi-Jarlskog mechanism \cite{Georgi:1979df}, is to add one additional $\bf 45$ Higgs coupling, $ \psi \chi H_{\bf 45}$. Its original realization was through Yukawa couplings of the form, 

$$
Y^{(\bf \bar{5})}_{i j } \psi^i \chi^j H_d , ~~ Y^{(\bf \bar{5})}_{i j } = \begin{pmatrix}
0 & a' & 0 \\ a & 0 & 0 \\ 0 & 0 & c \end{pmatrix}, ~~ ~~ Y^{(\bf 45)}_{i j } \psi^i \chi^j H_{\bf 45} , ~~ Y^{(\bf 45)}_{i j } = \begin{pmatrix} 
0 & 0 & 0 \\ 0 & b & 0 \\ 0 & 0 & 0 
 \end{pmatrix}.
$$
The Clebsch-Gordan coefficients of the $\bf 45$ coupling  give an extra factor of $-3$ in the (22) position of $Y^{(-1)}$. With the assumption that $a \sim a'\ll b \ll c$, this reproduces Eq.(\ref{gj}). 

Although it predicts the correct light charged-lepton mass ratio, this simple model makes another prediction, namely that the mixing between the light charged leptons is given by $\theta^e_{12} = \lambda / 3$
; with TBM or BM seesaw mixing, this corresponds to  

$$ \theta_{13}^{} = \frac{\lambda}{3 \sqrt{2}} \approx 3^\circ , $$
a value too small given the current data. 

However, $\bf 45$ couplings can in principle occur in any matrix element; in this work we explore all such possible couplings. As an example, if the ${\bf 45}$ coupling dominates the $(12)$ and $(23)$ entries, we would have

\begin{equation}
Y^{(-1/3)}_{} \sim \begin{pmatrix}
 & G &   \\   &   & J \\   &   &  \end{pmatrix}, ~~{\rm then}~~ Y^{(-1)}_{} \sim \begin{pmatrix}
 &  &   \\ -3G &   &   \\   & -3J & \, \end{pmatrix}, 
\end{equation}
In the following, we use this mechanism to relate $Y^{(-1/3)}_{}$ to $Y^{(-1)}_{}$ and close the flavor ring.



\section{ Numerical Search for Allowed $Y^{(-1/3)}_{}$ and $Y^{(-1)}_{}$}
\label{sec:search}
In this section, we describe our numerical search  for experimentally allowed charge $({-1}/{3})$  and  $(-1)$ Yukawa matrices, constrained by the following set of assumptions: 

\begin{itemize}
\vskip .2cm
\item Since we assume the up-quark Yukawa matrix $Y^{(2/3)}_{}$ to be diagonal, the CKM mixing matrix comes from the diagonalization of $Y^{(-1/3)}_{}$, 

\begin{equation} \label{eq:Y_d}
 Y^{(-1/3)}_{} = \m U_{\rm CKM}^{} \m D^{}_d\, \m V_{}^{\dagger},
\end{equation}
where $\m D_d$ is the diagonal down-quark mass matrix, and $\m V$ is an hitherto arbitrary unitary matrix.  

\vskip .2cm
\item
To simplify the search, we assume that this unitary matrix contains no phases,\footnote{Although we neglect phases in $\m V$, we include for completeness a general phase analysis in Section \ref{sec:analysis}.} and is parametrized as,  

\begin{equation} \label{eq:V}
\m V = \begin{pmatrix}
1 & 0 & 0 \\ 0 & c_{23} & s_{23} \\ 0 & -s_{23} & c_{23}  
\end{pmatrix} \begin{pmatrix}
c_{13} & 0 & s_{13} \\ 0 & 1 & 0 \\ -s_{13} & 0 & c_{13}
\end{pmatrix} \begin{pmatrix}
c_{12} & s_{12} & 0 \\ -s_{12} & c_{12} & 0 \\ 0 & 0 & 1
\end{pmatrix},
\end{equation}
where $c_{ij} = \cos \beta_{ij}$, $s_{ij} = \sin \beta_{ij}$.

\vskip .2cm 
\item 
As suggested by $SU(5)$, the charged-lepton Yukawa matrix $Y^{(-1)}_{}$, is the transpose of $Y^{(-1/3)}_{}$, up to GJ entries. 
\vskip .2cm
\item 
We assume seesaw neutrino mixing matrices, with no phases as well, of the Tribimaximal or Bimaximal forms:


\begin{eqnarray} \label{eq:Useesaw}
\m U_{\rm seesaw} = \begin{pmatrix}
~~~ \cos \theta_{12}^\nu & -\sin \theta_{12}^\nu & 0 \\
\frac{1}{\sqrt{2}} \sin\theta_{12}^\nu & \frac{1}{\sqrt{2}} \cos\theta_{12}^\nu & -\frac{1}{\sqrt{2}} \\
\frac{1}{\sqrt{2}} \sin\theta_{12}^\nu & \frac{1}{\sqrt{2}} \cos\theta_{12}^\nu & \frac{1}{\sqrt{2}}
\end{pmatrix},
\end{eqnarray}
where $\theta^\nu_{12} = \arctan (1/\sqrt{2})$ for TBM mixing, and $\theta^\nu_{12} = 45^\circ $ for BM mixing. In this parametrization, all TBM angles lie in the fourth quadrant. This choice is arbitrary; for more details, see Appendix D.

In absence of phases in $\m V$ and $\m U_{\rm seesaw}$, phases in $Y^{(-1)}_{}$, obtained from the CKM matrix, lead to the CP-violating phase in the MNSP matrix. However, when searching for solutions with a \emph{symmetric} $Y^{(-1/3)}$, we consider in Section 3.3 all possible phases in $\m U_{\rm seesaw}$.

 
 \end{itemize}
\vskip .3cm
\subsection{Search Preliminaries}
\vskip .2cm
\noindent The search inputs, expressed in terms of the Wolfenstein parametrization, are:

\begin{itemize}

\item The CKM matrix,

\be\label{ckm}
\m U^{}_{\rm CKM} ~=~\begin{pmatrix}
1-\lambda^2/2 & \lambda & A \lambda^3 (\rho - i \eta) \\ -\lambda & 1 - \lambda^2/2 & A \lambda^2 \\  A \lambda^3 (1 - \rho - i \eta) & -A \lambda^2 & 1 
\end{pmatrix} + \m O(\lambda^4),
\ee
with GUT scale values \cite{ross&serna},  $\lambda = 0.227$,  $\rho = 0.22$, and $\eta = 0.33$. In supersymmetry, $A$ is the parameter most sensitive to $\tan\beta$. Its benchmark value is  $A = 0.77$, which corresponds to $\tan\beta=10$. For  reference, $A=0.72$ for $\tan\beta=50$. In most cases we fix $A = 0.77$, but we provide some examples of how varying $A$ can affect the search results. 

\item The down-quark mass ratios, allowing for the different signs of $(m_d,m_s)$, labelled as $(\pm,\pm)$,

\be\label{downratio}
\m D^{}_d~=~m_b \begin{pmatrix}
\pm \frac{\lambda^4}{3} & & \\ &  \pm\frac{\lambda^2}{3} & \\ & & 1
\end{pmatrix},\ee
which ensures the Gatto-Sartori-Tonin relation \cite{Go_Gatto!} 

$$\lambda~=~\sqrt{\frac{m_d}{m_s}}.$$ 
A detailed discussion of the validity of this parametrization and its effects on our search results can be found in Section \ref{sec:massratio}.

\end{itemize}

\vskip .2cm
\subsection{Search Procedure} 

\vskip .2cm
\noindent The numerical search, using Mathematica,\footnote{The interested reader may find the corresponding notebook and their documentation as ancillary files on the arXiv.} proceeds as follows:

\vskip .2cm
\begin{itemize}

\item 

The first step consists of providing an input $Y^{(-1/3)}_{}$. We search by the number of non-zero angles $\beta_{ij}$ in its right-handed rotation matrix $\m V$, see Eq.(\ref{eq:V}). 

\item 

With $\m V$ and thus $Y^{(-1/3)}$ specified, we consider all $Y^{(-1)}$'s obtained from $Y^{(-1/3)}$ by transposition and given assignments of GJ factors. These $Y^{(-1)}$'s are organized in terms of the number of GJ factors they contain.

\item 
The outputs of the search are the charged lepton mass ratios, the neutrino mixing angles, and the MNSP Jarlskog invariant, obtained from the diagonalization of $Y^{(-1)}$.\footnote{Another possible output to check is $m_b / m_{\tau}$. However, upon performing the search, we found no solutions with a GJ insertion in the (33) entry, implying that, up to small Cabibbo corrections, $m_b \approx m_{\tau}$ at unification. }

\item 
A ``successful solution" is defined as one that yields both mass ratios within their allowed ranges at the GUT scale \cite{ross&serna},

\be
0.0046 \leq  |m_e / m_\mu | \leq 0.0050,\qquad 
0.048 \leq  |m_\mu / m_\tau | \leq 0.06, 
\ee
\vskip .2 cm
and leptonic mixing angles which fall within the ``best fit" ranges \cite{fit}, shown in Table 1.\footnote{Our search is for acceptable GUT scale parameters. As explored in \cite{RGE}, RG effects can be appreciable depending on the neutrino mass spectrum. In our previously considered model, with TBM mixing, the lightest neutrino mass is sufficiently small so that these effects are negligible; for a more general neutrino spectrum, additional care should be given.} 

This table shows that at 1$\sigma$, some but not all solutions may discriminate between normal and inverted neutrino hierarchies; however, at 3$\sigma$ there is substantial overlap.  

\begin{table}
\centering
\begin{tabular*}{\textwidth}{ c c c c }
\hline
\hline
\noalign{\smallskip}
~~~~Parameter~~~~ &~~Best fit~~~& ~~~1$\sigma$ range~~~ &  ~~~3$\sigma$ range~~ \\
\noalign{\smallskip}
\hline
\hline
\noalign{\smallskip}\noalign{\smallskip}
$\theta_{12}$~~~~ & $33.7^\circ$ & $32.6^\circ$ - $34.8^\circ$ & $30.6^\circ$ - $36.8^\circ$ \\
\noalign{\smallskip}
\noalign{\smallskip}
$\theta_{23}$ (NH) & $40.7^\circ$ & $39.1^\circ$ - $42.4^\circ$ & $36.7^\circ$ - $53.2^\circ$ \\
\noalign{\smallskip}
\noalign{\smallskip}
$\theta_{23}$ ~(IH) & $41.4^\circ$ & $39.7^\circ$ - $44.8^\circ$ $\oplus$ $46.8^\circ$ - $51.4^\circ$ & $37.0^\circ$ - $54.3^\circ$ \\
\noalign{\smallskip}
\noalign{\smallskip}
$\theta_{13}$ (NH) & $8.80^\circ$ & $8.45^\circ$ - $9.21^\circ$ & $7.65^\circ$ - $9.92^\circ$ \\
\noalign{\smallskip}
\noalign{\smallskip}
$\theta_{13}$ ~(IH) & $8.89^\circ$ & $8.49^\circ$ - $9.23^\circ$ & $7.67^\circ$ - $9.97^\circ$\\
\noalign{\smallskip}
\hline
\hline
\end{tabular*}
\caption{Neutrino mixing angles from global fit results \cite{fit}. NH stands for normal hierarchy, IH for inverted hierarchy.}
\label{tb:fit}
\end{table} 

Solutions may be characterized by how close their numerical outputs are to these best fit values. At both extremes, an ``excellent" solution is one for which all mixing angles fall within their 1$\sigma$ range, a ``poor" solution one for which the mixing angles only agree at 3$\sigma$. In between, subtleties arise, as there may be solutions for which two of the mixing angles are within 1$\sigma$ but the third agrees only at 3$\sigma$.

The next stage in the search is guided by the reasonable expectation that as we increase the number of angles in $\m V$, a better fit to the experimental values may be found. Accordingly, when one or two of the three input parameters $\beta_{ij}$ are zero, we consider a ``solution" to be one for which the neutrino mixing angles fall within the 3$\sigma$ range. If all three input angles are non-zero, we instead demand that the neutrino mixing angles fall within the 1$\sigma$ range.  

\item 
A technical remark. 

We first perform coarse scans over the input angles $\beta_{ij}$, with a step size of $1^\circ$ to search for solutions. The results fall into three categories. 

The first category are solutions, for which all output parameters fall within the appropriate ranges described above. 

The next are ``close solutions'' for which the output mass ratios and mixing angles agree within some tolerance beyond their bounds. For these ``close solutions'', a finer scan is performed by varying $\beta_{ij}$ with a step size of $0.1^\circ$ to see if agreement with the appropriate bounds can be achieved.     

In the last category are the null results, whose outputs lie outside of the appropriate bounds even when a tolerance is included. For these, no further tuning of the $\beta_{ij}$ may bring the outputs within their appropriate ranges.   

\end{itemize}

\subsection{First Result: $Y^{(-1/3)}$ is Not Symmetric}

We begin by searching for symmetric $Y^{(-1/3)}_{}$, which, under our assumption of a diagonal $Y^{(2/3)}$, is totally determined,

\be
 Y^{(-1/3)}_{} = \m U_{\rm CKM}^{}\m D^{}_d\, \m U_{\rm CKM}^{T},
\ee
up to different signs for $m_d$ and $m_s$. 

When we conduct the numerical search using this matrix as an input, we find \emph{no} pattern of GJ insertions that reproduces the lepton masses and mixings. The same conclusion holds even if we relax two of our previous assumptions; varying the $A$ parameter from $0.6$ to $0.8$, and allowing for additional phases to appear in $\m U_{\rm seesaw}$ for both TBM and BM mixings produce no additional solutions; see Section \ref{sec:analysis}.
 
As a curiosity, we investigated further the reason why there are no solutions; for those $Y^{(-1)}$ with a satisfactory value of $m_e / m_\mu$, $\theta_{13}$ is around $ 3^\circ$. This illustrates the constraining and important role of the recent measurement of $\theta_{13}$, and its tension with $m_e / m_\mu$.

\subsection{One-Angle Solutions}
We therefore look for \emph{asymmetric} $Y^{(-1/3)}$,  beginning with the simplest case, where only one of the rotation angles in $\m V$ is non-zero. 
\begin{itemize} 

\item There are no one-angle solutions for Bimaximal mixing.

\item There are one-angle solutions for Tri-bimaximal mixing only for $\beta_{13}\ne 0$, and with two and three GJ insertions.

These solutions are narrowly centered around $\beta_{13}=183^\circ$ for $(++)$, and $\beta_{13}=3^\circ$ for $(-+)$. Solutions for $(--)$ and  $(+-)$ are the same as $(-+)$ and  $(++)$ respectively, with the value of $\beta_{13}$ shifted by $180^\circ$.  

Below we give two examples of such solutions.

\end{itemize}
\vskip .2cm
 
\noindent  {\bf Example I}: (seesaw) TBM mixing, $(+,+),~\beta_{13}=183^\circ$,  and  two GJ factors in the (21) and (22) elements of $Y^{(-1/3)}$ yield

\begin{eqnarray} \label{eq:sol1}
\theta_{12} = 31.2^\circ , \quad \theta_{23} = 44.6^\circ , \quad \theta_{13} = 8.29^\circ, ~~~~~~~~~~\nonumber \\
\quad
\frac{m_e}{m_\mu} = 0.0051, \quad \frac{m_\mu}{m_\tau} = 0.051, \quad J_{\rm MNSP}=3.6 \times 10^{-7},
\end{eqnarray}
where $J_{\rm MNSP}$ is the predicted Jarlskog invariant in the MNSP matrix. There exists a similar solution with the same $\beta_{13}$ and an additional GJ factor in the small $(32)$ element of $Y^{(-1/3)}$. 

They are ``close solutions", with $m_e /m_\mu$ slightly above its upper bound. Their predicted value of $\theta_{12}$ is lower than the best fit, falling slightly outside of the 2$\sigma$ range, but agreeing at 3$\sigma$. The other two mixing angles also fall outside the 1$\sigma$ range, with a small caveat; if the neutrino hierarchy is inverted, $\theta_{23}$ agrees at 1$\sigma$. These are examples of solutions that discriminate between the two hierarchies, see Table 1.

A finer scan of $\beta_{13}$ around $183^\circ$ with a step size of $0.01^\circ$ reveals that, for a $1^\circ$ variation in $\beta_{13}$, the ratio $m_e / m_\mu$ varies by at most $\pm 0.003$. For $183.03^\circ<\beta_{13}<183.12$, ($A=0.77$), we find

\bean
 && 31.0^\circ <\theta_{12}< 31.1^\circ , \qquad \theta_{23} = 44.6^\circ , \qquad  8.37^\circ <\theta_{13}< 8.63^\circ , \\ 
 &&\\
 &&  0.0046<\frac{m_e}{m_{\mu}}<0.0050 , \qquad  \qquad \frac{m_\mu}{m_{\tau}}=0.051.
\eean

A second possibility is to fix $\beta_{13}$ and instead vary $A$. For example, setting $\beta_{13}=183^\circ$, and  $0.79 <A< 0.81$, we find,  

\bean
 && 30.9^\circ <\theta_{12}< 31.1^\circ , \qquad \theta_{23} = 44.6^\circ , \qquad  8.32^\circ <\theta_{13}< 8.48^\circ , \\ 
 &&\\
 &&  0.0047<\frac{m_e}{m_{\mu}}<0.0050 , \qquad \qquad  \frac{m_\mu}{m_{\tau}}=0.051,
\eean
and the higher values for $A$ imply $\tan\beta<10$.

\vskip .3cm
\noindent {\bf Analysis}
\vskip .3cm

\noindent The existence of a one-angle solution is of special interest, since all mass ratios and mixing angles are fit by one extra input parameter, $\beta_{13}$.  In addition,  its numerical value suggests a parametrization in terms of $\lambda$, with the angle in radians,

$$
\beta_{13} = \pi + B \lambda^2,
$$
where $B\approx 1$. $\m V$ is then given by

\begin{eqnarray}
\m V = \begin{pmatrix}
-1+\frac{B^2\lambda^4}{2} & 0 & -B\lambda^2 \\
0 & 1 & 0 \\
B\lambda^2 & 0 & -1+\frac{B^2\lambda^4}{2}
\end{pmatrix} + \m O(\lambda^6).
\end{eqnarray} 

Accordingly, $Y^{(-1/3)}$ takes the form

\begin{equation}
Y^{(-1/3)}= \left(
\begin{array}{ccc}
 -\frac{1}{3} \lambda ^4  -A B \lambda ^5 (-i \eta +\rho )  & \frac{\lambda ^3}{3} &  -A \lambda ^3 (-i \eta +\rho )  \\
 -A B \lambda ^4+\frac{\lambda ^5}{3} & \frac{1}{3} \lambda ^2 \left(1-\frac{\lambda ^2}{2}\right) & -A \lambda ^2 \\
 -B \lambda^2  & -\frac{A \lambda ^4}{3} & -1+\frac{B^2\lambda^4}{2}
\end{array}
\right)+ \m O(\lambda^6), \nonumber \\
\end{equation}
and $Y^{(-1)}$ is given by

\begin{equation}
Y^{(-1)}= \begin{pmatrix}
 -\frac{1}{3} \lambda ^4 -A B \lambda ^5 (-i \eta +\rho ) & 3 A B \lambda ^4-\lambda ^5 & -B \lambda^2  \\
 \frac{\lambda ^3}{3} & -\lambda ^2 \left(1-\frac{\lambda ^2}{2}\right) & -\frac{A \lambda ^4}{3} \\
-A \lambda ^3 (-i \eta +\rho ) & -A \lambda ^2 & -1+\frac{B^2\lambda^4}{2}
\end{pmatrix}
+ \m O(\lambda^6).
\end{equation}

Its diagonalization yields the charged-lepton mass ratios and the mixing angles $\theta_{ij}^e$ in the left-handed unitary matrix $\m U_{-1}$:\footnote{Some caution must be taken when obtaining $\theta_{12}^e$. It is not simply the ratio of the (12) and (22) entries of $Y^{(-1)}$; an $``AB\lambda^4"$ term coming from the mixing of $\theta_{13}^e$ must be taken out of the (12) entry before evaluating the ratio.}

\begin{eqnarray}
m_{\mu}/m_{\tau} &=& \lambda^2 + \m O(\lambda^4), \nonumber\\
m_e/m_\mu &=& \frac{1}{3}\lambda^2 - \frac{4}{3} A B \lambda^3 + \m O(\lambda^4), \nonumber \\
\theta_{12}^e &=& 4 AB\lambda^2+\m O(\lambda^3), \nonumber \\
\theta_{13}^e &=& -B\lambda^2+\m O(\lambda^4), \nonumber \\ 
\theta_{23}^e &=& \frac{A\lambda^4}{3}+\m O(\lambda^8).
\end{eqnarray}
The GJ factor in the (22) entry results in the correct value for $m_\mu/m_\tau$, while for $m_e/m_\mu$ the input angle $\beta_{13}$ also plays an important role. Since the Tri-bimaximal seesaw mixing matrix is a good approximation to the data, we expect the three mixing angles $\theta_{ij}^e$ in $\m U_{-1}$ to be small. Note that $\theta_{12}^e$ now differs from $\beta_{12}=0$ because of the placements of GJ factors. 

Using Eq.(\ref{eq:MNSP}), we obtain the MNSP mixing angles,

\begin{eqnarray}
\theta_{23} &= & \frac{\pi}{4}+\m O(\lambda^4) , \nonumber \\
\theta_{13} &= &\frac{(4A+1)B\lambda^2}{\sqrt{2}} +\m O(\lambda^3) , \nonumber \\
\theta_{12} &= & \arctan\frac{1}{\sqrt{2}} - \delta\theta_{12} , 
\end{eqnarray}
where
\begin{eqnarray}
\delta\theta_{12} = \frac{(4A-1)B\lambda^2}{\sqrt{2}} +\m O(\lambda^3).
\end{eqnarray}
These analytical expressions for the mixing angles agree with the  numerical values,

$$
\theta_{23}~\approx 45^\circ,~~\theta_{13}~\approx 8.6^\circ,
~~\theta_{12}~\approx 30.9^\circ.
$$
Since $m_e/m_\mu$, $\theta_{13}$ and $\theta_{12}$ all depend on a single input angle $\beta_{13}$, one can derive two sum rules among them, 

\begin{eqnarray}\label{sumrules}
\arctan\frac{1}{\sqrt{2}}&\approx & \theta_{12}+\frac{4A-1}{4A+1}\theta_{13}\approx\theta_{12}+\frac{1}{2}\theta_{13}, \nonumber \\
\frac{m_e}{m_\mu} &\approx & \frac{\lambda^3}{3}+\frac{\sqrt{2}\lambda}{3} \left( \frac{\lambda}{\sqrt{2}}-\theta_{13}\right ).
\end{eqnarray}
The second relation between $m_e/m_\mu$ and $\theta_{13}$ is quite interesting; since  the measured value of $\theta_{13}$ is close to $\lambda/\sqrt{2}$, the order $\lambda^2$ terms in $m_e/m_\mu$ cancel, resulting in a single $\lambda^3/3$ term, which brings $m_e/m_\mu$ within the allowed range.

A naive use of $SU(5)$ and $SO(10)$ relations, assuming a symmetric $Y^{(-1/3)}$, and neglecting the Georgi-Jarlskog contributions,  was found to yield $\theta_{13}=\lambda/\sqrt{2}$ \cite{lamsqrt2_1,lamsqrt2_2} in the TBM case, but at the expense of a much too large electron mass. Conversely, many subsequent models found that starting from the observed electron mass, the same assumptions yielded $\theta_{13}=\lambda/3\sqrt{2}$. Only an asymmetric $Y^{(-1/3)}$ can relieve the tension between $m_e/m_\mu$ and $\theta_{13}$. 

\vskip .3cm
\noindent  {\bf Example II}: (seesaw) TBM mixing, $(-,+),~\beta_{13}=3^\circ$, and  two GJ insertions in the $(22)$ and $(23)$ elements of $Y^{(-1/3)}$, yield:

\begin{eqnarray}
\theta_{12} = 30.5^\circ , \quad \theta_{23} = 44.7^\circ , \quad \theta_{13} = 8.88^\circ, ~~~~~~~~~~\nonumber \\
\quad
\frac{m_e}{m_\mu} = 0.0039, \quad \frac{m_\mu}{m_\tau} = 0.050, \quad J_{\rm MNSP}=4.3 \times 10^{-7}.
\end{eqnarray}
It is another ``close" solution, for which  agreement with the allowed range of $m_e / m_\mu$ is attained with $0.72<A<0.74$, this time finding: 

\begin{eqnarray}
 && 30.9^\circ <\theta_{12}< 31.0^\circ , \qquad \theta_{23} = 44.7^\circ , \qquad  8.34^\circ< \theta_{13} < 8.50^\circ , \\ 
 &&\\
 && 0.0046<\frac{m_e}{m_{\mu}}<0.0050 , \qquad \qquad \frac{m_\mu}{m_{\tau}}=0.050.
\end{eqnarray}
This solution tends to favor lower values of $A~(\tan \beta>10)$. It also predicts a low value of $\theta_{12}$, very close to the 3$\sigma$ bound, and favors the inverted hierarchy; its value for $\theta_{13}$ however, lies very close to the best fit value.  As in the previous example, and for the same reason, a solution with three GJ insertions exists with the additional factor in the $(32)$ element of $Y^{(-1/3)}$. 

For this example, the numerical value of $\beta_{13}$ suggests a slightly different parametrization,

$$
\beta_{13} =  B \lambda^2,
$$
with $B \approx 1$. However, we find that a similar analysis to that given for Example I leads to exactly the same sum rules Eqs. (\ref{sumrules}). 

It is quite remarkable that given our assumptions, a single free parameter is able to reasonably fit all three mixing angles and the two mass ratios.

We next turn to the cases when two of the rotation angles in $\m V$ are non-zero.

\subsection{Two-Angle Solutions}

 With two input parameters, we find a larger set of solutions with a variety of GJ factors. Solutions now exist for both Tri-bimaximal and Bimaximal seesaw mixing schemes. All four sign combinations of ($m_d$, $m_s$) can be related to one another by shifting input angles, and only one case is independent. We focus on the (+,+) case. 

\vskip .3cm 
\noindent {\bf Tri-bimaximal  Seesaw Mixing}
\vskip .2cm
\noindent For TBM mixing, (+,+), and $A=0.77$, we find solutions with $\beta_{23}=0$ and $\beta_{12}=0$ only; there are no two-angle solutions with  $\beta_{13}=0$. Compared to the one angle case, the solutions occur for a wide range of angles and for a diverse set of GJ patterns. Surprisingly,  however, with two angles we are still unable to find a solution for which all mixing angles agree at the $1\sigma$ level.  

A complete list of solutions with two angles and TBM mixing may be found in the tables of Appendix A.



\begin{figure}[h!]
 \centering
\scalebox{0.58}{\includegraphics*{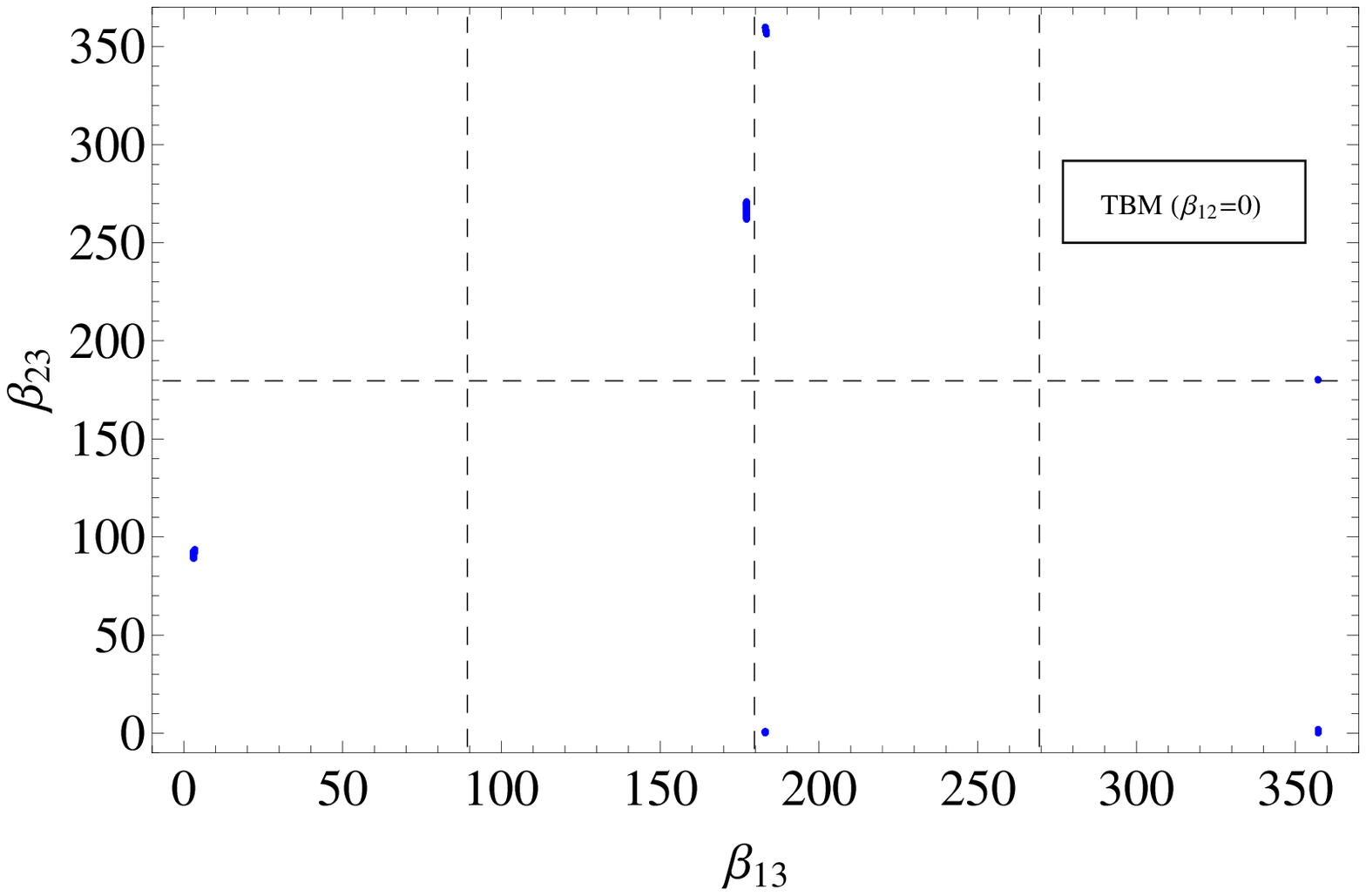}} \scalebox{0.58}{\includegraphics*{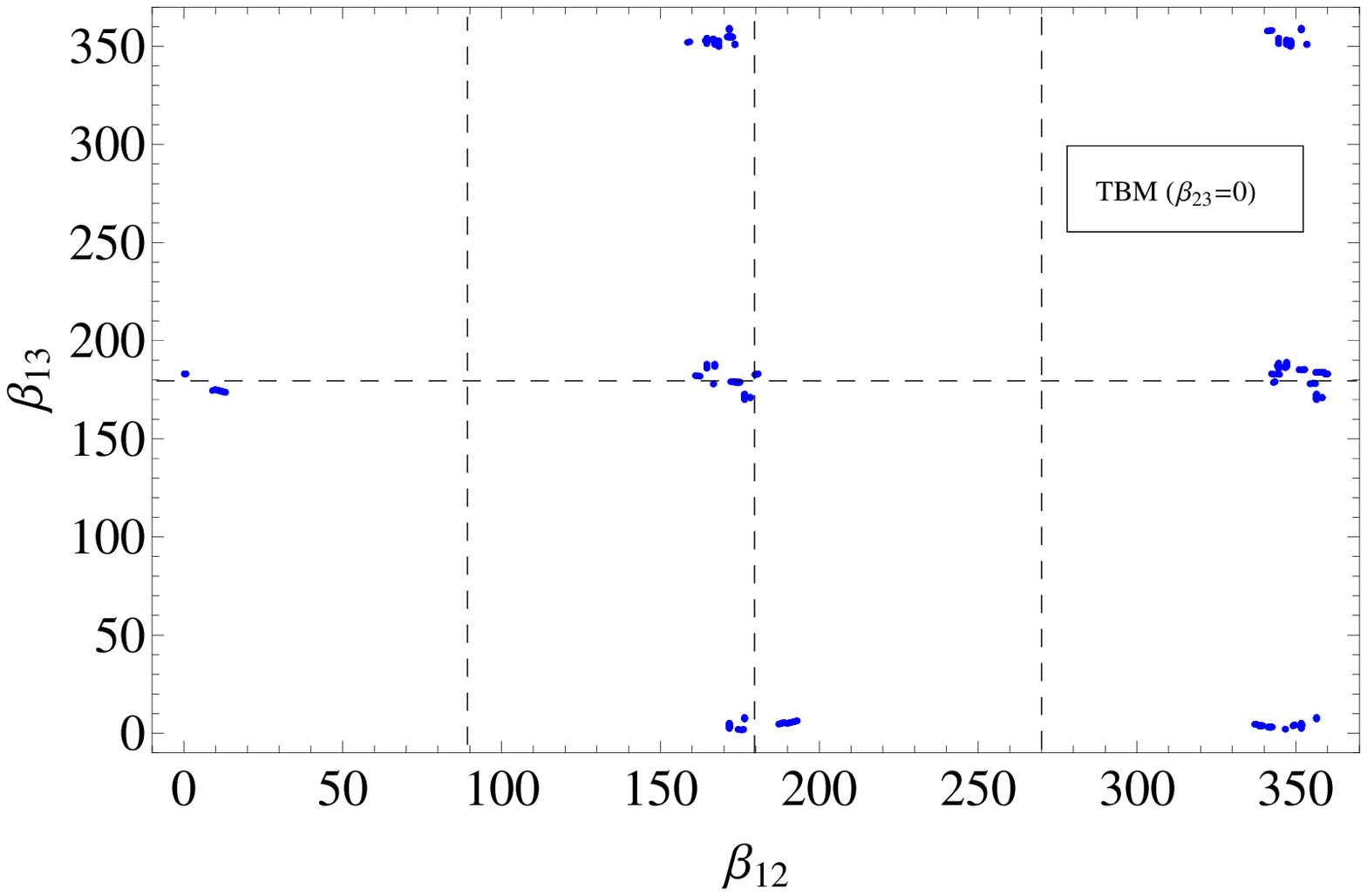}} 
\caption{Allowed parameter space for TBM mixing with two non-zero input angles $\beta_{ij}$. The first figure gives the allowed values of $\beta_{23}$ and $\beta_{13}$ with $\beta_{12} = 0$, while the bottom figure gives the allowed values of $\beta_{12}$ and $\beta_{13}$ with $\beta_{23} = 0$. As noted in the text, no solutions exist with $\beta_{13} = 0$.}
\label{fg:TBM2angle}
\end{figure}

Even for two angles we find the parameter space of solutions to be ``small", with specific regions singled out, as can be seen in Fig.\ref{fg:TBM2angle}. A close inspection of the figure shows that solutions occur for $\beta_{ij}$ close to an axis, and there are more solutions  with $\beta_{23} = 0$, than with $\beta_{12}=0$. 

Since the allowed patterns of inserting GJ factors are relevant for model building, we provide a list below with respect to the number of GJ factors for both $\beta_{23}=0$ and $\beta_{12}=0$.

\begin{itemize}
\item When $\beta_{23}=0$, we find solutions with up to six GJ factors.

\begin{itemize}

\item One GJ factor: 
\begin{eqnarray}
\begin{pmatrix}
 & & \\ & \times & \\ & & 
\end{pmatrix}.
\end{eqnarray}

\item Two GJ factors: 
\begin{eqnarray}
\begin{pmatrix}
 & & \\ \times & \times & \\ & & 
\end{pmatrix}, \quad \quad
\begin{pmatrix}
 & & \\  & \times & \\ & \times & 
\end{pmatrix}, \quad \quad
\begin{pmatrix}
 & & \\ & \times & \times \\ & & 
\end{pmatrix}.
\end{eqnarray}

\item Three GJ factors: 
\begin{eqnarray}
\begin{pmatrix}
 & & \\ \times & \times & \\ & \times & 
\end{pmatrix}, \quad \quad
\begin{pmatrix}
 & \times & \\ \times & \times & \\ & & 
\end{pmatrix}, \quad \quad
\begin{pmatrix}
\times & & \times \\ & \times & \\ & &  
\end{pmatrix}, \nonumber \\
\begin{pmatrix}
 & & \\ & \times & \times \\ \times & & 
\end{pmatrix}, \quad \quad
\begin{pmatrix}
 &  & \\  & \times & \times \\ & \times & 
\end{pmatrix}, \quad \quad
\begin{pmatrix}
\times & &  \\ & \times & \\ \times & & 
\end{pmatrix}.
\end{eqnarray}

\item Four GJ factors:
\begin{eqnarray}
\begin{pmatrix}
 & \times & \\ \times & \times & \\ & \times & 
\end{pmatrix}, \quad \quad
\begin{pmatrix}
\times &  & \times \\  & \times & \\ & \times & 
\end{pmatrix}, \quad \quad
\begin{pmatrix}
\times & & \times \\ \times & \times & \\ & & 
\end{pmatrix}, \nonumber\\
\begin{pmatrix}
 & & \\ & \times & \times \\ \times & \times & 
\end{pmatrix}, \quad \quad
\begin{pmatrix}
\times &  & \\   \times & \times & \\ \times & & 
\end{pmatrix}, \quad \quad
\begin{pmatrix}
\times & \times &  \\ & \times & \\ \times & & 
\end{pmatrix}, \nonumber \\
\begin{pmatrix}
 & \times & \\ \times & \times & \times \\ &  & 
\end{pmatrix}, \quad \quad 
\begin{pmatrix}
\times &  & \\  & \times & \\ \times & \times & 
\end{pmatrix}, \quad \quad
\begin{pmatrix}
\times & & \times \\  & \times & \times \\ & & 
\end{pmatrix}, \nonumber \\
\begin{pmatrix}
\times & &  \\  & \times & \times \\ \times & &  
\end{pmatrix}. ~~~~~~~~~~~~~~~~~~~~~~~~~~~~~~~~
~~~~~~~~~~~~~~~~~~~~~
\end{eqnarray}

\item Five GJ factors:
\begin{eqnarray}
\begin{pmatrix}
\times & \times & \\  & \times & \times \\ \times & & 
\end{pmatrix}, \quad \quad
\begin{pmatrix}
\times &  & \times \\ \times & \times & \\ & \times & 
\end{pmatrix}, \quad \quad
\begin{pmatrix}
\times & &  \\ \times & \times & \times \\  \times & & 
\end{pmatrix}, \nonumber\\
\begin{pmatrix}
\times & \times & \\ & \times &  \\ \times & \times & 
\end{pmatrix}, \quad \quad
\begin{pmatrix}
\times & \times & \times \\ \times & \times &  \\ & & 
\end{pmatrix}, \quad \quad
\begin{pmatrix}
\times &  & \times  \\ & \times & \times \\ & \times &  
\end{pmatrix}, \nonumber \\
\begin{pmatrix}
\times &  & \times \\ \times & \times & \times \\  & & 
\end{pmatrix}, \quad \quad
\begin{pmatrix}
\times & &  \\ \times & \times &  \\ \times & \times & 
\end{pmatrix}, \quad \quad
\begin{pmatrix}
\times &  &   \\ & \times & \times \\ \times & \times &  
\end{pmatrix}, \nonumber \\
\begin{pmatrix}
 & \times &  \\ \times & \times & \times  \\ & \times & 
\end{pmatrix}, \quad \quad
\begin{pmatrix}
\times & \times & \times \\ & \times & \times \\  &  & 
\end{pmatrix}, \quad \quad
\begin{pmatrix}
 & \times &  \\ & \times & \times \\  \times & \times  & 
\end{pmatrix}.
\end{eqnarray}

\item Six GJ factors:
\begin{eqnarray}
\begin{pmatrix}
\times & \times & \\  & \times & \times \\ \times & \times & 
\end{pmatrix}, \quad \quad
\begin{pmatrix}
\times & \times  & \times \\ \times & \times & \\ & \times & 
\end{pmatrix}, \nonumber \\
\begin{pmatrix}
\times & \times & \times \\  & \times & \times  \\  & \times & 
\end{pmatrix}, \quad \quad
\begin{pmatrix}
\times & & \times  \\ \times & \times & \times  \\ & \times & 
\end{pmatrix}.
\end{eqnarray}

\end{itemize}

\item When $\beta_{12}=0$, there are solutions with only two and three GJ factors.

\begin{itemize}

\item Two GJ factors:
\begin{eqnarray}
\begin{pmatrix}
 &  & \\ \times  & & \times \\ & & 
\end{pmatrix}, \quad \quad
\begin{pmatrix}
 &  &  \\ & \times & \times \\ & & 
\end{pmatrix}, \quad \quad
\begin{pmatrix}
 & &  \\ \times & \times & \\ & &  
\end{pmatrix}.
\end{eqnarray}

\item Three GJ factors:
\begin{eqnarray}
\begin{pmatrix}
 &  & \\   \times & \times & \\ & \times & 
\end{pmatrix}, \quad \quad
\begin{pmatrix}
 &  &  \\  & \times & \times \\ & \times & 
\end{pmatrix},\\ 
\begin{pmatrix}
 &  &  \\  \times & & \times \\ & & \times
\end{pmatrix}, \quad \quad
\begin{pmatrix}
 &  &  \\  & \times & \times \\ & & \times
\end{pmatrix}.
\end{eqnarray}

\end{itemize}

\end{itemize}

Among these patterns, two are noteworthy. The first one is for $\beta_{23}=0$, with one GJ factor in the (22) element of $Y^{(-1/3)}$. Surprisingly, the placement of the GJ factor coincides with the original realization of the GJ relations, but this time with a compatible small value of $\theta_{13}$ .  The second one has two GJ factors in off-diagonal entries, and $\beta_{12}=0$. 

For these reasons, we provide numerical details for these two cases.

\begin{itemize}

\item {\bf Example III}: (seesaw) TBM mixing, (+,+), $A=0.77$, and
\begin{eqnarray}
\beta_{12}=168^\circ, \quad \quad \beta_{13}=352^\circ, \quad \quad \beta_{23}=0^\circ,
\end{eqnarray}
with one GJ factor in the (22) element of $Y^{(-1/3)}$, yields,
\begin{eqnarray}
\theta_{12}=32.6^\circ, \quad \theta_{23}=45.1^\circ, \quad \theta_{13}=8.68^\circ, ~~~~~~~~~~~~\nonumber \\
\quad \frac{m_e}{m_\mu}=0.0054, \quad \frac{m_\mu}{m_\tau}=0.049, \quad J_{\rm MNSP}=9.6\times 10^{-8}.
\end{eqnarray}
We note that for this example, better agreement with the best fit values of $\theta_{12}$ and $\theta_{13}$ is achieved. Only $\theta_{23}$, the mixing angle with the largest uncertainty, is outside the 1$\sigma$ range. 

\vskip 0.3cm
\noindent {\bf Analysis}
\vskip 0.2cm

Similar to the one-angle solution, we want to derive analytical formulae for the charged lepton mass ratios and neutrino mixing angles in terms of the input angles.  With $\beta_{12}$ and $\beta_{13}$ close to the x-axis, one can parametrize $\beta_{12}=\pi-C\lambda$ and $\beta_{13}=B\lambda$, and write $\m V$ as

\begin{eqnarray}
\m V = \begin{pmatrix}
1-\frac{B^2\lambda^2}{2} & 0 & B\lambda \\
0 & 1 & 0 \\
-B\lambda & 0 & 1-\frac{B^2\lambda^2}{2} \\
\end{pmatrix} \begin{pmatrix}
-1+\frac{C^2\lambda^2}{2} & C\lambda & 0 \\
-C\lambda & -1+\frac{C^2\lambda^2}{2} & 0 \\
0 & 0 & 1
\end{pmatrix}+\m O(\lambda^3),
\end{eqnarray}
where numerically $C = 0.92$ and $B= -0.62$. The down-quark and charged lepton Yukawa matrices $Y^{(-1/3)}$ and $Y^{(-1)}$ follow, and diagonalizing $Y^{(-1)}$ yields the mass ratios of the charged leptons,

\begin{eqnarray}
m_\mu /m_\tau &=& \lambda^2 + \m O(\lambda^4), \nonumber \\
m_e/m_\mu & =& (\frac{4}{9}C-\frac{1}{3})\lambda^2 + \m O(\lambda^4)\approx \frac{1}{9}\lambda^2,
\end{eqnarray}
coincidentally the Georgi-Jarlskog value. The mixing angles in $\m U_{-1}$ can also be found, and one can derive the following neutrino mixing angles in the MNSP matrix by using Eq.(\ref{eq:MNSP}),

\begin{eqnarray}
\theta_{23} &= & \frac{\pi}{4}+\m O(\lambda^4) , \nonumber \\
\theta_{13} &= &\frac{(C-3B)\lambda}{3\sqrt{2}}+\m O(\lambda^3), \nonumber \\
\theta_{12} &= & \arctan\frac{1}{\sqrt{2}} - \delta\theta_{12} , 
\end{eqnarray}
where
\begin{eqnarray}
\delta\theta_{12} =- \frac{(C+3B)\lambda}{3\sqrt{2}}+ \m O(\lambda^3), 
\end{eqnarray}
corresponding to

\begin{equation}
\theta_{23}~\approx 45^\circ,~~\theta_{13}~ \approx 8.5^\circ,
~~\theta_{12}~\approx 32.4^\circ.
\end{equation}

$m_e/m_\mu$ and $\theta_{13}$ are no longer correlated, and only one sum rule can be derived among $m_e/m_\mu$, $\theta_{13}$ and $\theta_{12}$,

\begin{eqnarray}
\frac{m_e}{m_\mu} \approx -\frac{\lambda^2}{3} + \frac{2\sqrt{2}\lambda}{3} (\theta_{12}+\theta_{13}-\arctan\frac{1}{\sqrt{2}}).
\end{eqnarray}
One also notices that the order $\m O (\lambda^2)$ terms in $m_e/m_\mu$ are not fully cancelled by taking $\theta_{13} \approx \lambda/\sqrt{2}$. Instead, an order $\lambda^2$ term, $\sim \lambda^2/9$, is left over, and causes $m_e/m_\mu$ to fit within the allowed range. This way of obtaining the correct value for $m_e/m_\mu$ is quite distinct from what we found in the one-angle solution, where the correct value for $m_e/m_\mu$ stems from  an $\m O (\lambda^3)$ term.

\item {\bf Example IV}: (seesaw) TBM mixing, (+,+), $A=0.77$, and 
\begin{eqnarray}
\beta_{12}=0^\circ, \quad \quad \beta_{13}=3^\circ, \quad \quad \beta_{23}=90^\circ,
\end{eqnarray}
with two GJ factors in the (21) and (23) element of $Y^{(-1/3)}$, yields,
\begin{eqnarray}
\theta_{12}=31.2^\circ, \quad \theta_{23}=44.6^\circ, \quad \theta_{13}=8.29^\circ, ~~~~~~~~~~~\nonumber \\
 \frac{m_e}{m_\mu}=0.0051, \quad \frac{m_\mu}{m_\tau}=0.051, \quad J_{\rm MNSP}=3.6\times 10^{-7}.
\end{eqnarray}
This output is exactly the same as the one angle case with $\beta_{13}=183^\circ$ and with two GJ factors in the (21) and (22) elements of $Y^{(-1/3)}$ (see Eq.(\ref{eq:sol1})). This is because $\beta_{23}$, whose value is $90^\circ$, plays the role of switching the second and third column of $Y^{(-1/3)}$. It is also the reason why one can have a GJ factor in the (33) element of $Y^{(-1/3)}$ for some patterns in the case of $\beta_{12}=0$.
\end{itemize}

\begin{figure}[h!]
 \centering
\scalebox{0.5}{\includegraphics*{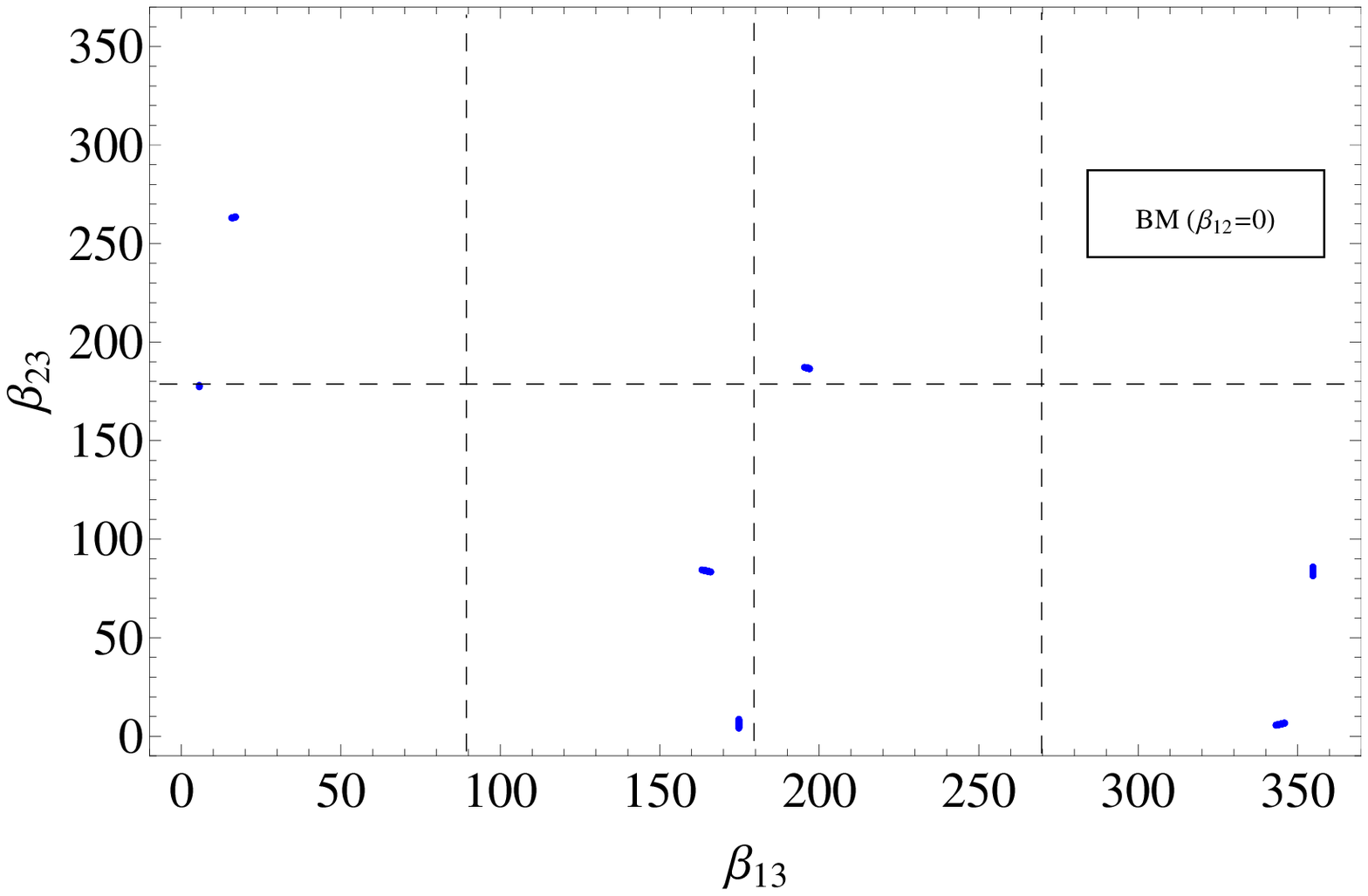}} \scalebox{0.5}{\includegraphics*{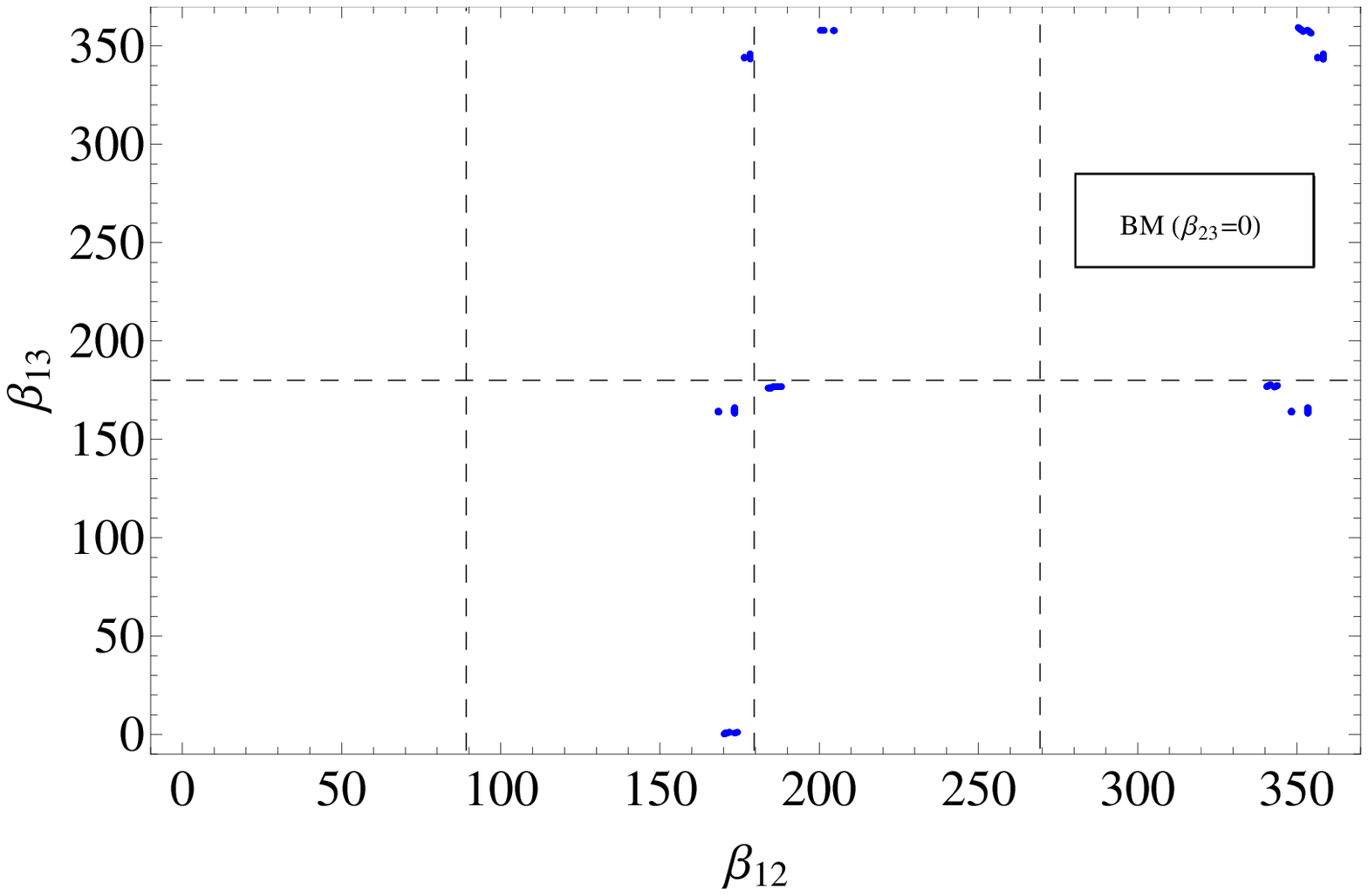}} \scalebox{0.5}{\includegraphics*{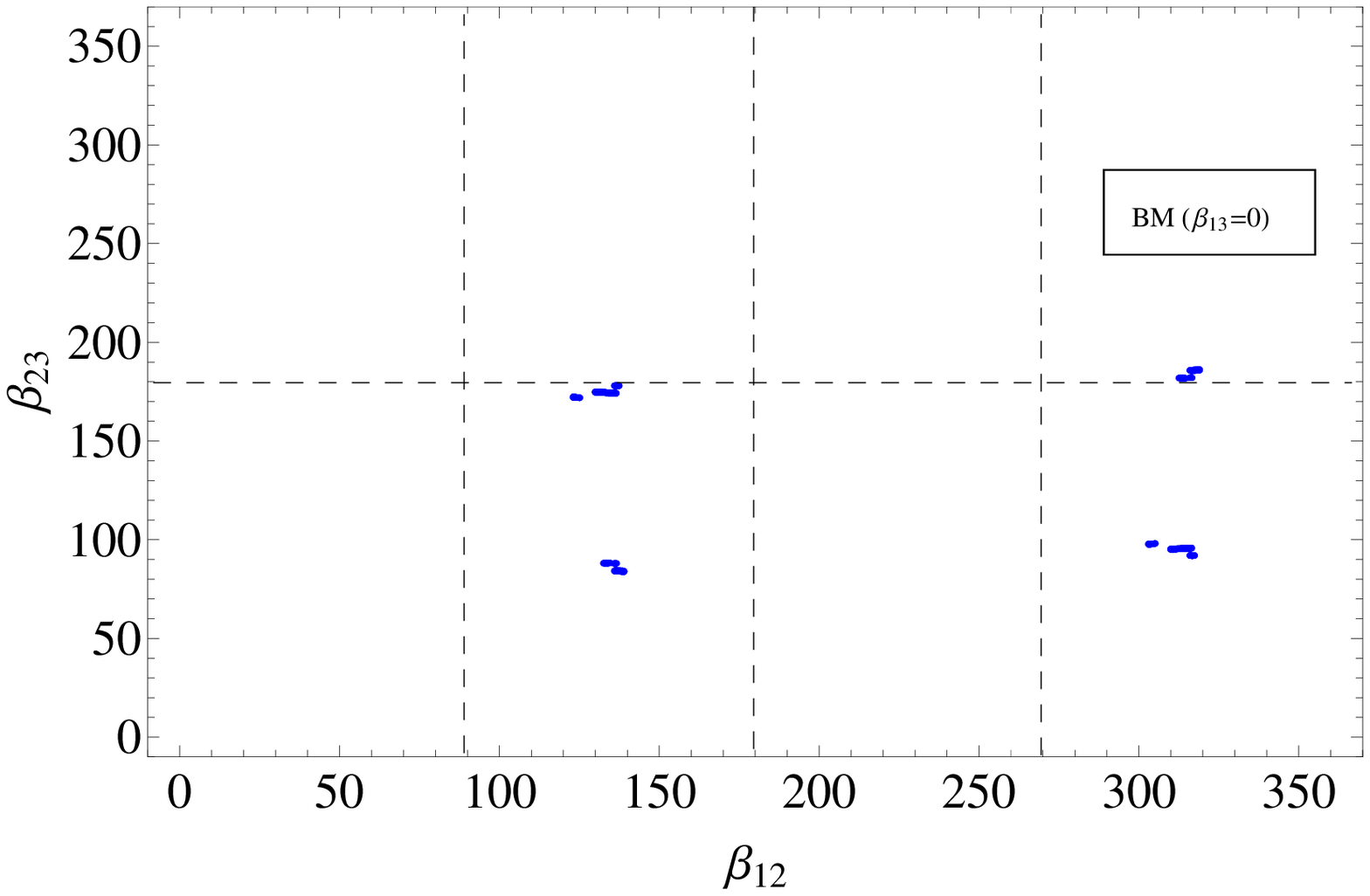}} 
\caption{Allowed parameter space for BM mixing with two non-zero input angles $\beta_{ij}$. }
\label{fg:BM2angle}
\end{figure}

\vskip .3cm
\noindent{\bf Bimaximal Seesaw Mixing}
\vskip .3cm
\noindent For the Bimaximal case, the patterns of two-angle solutions are quite different, and come in three types with  $\beta_{23}=0$, $\beta_{12}=0$, and $\beta_{13}=0$, ($A=0.77$). Likewise the patterns of GJ insertions are distinct from those of TBM mixing. We plot the parameter space for all three  cases in Fig.\ref{fg:BM2angle}.  A complete detailed list of these solutions  and their corresponding GJ patterns can be found in the tables of Appendix A. 

As for TBM mixing, we find the parameter space for two angles to be quite restricted, with input angles $\beta_{ij}$  near an axis, although we note several solutions with $\sin\beta_{ij}$ significantly larger and off-axis.

From this set of solutions we give two interesting examples.

\begin{itemize}

\item {\bf Example V}: (seesaw) BM mixing, (+,+), $A=0.77$, and
\begin{eqnarray}
\beta_{12}=173^\circ, \quad \quad \beta_{13}=165^\circ, \quad \quad \beta_{23}=0^\circ,
\end{eqnarray}
with one GJ factor in the (22) element of $Y^{(-1/3)}$, yields,
\begin{eqnarray}
\theta_{12}=32.4^\circ, \quad \theta_{23}=46.1^\circ, \quad \theta_{13}=8.67^\circ, ~~~~~~~~~\nonumber \\
 \frac{m_e}{m_\mu}=0.0040, \quad \frac{m_\mu}{m_\tau}=0.050, \quad J_{\rm MNSP} = 9.9 \times 10^{-7}.
\end{eqnarray}

The analysis of this example follows closely that of Example III, with similar conclusions about $m_e/m_\mu$ and $\theta_{13}$.

\item {\bf Example VI}: (seesaw) BM mixing, (+,+), $A=0.77$, and 

\begin{eqnarray}
\beta_{12}=314.3^\circ, \quad \quad \beta_{13}=0^\circ, \quad \quad \beta_{23}=181.8^\circ,
\end{eqnarray}
where $\beta_{12}$ is not near an axis, with four GJ factors in the (12), (22), (23) and (32) matrix elements of $Y^{(-1/3)}$. It yields,
\begin{eqnarray}
\theta_{12}=34.3^\circ, \quad \theta_{23}=39.2^\circ, \quad \theta_{13}=8.70^\circ, ~~~~~~~~~\nonumber \\
 \frac{m_e}{m_\mu}=0.0049, \quad \frac{m_\mu}{m_\tau}=0.051, \quad J_{\rm MNSP}=2.6 \times 10^{-6}.
\end{eqnarray}
This solution is noteworthy because all mixing angles now fall within the 1$\sigma$ range, remarkably close to their best fit values. 
We may contrast it with TBM mixing, for which no such solutions exist. 

\vskip 0.3cm
\noindent {\bf Analysis}
\vskip 0.2cm 
To investigate this solution analytically, we first parametrize the two input angles in terms of $\lambda$ as,

\begin{eqnarray}
\beta_{12} = -\frac{\pi}{4}-C \lambda^3, \qquad \beta_{23} = \pi+D\lambda^2,
\end{eqnarray}
trading them for two order-one parameters, $C = 1.04$ and $D = 0.61$. As before,

$$
\frac{m_\mu}{m_\tau} = \lambda^2 + \m O(\lambda^4). 
$$
Similarly,

\be
m_e/m_\mu = -\frac{\lambda ^2}{3}+2 \sqrt{2} A D \lambda ^3-2 \sqrt{2} A D \lambda ^4 + \m O(\lambda^5).
\ee
The $\m O (\lambda^4)$ term has to be kept because, with $D=0.61$ and $A=0.77$  the first two terms in this expansion nearly cancel, in an apparent numerical coincidence.  We also have 

\be
\theta_{23} = \frac{\pi}{4} - 3D\lambda^2 + \m O(\lambda^4),\ee
but it does not appear easy to derive analytical  expressions for the other two MNSP mixing angles $\theta_{12}$ and $\theta_{13}$, although by numerical methods, we find

\be
\theta_{13} = \frac{1}{6} + \m O(\lambda^2), \qquad
\theta_{12} = \arcsin \frac{5}{6\sqrt{2}}+ \m O(\lambda^2).
\ee

\end{itemize}

Finally, we come to the most computationally challenging case, where all three angles in $\m V$ are non-zero.

\subsection{Three-Angle Solutions}

As we increase the number of input angles to three, the number of solutions increases, as does the time required to scan the full parameter space. Luckily, we may use analytical arguments to guide our search, and restrict the scan to specific regions of the $\beta_{ij}$ within their full $(0,2 \pi)$ range. 

That such an analysis is possible follows from the fact that the TBM and BM mixing angles are reasonably close to the best fit values of the MNSP mixing angles. Aside from subtleties involving the quadrants of the angles (see Appendix D), this allows us to conclude that the mixing angles $\theta_{ij}^e$ in $\m U_{-1}$ need to be close to one of the axes.

Knowing the required size of the corrections, one may expand Eq.(\ref{eq:MNSP}) in terms of the small mixing angles of $\m U_{-1}$, obtaining bounds on their magnitudes.  

Translating them to bounds on the mixing angles in $\m V$ requires more care, as going from $Y^{(-1/3)}$ to $Y^{(-1)}$ involves both a transposition and modification by GJ factors. The complete details on how these bounds are obtained may be found in Section \ref{sec:analysis}. 

To summarize our results, we find that to obtain suitable corrections, $\beta_{23}$ must be at most $10^\circ$ away from the axes for both TBM and BM mixing, while $\beta_{13}$  can be $10^\circ$ ($20^\circ$) for TBM (BM); $\beta_{12}$ however is unrestricted, and can assume its full $(0,2\pi)$ range.    

Following the two-angle case, we present our results in table form at the end of  Appendix A. With the addition of the third input angle, we find that while the number of 1$\sigma$ solutions increases, they are surprisingly still quite rare. 

\begin{figure}[h!]
 \centering
\scalebox{0.60}{\includegraphics*{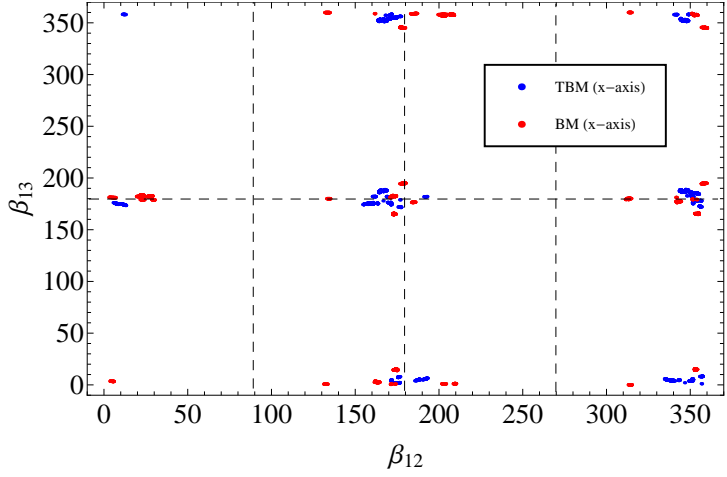}} 
\caption{A plot of the allowed values of $\beta_{13}$ and $\beta_{12}$ for $\beta_{23}$ close to the x-axis for both TBM and BM mixing.}
\label{fg:TBM_vs_BM_x}
\end{figure}

\begin{figure}[h!]
 \centering
\scalebox{0.60}{\includegraphics*{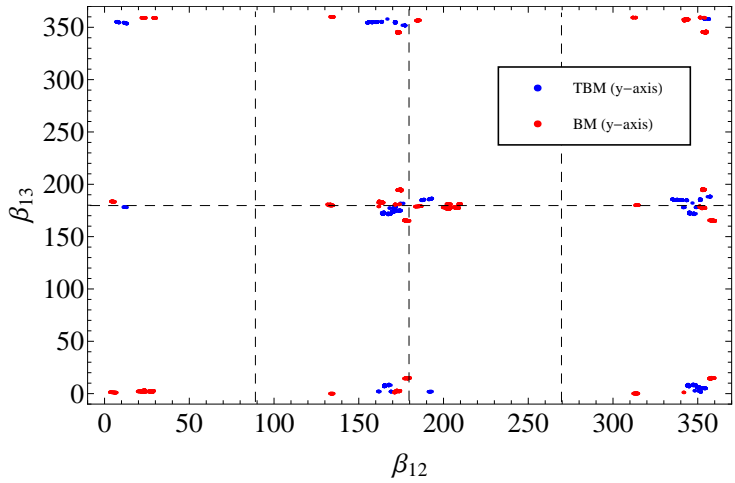}} 
\caption{A plot of the allowed values of $\beta_{13}$ and $\beta_{12}$ for $\beta_{23}$ close to the y-axis for both TBM and BM mixing.}
\label{fg:TBM_vs_BM_y}
\end{figure}

To get a feel for how the parameter space looks for three angles, we plot the allowed regions of $\beta_{12}$ and $\beta_{13}$ for both TBM and BM mixing in Figs.\ref{fg:TBM_vs_BM_x}-\ref{fg:TBM_vs_BM_y}; we distinguish the two cases by whether $\beta_{23}$ is close to the x- or y-axes.

As in the two-angle case,  the number of solutions is quite small, tending to cluster around specific regions close to one of the axes. We also find that the BM solutions tend to cover a wider spread of input angles, similar to the two-angle solutions.  

To close, we give an example of a solution with three angles that produces outputs in good agreements with the best fit values.

\begin{itemize}

\item {\bf Example VII}: (seesaw) TBM mixing, (+,+), $A=0.77$, and
\begin{eqnarray}
\beta_{12}=169.7^\circ, \quad \quad \beta_{13}=354.0^\circ, \quad \quad \beta_{23}=176.5^\circ,
\end{eqnarray}
with two GJ factor in the (21) and (22) elements of $Y^{(-1/3)}$, yields,
\begin{eqnarray}
\theta_{12}=34.4^\circ, \quad \theta_{23}=41.3^\circ, \quad \theta_{13}=8.73^\circ, ~~~~~~~~~\nonumber \\
 \frac{m_e}{m_\mu}=0.0050, \quad \frac{m_\mu}{m_\tau}=0.059, \quad J_{\rm MNSP} = 1.1 \times 10^{-8}.
\end{eqnarray}
\end{itemize}


\section{Analysis of the Results}
\label{sec:analysis}

Although our solutions were found numerically, we would like to try and understand them analytically so as to answer two questions: How are the particular mixing angles in $\m V$ singled out? What are the effects of the insertions of GJ factors? We investigate these questions in two steps: first, find out what the needed mixing angles in $\m U_{-1}$ are, given a particular form of $\m U_{\rm seesaw}$; second, connect them with the input angles in $\m V$ through the influences of GJ factors.

\subsection*{Mixing angles in $\m U_{-1}$}

We start with a useful parametrization of $\m U_{-1}$ and $\m U_{\rm MNSP}$ that may be used to calculate the deviations of the mixing angles and phase in $\m U_{\rm MNSP}$ from their values in $\m U_{\rm seesaw}$. Following \cite{King}, an Iwasawa decomposition of the unitary matrices $\m U_{-1}$ and $\m U_{\rm seesaw}$ is given by, 

\begin{eqnarray} \label{eq:parametrization}
\m U_{-1} &=& P_L^e~ R_{23}^e~ U_{13}^{e\prime} ~R_{12}^e~ P_R^e, \nonumber \\
\m U_{\rm seesaw} &=& P_L^\nu~ R_{23}^\nu~ U_{13}^{\nu\prime} ~ R_{12}^\nu~ P_R^\nu, 
\end{eqnarray}  
where $P_L$ and $P_R$ are phase matrices, $R_{ij}$ is a rotation matrix, and $U_{ij}$ are unitary matrices defined by,

\begin{eqnarray}
P_L &=& \begin{pmatrix}
e^{i\varphi_1} & & \\ & e^{i\varphi_2} & \\ & & e^{i\varphi_3}
\end{pmatrix}, \quad \quad P_R = \begin{pmatrix}
e^{i\beta_1} & & \\ & e^{i\beta_2} & \\ & & 1
\end{pmatrix}, \nonumber \\
R_{12} &=& \begin{pmatrix}
c_{12} & s_{12} & 0 \\ -s_{12} & c_{12} & 0 \\ 0 & 0 & 1
\end{pmatrix}, \quad U_{12} = \begin{pmatrix}
c_{12} & s_{12} e^{-i \delta_{12}} & 0 \\ -s_{12}e^{i \delta_{12}} & c_{12} & 0 \\ 0 & 0 & 1
\end{pmatrix}, \nonumber  \\
R_{13} &=& \begin{pmatrix}
c_{13} & 0 & s_{13} \\ 0 & 1 & 0 \\ -s_{13} & 0 & c_{13}
\end{pmatrix}, \quad U_{13} = \begin{pmatrix}
c_{13} & 0 & s_{13} e^{-i \delta_{13}} \\ 0 & 1 & 0 \\ -s_{13}  e^{i \delta_{13}} & 0 & c_{13}
\end{pmatrix}, \nonumber 
\\
R_{23} &=& \begin{pmatrix}
1 & 0 & 0 \\ 0 & c_{23} & s_{23} \\ 0 & -s_{23} & c_{23}
\end{pmatrix}, \quad U_{23} = \begin{pmatrix}
1 & 0 & 0 \\ 0 & c_{23} & s_{23} e^{-i \delta_{23}} \\ 0 & -s_{23} e^{i \delta_{23}} & c_{23}
\end{pmatrix}. \nonumber 
\end{eqnarray} 
Inserting this parametrization into Eq.(\ref{eq:MNSP}), the MNSP matrix may then be written as, 

\begin{eqnarray}
\label{eq:MNSP_Product}
\m U_{\rm MNSP} &=& \m U_{-1}^\dagger \m U_{\rm seesaw} \nonumber \\
&=& (P_R^{e\dagger} P_L^{e\dagger} P_L^\nu P_R^\nu) ~ U_{12}^{e\dagger} U_{13}^{e\dagger} U_{23}^{e\dagger} U_{23}^\nu U_{13}^\nu U_{12}^\nu,
\end{eqnarray}  
by commuting the various phase matrices to the left. This shift of the phase matrices to the left has the effect of both placing phases in the original rotation matrices of Eq.(\ref{eq:parametrization}) and also of shifting the phases from their original values in $\m U_{ij}^\prime$. The relations between the old phases of Eq.(\ref{eq:parametrization}) and those of Eq.(\ref{eq:MNSP_Product}) are given by,

\begin{eqnarray}
\delta_{12}^\nu &=& \beta_1^\nu - \beta_2^\nu, \nonumber \\
\delta_{13}^\nu &=& \delta_{13}^{\nu\prime} + \beta_1^\nu, \nonumber \\
\delta_{23}^\nu &=& \beta_2^\nu, \nonumber \\
\delta_{12}^e &=& [(\varphi_2^\nu + \beta_2^\nu)-\varphi_2^e] - [ (\varphi_1^\nu + \beta_1^\nu) -\varphi_1^e], \nonumber \\
\delta_{13}^e &=& \delta_{13}^{e\prime} - (\varphi_3^\nu -\varphi_3^e)+ [(\varphi_1^\nu+\beta_1^\nu)-\varphi_1^e], \nonumber \\
\delta_{23}^e &=& (\varphi_3^\nu-\varphi_3^e) -[(\varphi_2^\nu+\beta_2^\nu)-\varphi_2^e]. \nonumber
\end{eqnarray}

Similarly, we decompose the MNSP matrix as,  

\begin{eqnarray}
\m U_{\rm MNSP} = P_L U_{23} U_{13} U_{12}.
\end{eqnarray}
The relations between the mixing angles in the MNSP matrix and those in $\m U_{-1}$ and $\m U_{\rm seesaw}$ can then be obtained by studying the expansion of the right hand side of Eq.(\ref{eq:MNSP_Product}) in terms of small angles, if any. 

In our case, fortunately, most angles are small, because the seesaw matrices are designed to account for the large angles in the data.
Thus the expansion parameters are $\theta_{13}^\nu$ and the $\theta_{ij}^e$, although the search results indicate that $\theta_{23}^e$ can sometimes be quite close to the y-axis. We therefore consider these two distinct cases separately:

\begin{itemize}

\item $|s_{13}^\nu|, |s_{12}^e|, |s_{13}^e|, |s_{23}^e| \ll 1$:

The first order relations between the mixing angles in the MNSP matrix and those in $\m U_{-1}$ and $\m U_{\rm seesaw}$ simplify to, 

\begin{eqnarray}
\label{eq:relations}
s_{23}e^{-i \delta_{23}} &\approx & s_{23}^\nu e^{-i \delta_{23}^\nu} -\theta_{23}^e c_{23}^\nu e^{-i \delta_{23}^e}, \nonumber \\
\theta_{13} e^{-i \delta_{13}} &\approx & \theta_{13}^\nu e^{-i \delta_{13}^\nu} - \theta_{13}^e c_{23}^\nu e^{-i \delta_{13}^e} - \theta_{12}^e s_{23}^\nu e^{-i(\delta_{23}^\nu-\delta_{12}^e)}, \nonumber \\
s_{12}e^{-i\delta_{12}} &\approx & s_{12}^\nu e^{-i \delta_{12}^\nu} + \theta_{13}^e c_{12}^\nu s_{23}^\nu e^{i (\delta_{23}^\nu - \delta_{13}^e)} - \theta_{12}^e c_{23}^\nu c_{12}^\nu e^{-i \delta_{12}^e}, 
\end{eqnarray}
with $\delta^{\rm MNSP} = \delta_{13}-\delta_{12}-\delta_{23}$. The above formulae will allow us to easily identify the sources of corrections to the mixing angles in $\m U_{\rm seesaw}$.

Further simplifications occur when restricting ourselves to TBM- and BM-compatible seesaw matrices of the form Eq.(\ref{eq:Useesaw}), whose Iwasawa decomposition gives
\begin{eqnarray}
\theta_{13}^\nu =0, \quad \theta_{23}^\nu = 45^\circ, \nonumber \\
\varphi_1^\nu =0, \quad \varphi_2^\nu = 180^\circ, \quad \varphi_3^\nu =0, \nonumber \\
\beta_1^\nu =0, \quad \beta_2^\nu =180^\circ.  
\end{eqnarray}

The relations in Eq.(\ref{eq:relations}) reduce to

\begin{eqnarray}
-s_{23}e^{-i\delta_{23}} &= & \sin \left(\frac{\pi}{4} + \delta \theta_{23} \right), \nonumber \\
\theta_{13}e^{-i\delta_{13}} &= & \delta \theta_{13}, \nonumber \\
-s_{12}e^{-i\delta_{12}} &= & \sin \left( \theta_{12}^\nu + \delta \theta_{12} \right),
\end{eqnarray}
where
\begin{eqnarray}
\label{eq:corrections}
\delta \theta_{23} &\approx & \theta_{23}^e e^{-i\delta_{23}^e}, \nonumber \\
\delta \theta_{13} &\approx & \frac{1}{\sqrt{2}} \left( \theta_{12}^e e^{-i \delta_{12}^e} - \theta_{13}^e e^{-i \delta_{13}^e} \right), \nonumber \\
\delta \theta_{12} &\approx & \frac{1}{\sqrt{2}}\left( \theta_{12}^e e^{-i \delta_{12}^e} + \theta_{13}^e e^{-i \delta_{13}^e} \right).
\end{eqnarray}

Because of the phases, these corrections $\delta \theta_{ij}$ are in general complex. Our  search assumes no phases in both $\m V$ and $U_{\rm seesaw}$, so that the $\delta \theta_{ij}$ are approximately real,\footnote{Small non-zero phases can be generated for two reasons: the CKM matrix is only unitary to the order of $\lambda^3$; the assignment of GJ factors.} and serve as corrections to the seesaw mixing angles. 

The required corrections to the seesaw mixing angles can be computed from their global fits \cite{fit}. For the TBM mixing ($\theta_{12}^\nu \approx 35.3^\circ$) and using the $3\sigma$ range of neutrino mixing angles, the range of the required corrections are

\begin{eqnarray}
\delta\theta_{23} \sim (-9^\circ\leftrightarrow 8^\circ), \quad |\delta\theta_{13}| \sim (8^\circ\leftrightarrow 10^\circ), \quad \delta \theta_{12}^{TBM} \sim (-4.8^\circ\leftrightarrow 1.7^\circ),
\end{eqnarray}
while for BM mixing ($\theta_{12}^\nu = 45^\circ$) the required $\delta \theta_{12}$ is instead 
\begin{eqnarray}
\delta \theta_{12}^{BM} \sim (8^\circ, 14.5^\circ),
\end{eqnarray}
much larger than that for TBM mixing.

These corrections can be further translated into the mixing angles $\theta_{ij}^e$ in $\m U_{-1}$ by using Eq.(\ref{eq:corrections}). Taking into account possible negative signs generated by the phases $\delta_{ij}^e$, the  magnitude range of $\theta_{23}^e$ is given by

\begin{eqnarray}
\label{eq:theta23e}
|\theta_{23}^e| \sim (0^\circ \leftrightarrow 9^\circ), 
\end{eqnarray}
and the allowed regions for the magnitudes of $\theta_{12}^e$ and $\theta_{13}^e$ are given in Fig.\ref{fig:corrections}. 

\begin{figure}[h!]
 \centering
\scalebox{0.8}{\includegraphics*{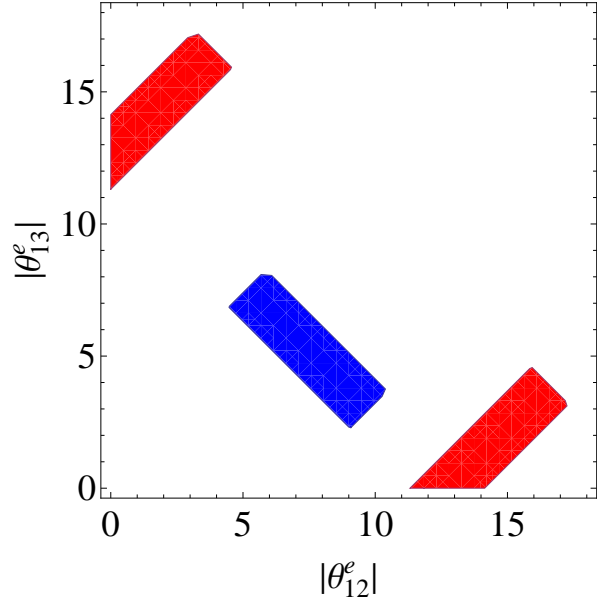}}  
\caption{Allowed regions for the magnitudes of $\theta_{12}^e$ and $\theta_{13}^e$. The blue and red regions correspond to the TBM and BM mixings respectively.}
 \label{fig:corrections}
\end{figure}

\begin{figure}[h!] 
 \centering
\scalebox{0.5}{\includegraphics*{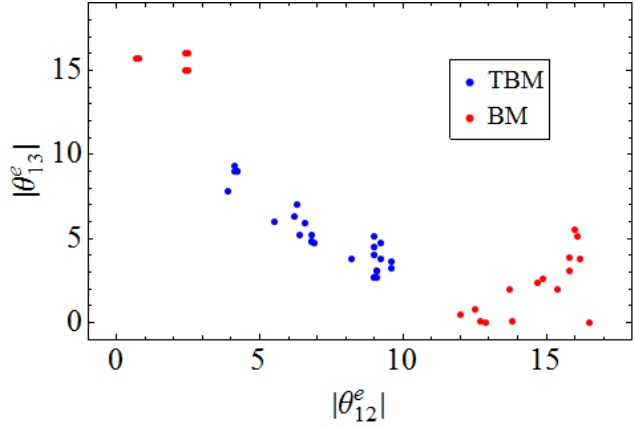}} 
\caption{A scatter plot of the magnitudes of $\theta_{12}^e$ and $\theta_{13}^e$ in the two-angle solutions.}
\label{fig:soln}
\end{figure}

In Fig.\ref{fig:corrections}, one notices that the regions for the TBM and BM mixings are disconnected. For the TBM case, the magnitudes of $\theta_{12}^e$ and $\theta_{13}^e$ tend to be close to each other, while they are quite apart for the BM case. This feature is indeed confirmed by the two-angle solutions, as can be seen in Fig.\ref{fig:soln}.

\item $|s_{13}^\nu|, |s_{12}^e|, |s_{13}^e|, |c_{23}^e| \ll 1$:

In this case, at the leading order the MNSP mixing angles can be written as

\begin{eqnarray}
s_{23} e^{-i\delta_{23}} &\approx & -c_{23}^\nu e^{-i\delta_{23}^e} + s_{23}^\nu c_{23}^e e^{-i\delta_{23}^\nu}, \nonumber \\
\theta_{13} e^{-i\delta_{13}} &\approx & \theta_{13}^\nu e^{-i\delta_{13}^\nu} + c_{23}^\nu \theta_{12}^e e^{-i (\delta_{12}^e + \delta_{23}^e)} - s_{23}^\nu \theta_{13}^e e^{-i (\delta_{23}^\nu + \delta_{13}^e -\delta_{23}^e)}, \nonumber \\
s_{12} e^{-i \delta_{12}} &\approx & s_{12}^\nu e^{-i \delta_{12}^\nu} - c_{12}^\nu s_{23}^\nu \theta_{12}^e e^{i(\delta_{23}^\nu - \delta_{12}^e - \delta_{23}^e)} - c_{12}^\nu c_{23}^\nu \theta_{13}^e e^{-i(\delta_{13}^e - \delta_{23}^e)}, \nonumber \\
& &
\end{eqnarray}

Applying the above general formulae to the previous form of $\m U_{\rm seesaw}$, one arrives at 
\begin{eqnarray}
-s_{23}e^{-i(\delta_{23}-\delta_{23}^e)} &= &  \sin \left(\frac{\pi}{4} + \delta \theta_{23} \right), \nonumber \\
\theta_{13}e^{-i\delta_{13}} &= & \delta\theta_{13}, \nonumber \\
-s_{12}e^{-i\delta_{12}} &= & \sin \left( \theta_{12}^\nu + \delta \theta_{12} \right),
\end{eqnarray}
where
\begin{eqnarray}
\label{eq:corrections90}
\delta \theta_{23} &\approx & c_{23}^e e^{i\delta_{23}^e}, \nonumber \\
\delta \theta_{13} &\approx & \frac{1}{\sqrt{2}} \left( \theta_{12}^e e^{-i (\delta_{12}^e+\delta_{23}^e)} - \theta_{13}^e e^{-i (\delta_{13}^e - \delta_{23}^e)} \right), \nonumber \\
\delta \theta_{12} &\approx & -\frac{1}{\sqrt{2}}\left( \theta_{12}^e e^{-i (\delta_{12}^e+\delta_{23}^e)} + \theta_{13}^e e^{-i (\delta_{13}^e - \delta_{23}^e)} \right).
\end{eqnarray}

The structure of the above corrections is quite similar to the previous case, except for different phases and the replacement of $\theta_{23}^e$ by $c_{23}^e$. As these phases are not relevant to our search, one can perform a similar analysis on the $\theta_{ij}^e$, and eventually we find the above formulae agree with the two-angle results quite well.

\end{itemize}

Having established the required mixing angles in $\m U_{-1}$, can they be obtained from $Y^{(-1/3)}$ by using $SU(5)$?

\subsection*{From $Y^{(-1/3)}$ to $Y^{(-1)}$}

In the search, $Y^{(-1)}$ is obtained by first transposing $Y^{(-1/3)}$ with GJ factors in some entries. As a result the mixing angles $\beta_{ij}$ in $\m V$ can differ from those $\theta_{ij}^e$ in $\m U_{-1}$. Their relations will be the focus of this section. The mass ratios of the charged leptons are also affected by the GJ factors, but they are harder to investigate analytically.

Although in the solutions the $\beta_{ij}$ can appear in all four quadrants, we consider an example where they are all in the first or fourth quadrant and close to the x-axis. The other cases can be studied similarly with possibly different signs and $\sin\beta_{ij}$ replaced by $\cos \beta_{ij}$. 

$\m V$ is parametrized as

\begin{eqnarray}
\m V \approx \begin{pmatrix}
1 & 0 & 0 \\
0 & 1 & \beta_{23} \\
0 & -\beta_{23} & 1 
\end{pmatrix}
\begin{pmatrix}
1 & 0 & \beta_{13} \\
0 & 1 & 0 \\
-\beta_{13} & 0 & 1
\end{pmatrix}
\begin{pmatrix}
1 & \beta_{12} & 0 \\
-\beta_{12} & 1 & 0 \\
0 & 0 & 1
\end{pmatrix},
\end{eqnarray}
resulting in the  transposed  $Y^{(-1/3)}$,

\begin{eqnarray}
Y^{(-1/3) T} \sim 
\begin{pmatrix}
\times & \beta_{13}A\lambda^2+\beta_{12}\lambda^2/3 & \beta_{13} \\
\times & \lambda^2/3 &  -A\lambda^4/3 + \beta_{23}\\
\times & A \lambda^2 & 1
\end{pmatrix},
\end{eqnarray}
where high order corrections are neglected, and the entries in the first column are not shown, as they have no impact on obtaining the mixing angle $\theta_{ij}^e$. 

Although any entries in the above matrix can be assigned GJ factors, not all of them have significant impacts on $\theta_{ij}^e$. Next we will first identify what the relevant GJ entries are, and then evaluate their consequences.

Since the (23) entry in this matrix is small, singling out $\theta_{23}^e$ will not affect the other two angles by much, and vice versa. Therefore the relevant GJ entry for $\theta_{23}^e$ would be the (32) entry of $Y^{(-1/3)}$, and its effect is to have $\theta_{23}^e$ nearly three times larger than $\beta_{23}$.

The role of  the other two mixing angles $\theta_{12}^e$ and $\theta_{13}^e$ is more complicated because of the relatively large (32) entry, $A\lambda^2$. In this case the important GJ entries are in the (21), (22), (23) and (31) elements of  $Y^{(-1/3)}$. Since the (22) entry needs to be multiplied in order to obtain a correct value of $m_\mu$, we are left with eight different ways of assigning GJ factors to the remaining three entries. Among these eight cases we choose one, where GJ factors appear only in the (22), (23) and (31) entries, as an example to show how the relations between $(\beta_{12}, \beta_{13})$ and $(\theta_{12}^e, \theta_{13}^e)$ are derived, and then list the results for the other cases in Table \ref{tb:betatheta}.

\begin{table}

\centering
\begin{tabular}{c | c | c}
\hline
\hline
GJ entries & $\theta_{12}^e$ & $\theta_{13}^e$ \\
\hline
(22) & $\beta_{12}/3$ & $\beta_{13}$ \\
(21), (22) & $4A\beta_{13}\pm \beta_{12}$ & $\beta_{13}$ \\
(22), (23) & $4A\beta_{13}\pm \beta_{12}/3$ & $\beta_{13}$\\
(22), (31) & $4A\beta_{13}\pm \beta_{12}/3$ & $3\beta_{13}$ \\
(21), (22), (23) & $\beta_{12}$ & $\beta_{13}$ \\
(21), (22), (31) & $\beta_{12}$ & $3\beta_{13}$ \\
(22), (23), (31) & $8A\beta_{13} \pm \beta_{12}/3$ & $3\beta_{13}$ \\
(21), (22), (23), (31) & $12A\beta_{13} \pm \beta_{12}$ & $3\beta_{13}$\\
\hline
\hline
\end{tabular}
\caption{Relations between $(\beta_{12}$, $\beta_{13})$ and $(\theta_{12}^e$, $\theta_{13}^e)$ under the influences of GJ factors when $\beta_{23}$ is small.}
\label{tb:betatheta}
\end{table}

\vskip .2cm
In this case, $Y^{(-1)}$ looks like

\begin{eqnarray}\label{abovematrix}
Y^{(-1)} \sim 
\begin{pmatrix}
\times & \beta_{13}A\lambda^2+\beta_{12}\lambda^2/3 & -3 \beta_{13} \\
\times & -\lambda^2 & \times \\
\times & -3 A \lambda^2 & 1
\end{pmatrix}.
\end{eqnarray}
While $\theta_{13}^e=-3 \beta_{13}$ is determined by the (13) entry in Eq.(\ref{abovematrix}),  $\theta^e_{12}$ is affected by the relatively large $(12)$ entry, yielding,

\begin{eqnarray}
\theta_{12}^e = -(8A\beta_{13} + \beta_{12}/3).
\end{eqnarray}

Note that in Table \ref{tb:betatheta} we also include the cases where $\beta_{12}$ and $\beta_{13}$ are in the second or third quadrant. These possible choices of quadrants also lead to the possibilities of the $``\pm"$ signs in the above table. Overall negative signs are omitted in the sense that all $\beta_{ij}$ and $\theta_{ij}$ should be interpreted as deviations from the axes. These formulae agree with the two-angle results quite well.

The formulae in the above table can be used to find out what the allowed parameter space of the input mixing angle $\beta_{ij}$ is, as the required $\theta_{ij}^e$ have been found previously. For $\beta_{23}$, Eq.(\ref{eq:theta23e}) indicates that it is at most $10^\circ$ away from the axes for both the TBM and BM cases, while the allowed parameter space of the other two angles has to be studied separately for the TBM and BM cases, as we notice that in Fig.\ref{fig:corrections} their parameter space has no overlap.

With the help of Fig.\ref{fig:corrections} and the formulae in Table \ref{tb:betatheta}, one then finds that $\beta_{13}$ is at most $10^\circ$ ($20^\circ$) away from the axes for the TBM(BM) case, while $\beta_{12}$ can take on all possible values if one takes into account possible cancellations between the terms $4A\beta_{13}$ and $\beta_{12}/3$ in Table \ref{tb:betatheta}. These results have been used to guide our three-angle search.  

As a final remark, the formulae in Table \ref{tb:betatheta} may also shed some light on understanding the null search result for a symmetric $Y^{(-1/3)}$. Given a symmetric $Y^{(-1/3)}$, $\beta_{13}$ and $\beta_{12}$ would be around $0.2^\circ$ and $13^\circ$ respectively. This immediately eliminates the TBM case, as $\theta_{13}^e$ is now too small, while for the BM case, one is still left with several possible assignments of GJ factors. In this case it may be that all such assignments are incompatible with the mass ratios of charged leptons, and are also ruled out eventually.

\section{Impact of the Down-type Quark Mass Ratios}
\label{sec:massratio}

In our search, we have fixed the down-type quark mass ratios $m_d/m_s$ in order to reproduce the well motivated Gatto relation. However, in reality there may exist uncertainties in both $m_d/m_s$ and $m_s/m_b$, due to both their low energy values and possible SUSY threshold corrections \cite{threshold} introduced during the RG running. We next study the effects of relaxing this assumption, and allowing the mass ratios to vary within acceptable ranges, on our search. This amounts to deviations from the Gatto relations. For $m_d/m_s$, as these SUSY threshold corrections very nearly cancel, and the uncertainty comes mainly from the low energy data. According to \cite{threshold}, it is found to be at most
\begin{eqnarray}
0.044 \leq |\frac{m_d}{m_s}| \leq 0.061,
\end{eqnarray}
for $5 \leq \tan \beta \leq 75$. The uncertainty on $m_s/m_b$, however, depends on the size of the SUSY threshold corrections, determined by some unknown SUSY parameters. To be consistent with the several SUSY scenarios explored in \cite{threshold,ross&serna}, we choose

\begin{eqnarray}
0.008 \leq |\frac{m_s}{m_b}| \leq 0.021,
\end{eqnarray}
for $5 \leq \tan \beta \leq 75$. 

It should be noted that this full range amounts to a deviation from the Gatto relation.

In order to study the impact of these uncertainties on our previous search results, we focus on two simple cases: the symmetric and one-angle asymmetric cases.

\begin{itemize}

\item Even with these uncertainties, we still find no acceptable solutions for a symmetric $Y^{(-1/3)}_{}$. 

\item There are novel solutions for the one-angle asymmetric case:

\begin{itemize}
\item In contrast to our previous results, we now find solutions for Bimaximal mixing, occurring for $\beta_{12}$ non-zero. The allowed GJ patterns and parameter space for $m_d/m_s$, $m_s/m_b$ and $\beta_{12}$ are displayed in Fig.\ref{fg:1ABM}, and one finds the allowed $\beta_{12}$ are clustered around $168^\circ$ and $348^\circ$, suggesting a $(\pi-\lambda)$ or $(2\pi-\lambda)$ parametrization. 

\item For Tri-Bimaximal mixing, we find new solutions with $\beta_{13}$ non-zero and clustered around $183^\circ$, and additional GJ patterns. Fig.\ref{fg:1ATBM} displays the allowed GJ patterns and parameter space for $m_d/m_s$, $m_s/m_b$ and $\beta_{13}$.

\end{itemize}

\end{itemize}

The existence of these new solutions may be due to the sensitivity of $m_e/m_\mu$ to the down-quark mass ratios, as allowing them to vary decreases the constraining power $m_e/m_\mu$. In Fig.\ref{fg:contours}, we give an example of how $m_e/m_\mu$ and some of the other output parameters depend on the input down-quark mass ratios. Although the more complicated two-angle and three-angle cases are not addressed here, one may expect that some additional GJ patterns can occur for these cases as well. However, an extensive numerical study of these cases is beyond the scope of this paper.


\begin{figure}[H]
\scalebox{0.5}{\includegraphics*[140,470][500,670]{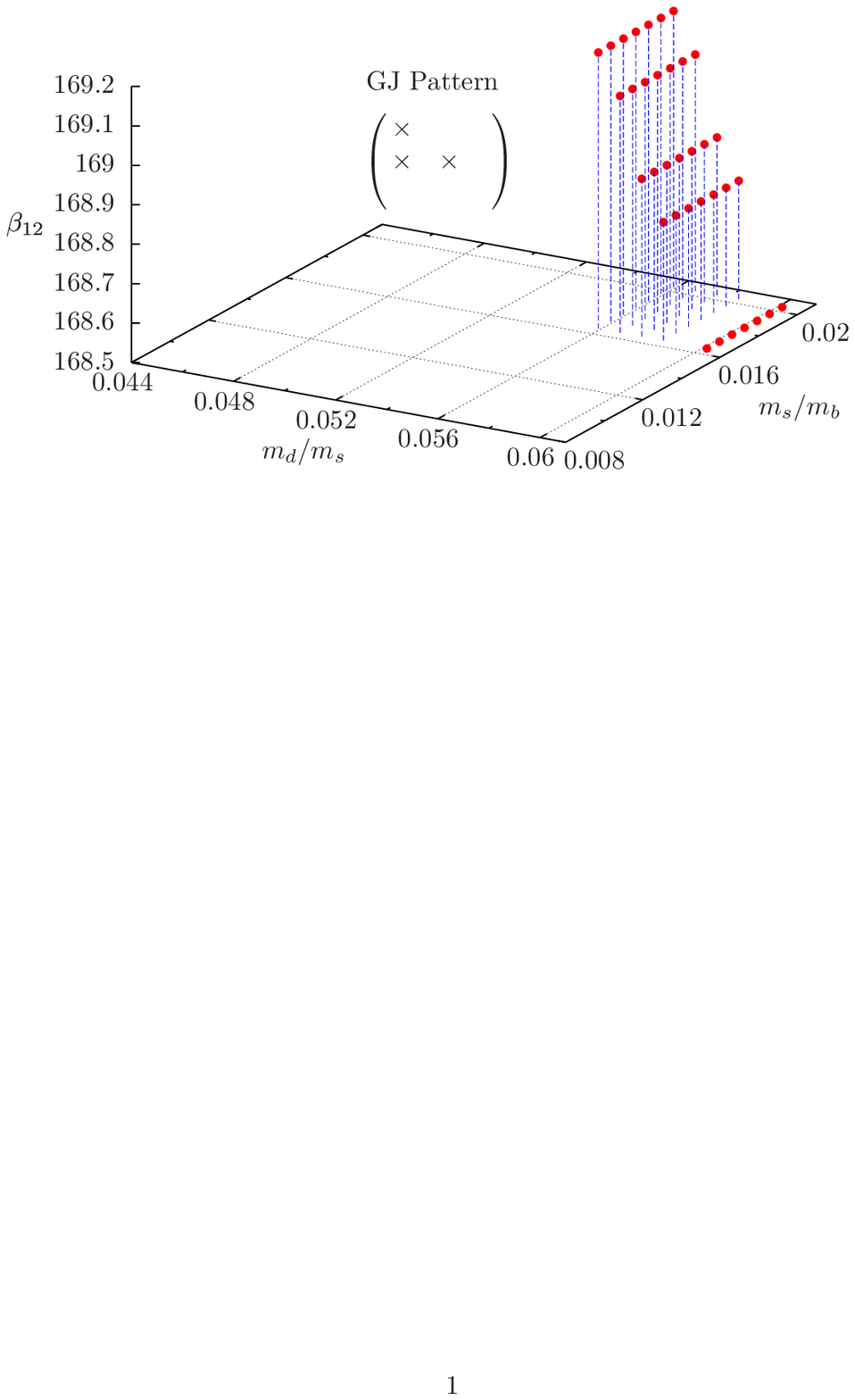}}
\scalebox{0.5}{\includegraphics*[140,470][500,670]{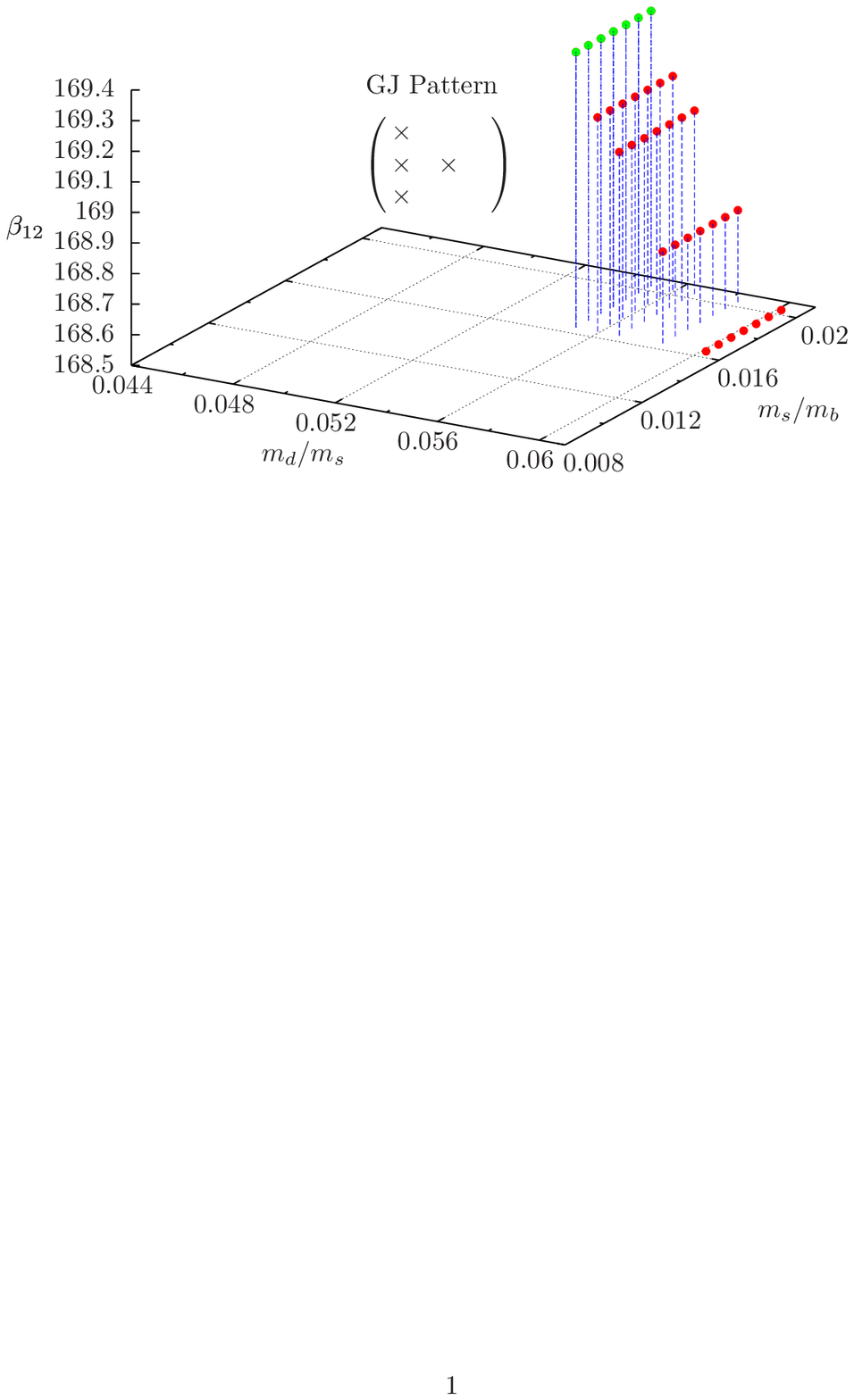}}
\scalebox{0.5}{\includegraphics*[140,470][500,670]{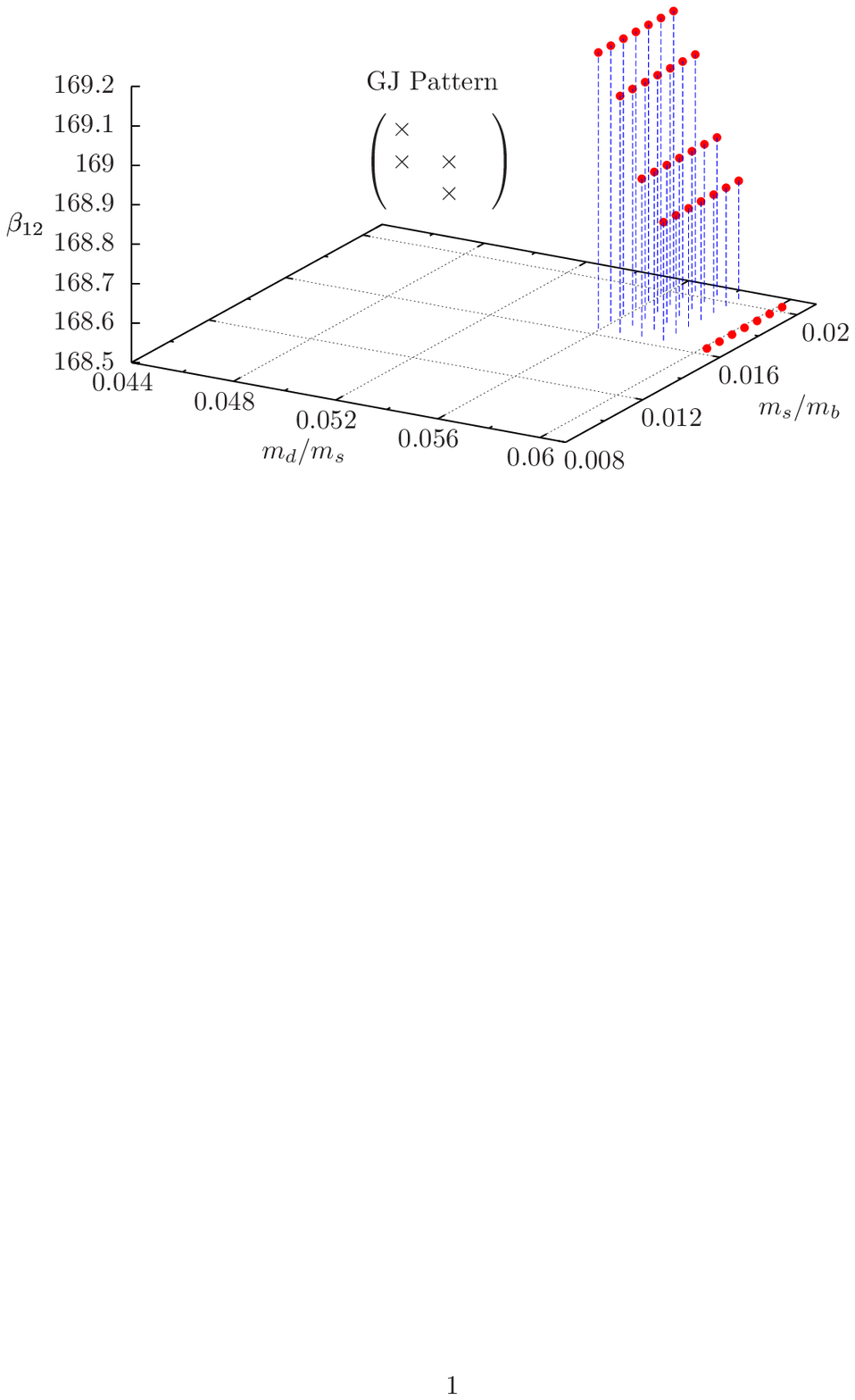}}
\scalebox{0.5}{\includegraphics*[140,470][500,670]{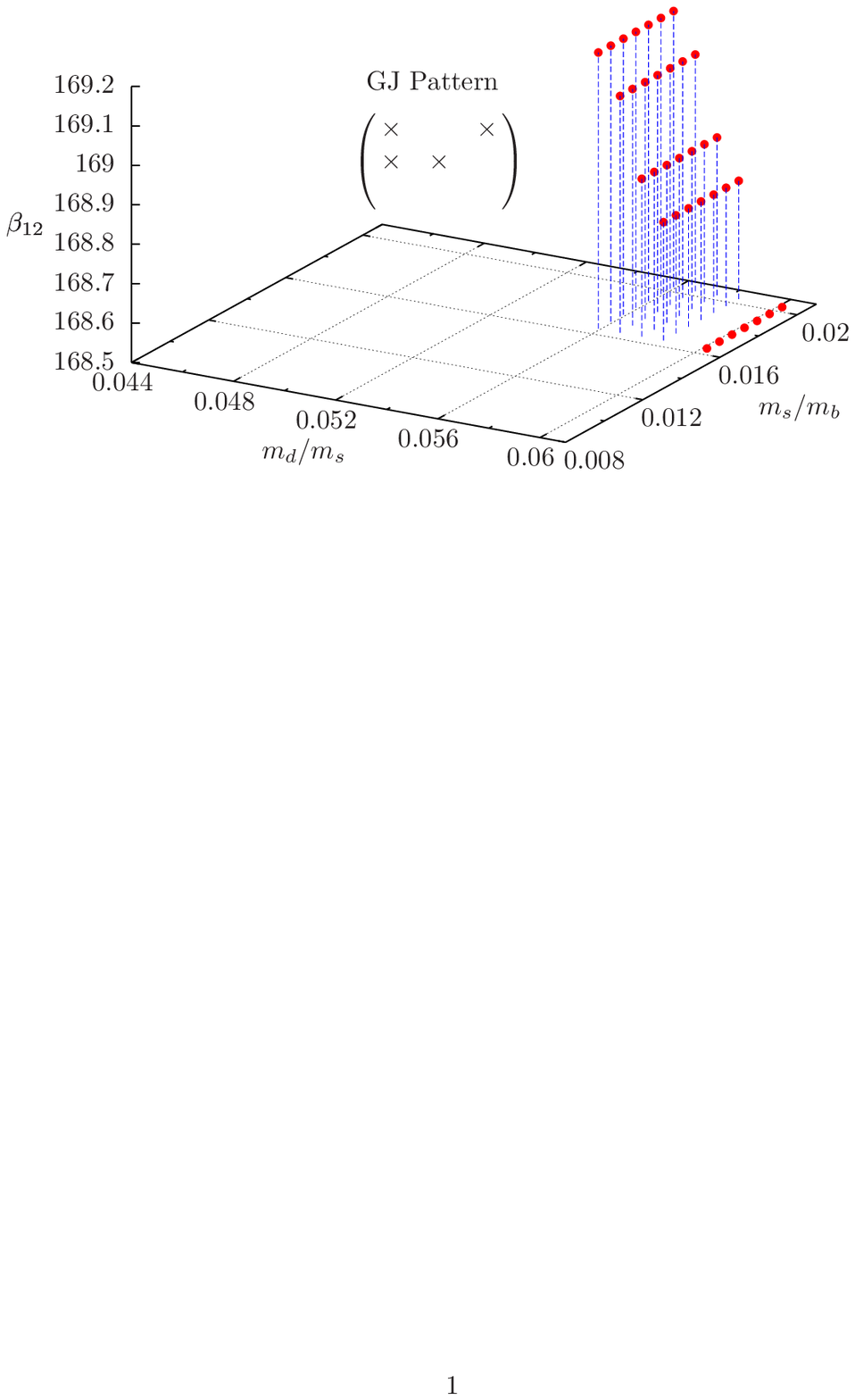}}
\scalebox{0.5}{\includegraphics*[140,470][500,670]{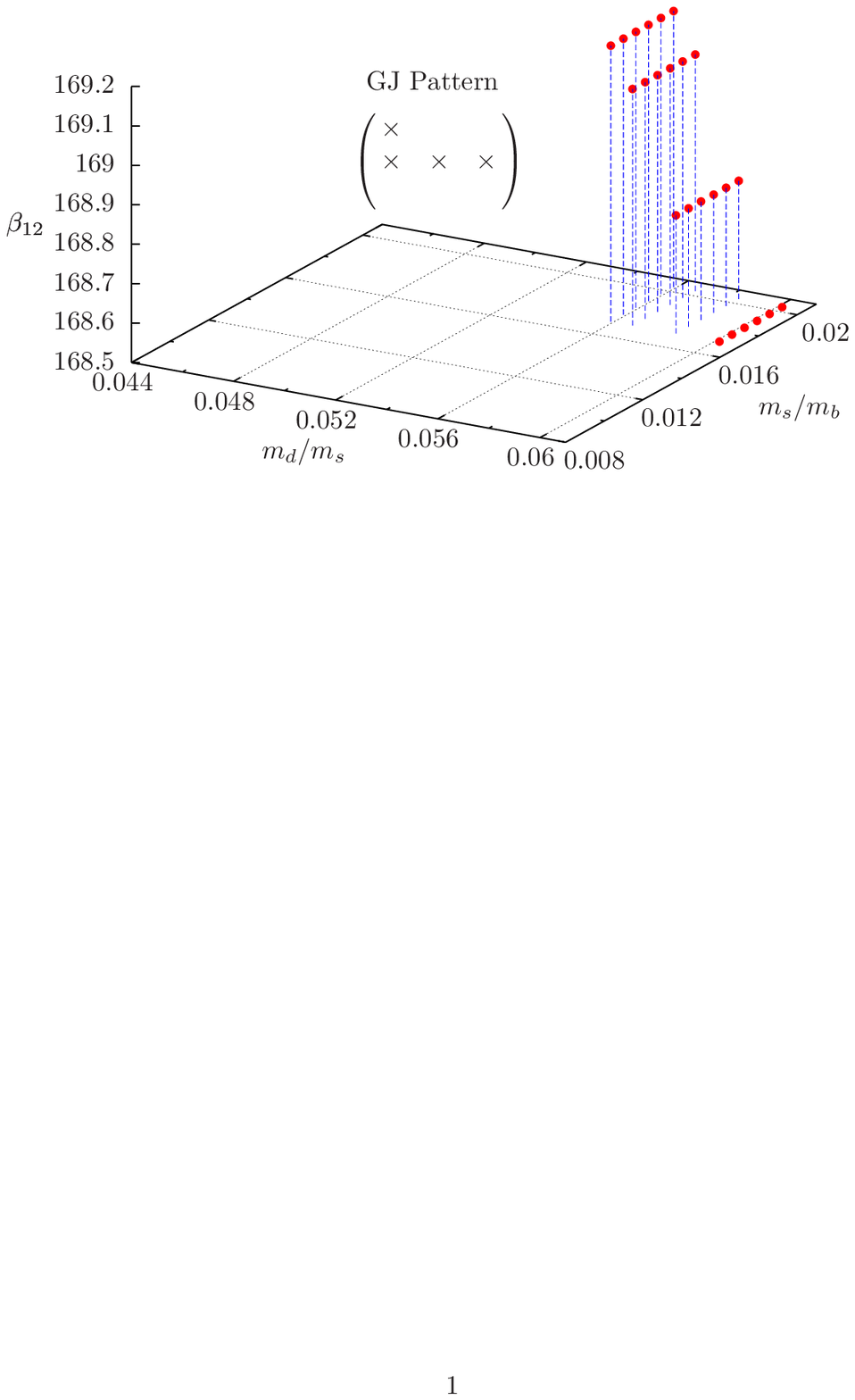}}
\scalebox{0.5}{\includegraphics*[140,470][500,670]{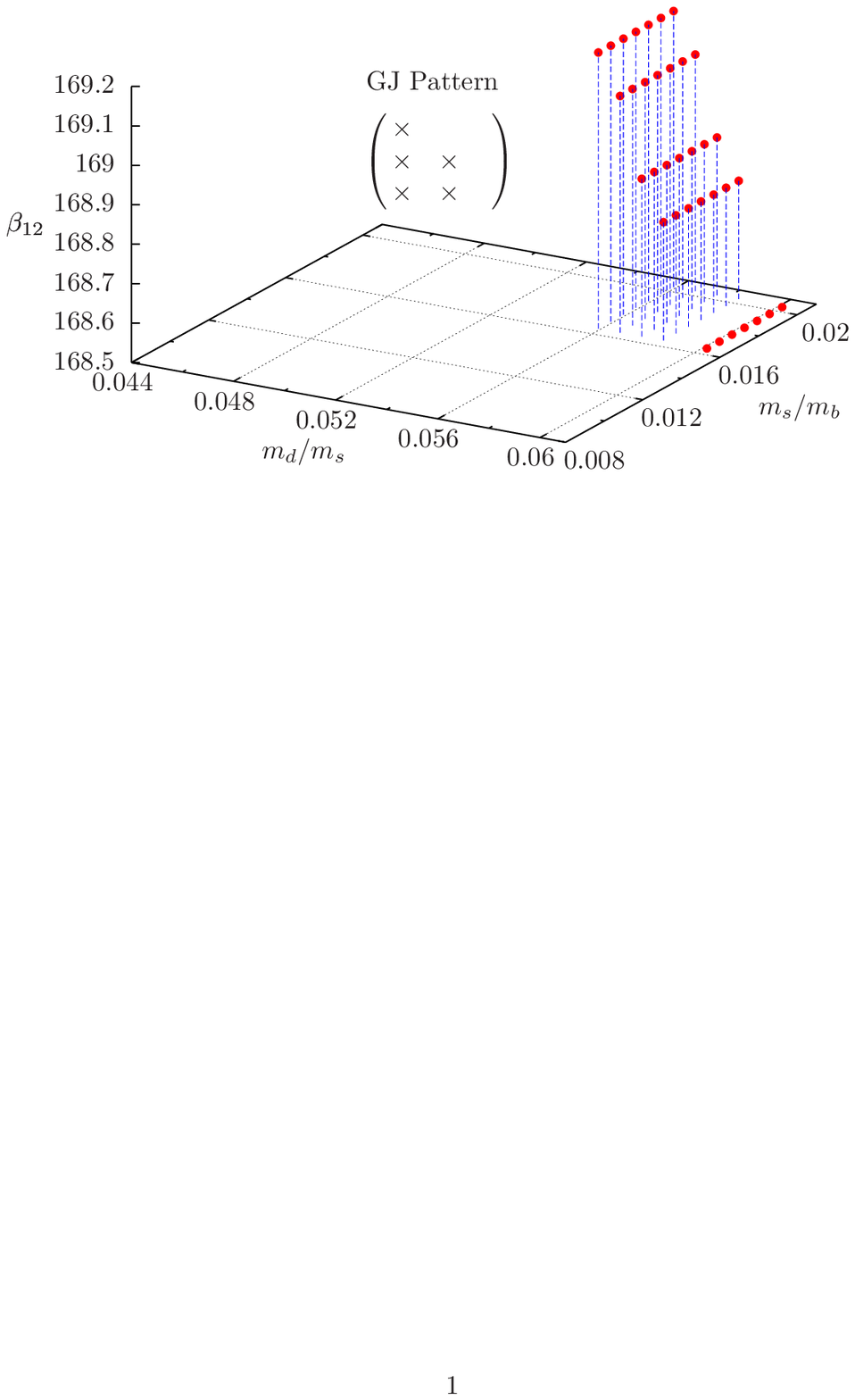}}
\scalebox{0.5}{\includegraphics*[140,470][500,670]{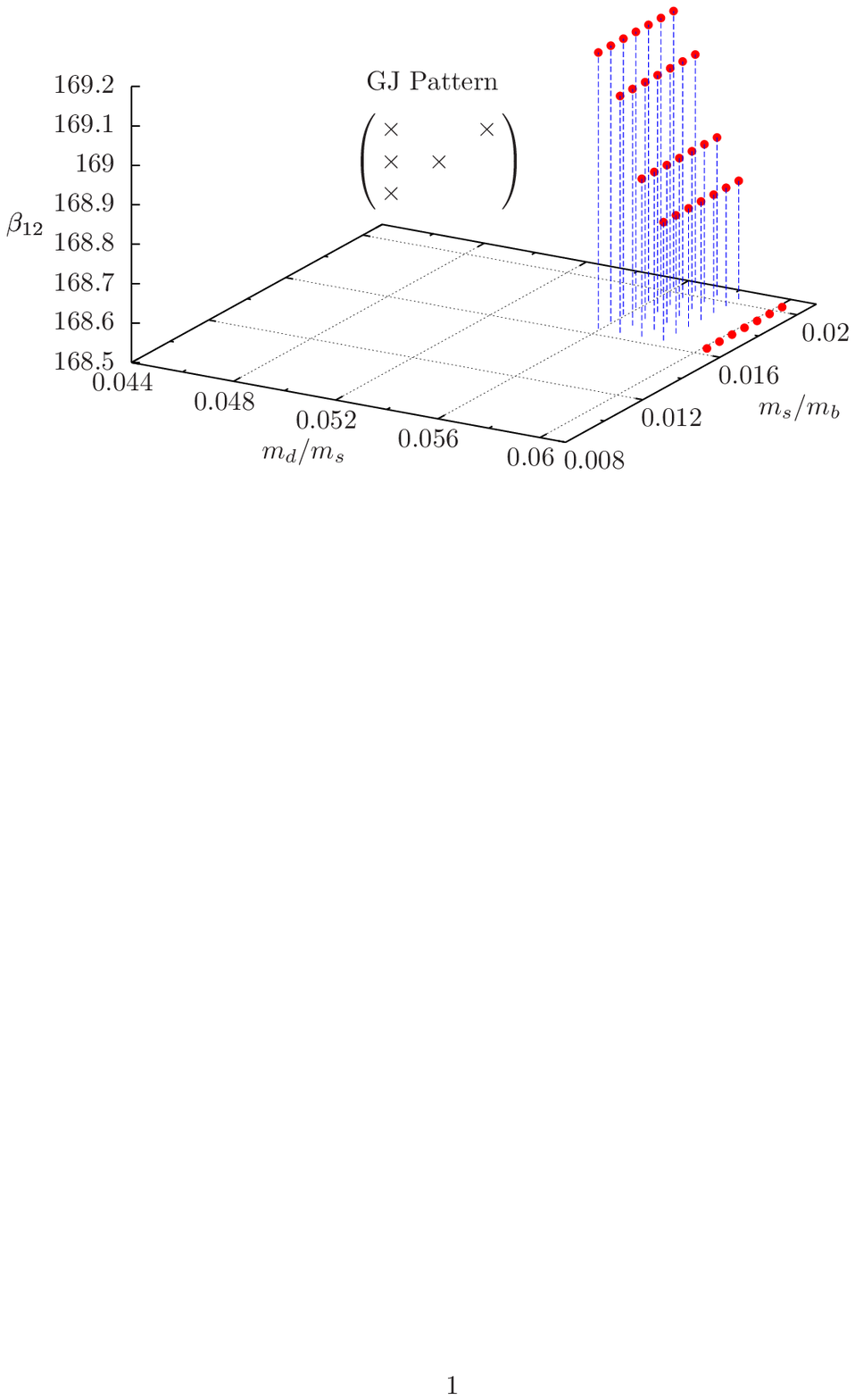}}
\scalebox{0.5}{\includegraphics*[140,470][500,670]{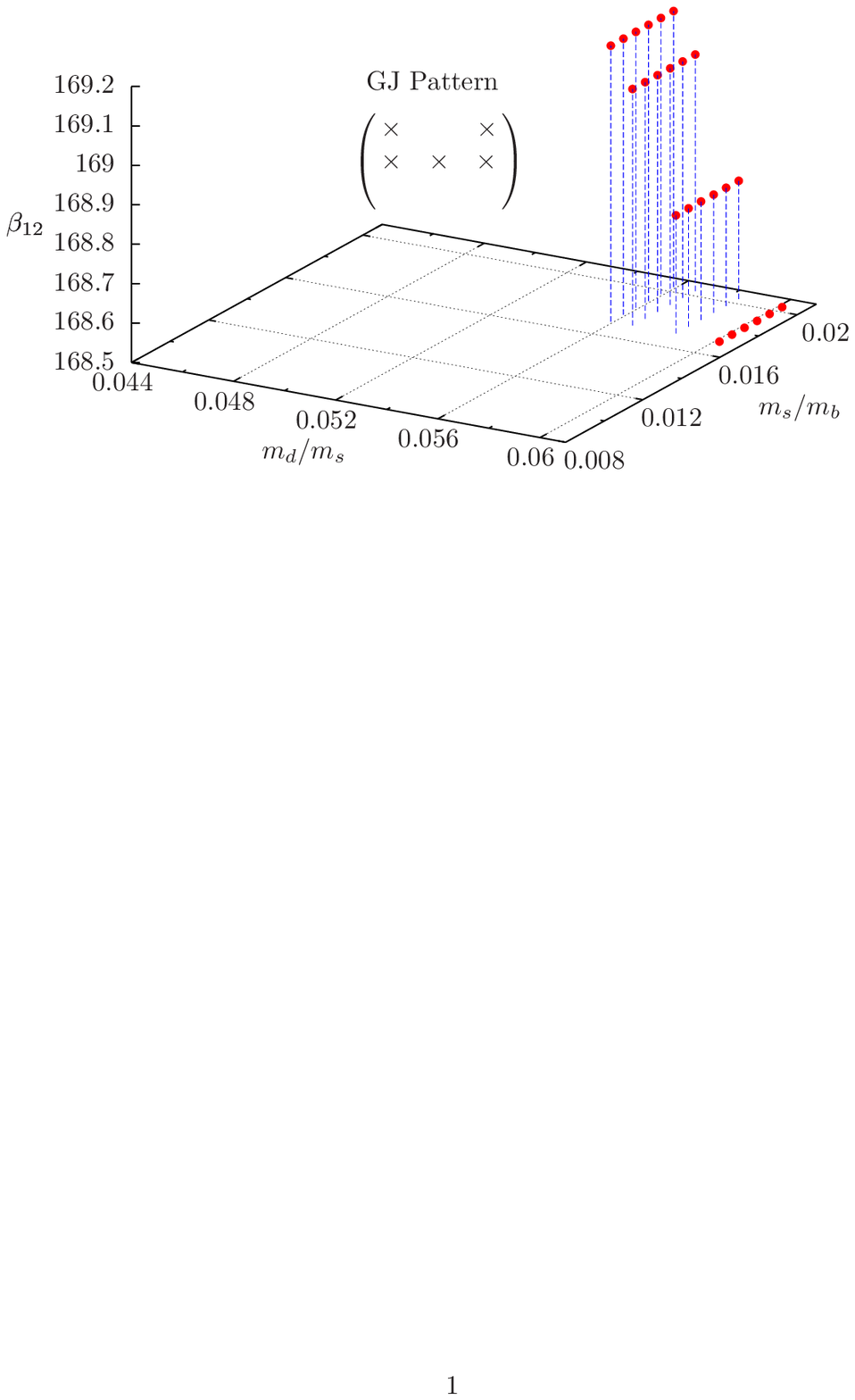}}
\caption*{(Fig.\ref{fg:1ABM} continued on the next page)}
\end{figure}

\begin{figure}[H]
\scalebox{0.5}{\includegraphics*[140,470][500,670]{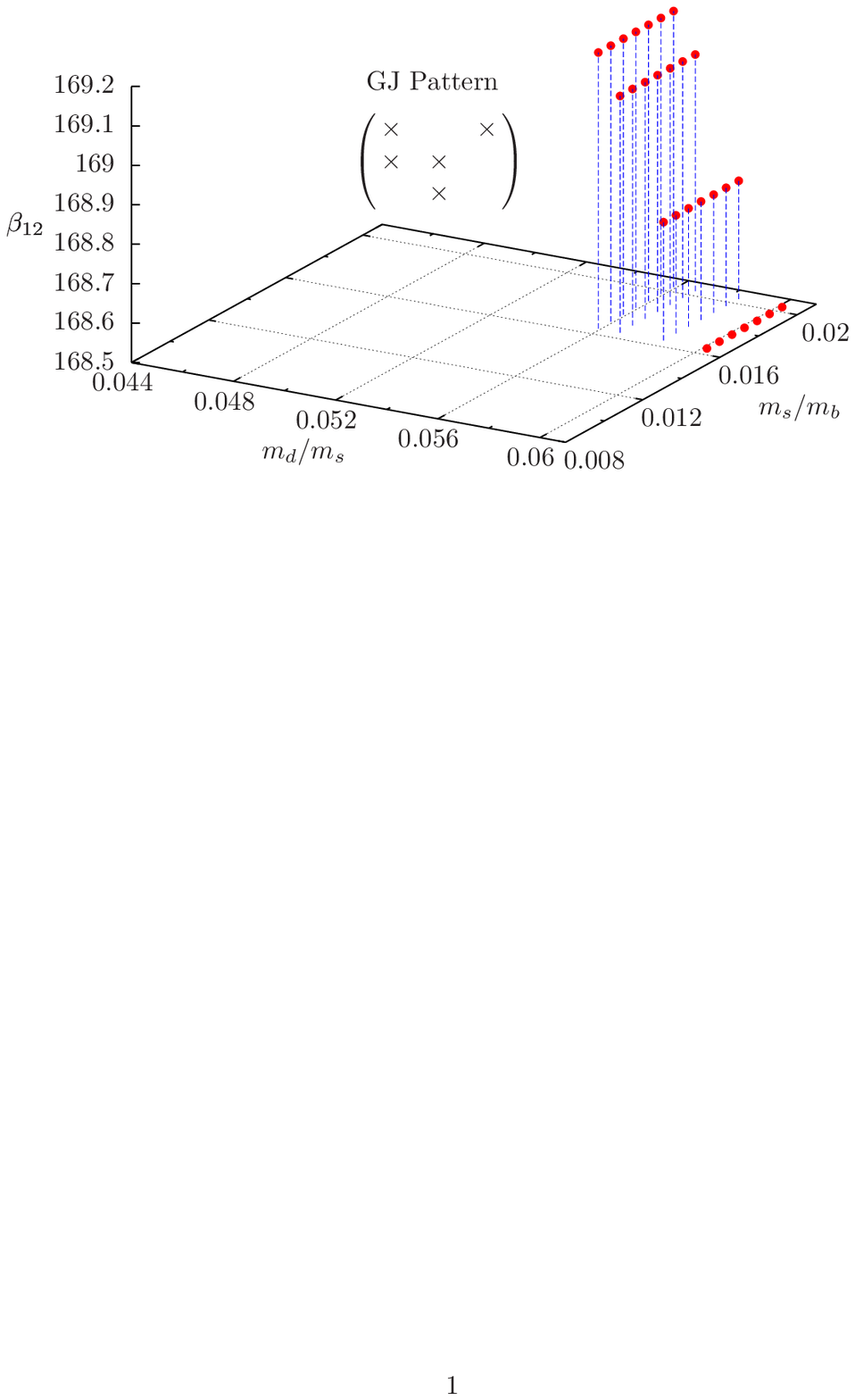}}
\scalebox{0.5}{\includegraphics*[140,470][500,670]{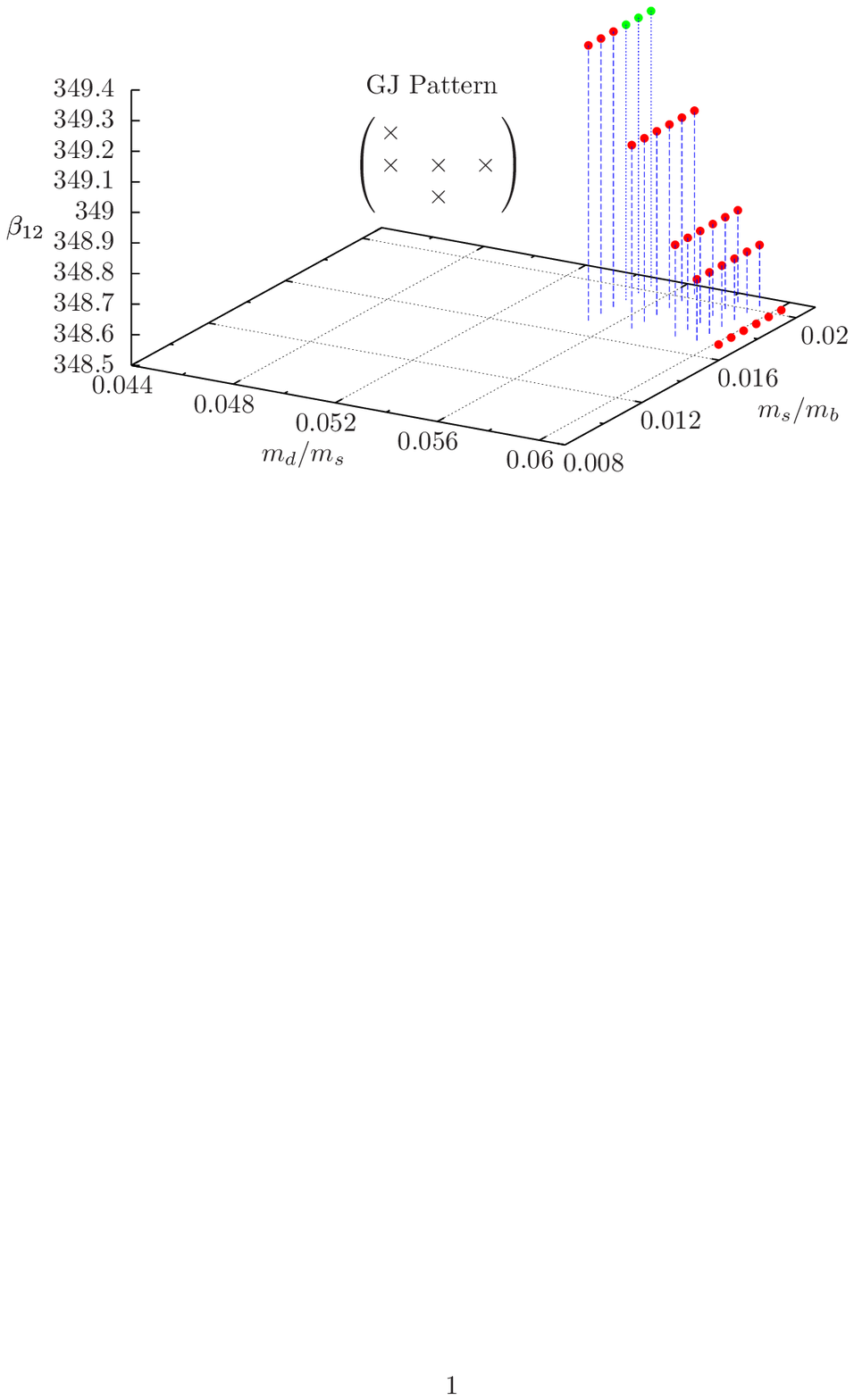}}
\scalebox{0.5}{\includegraphics*[140,470][500,670]{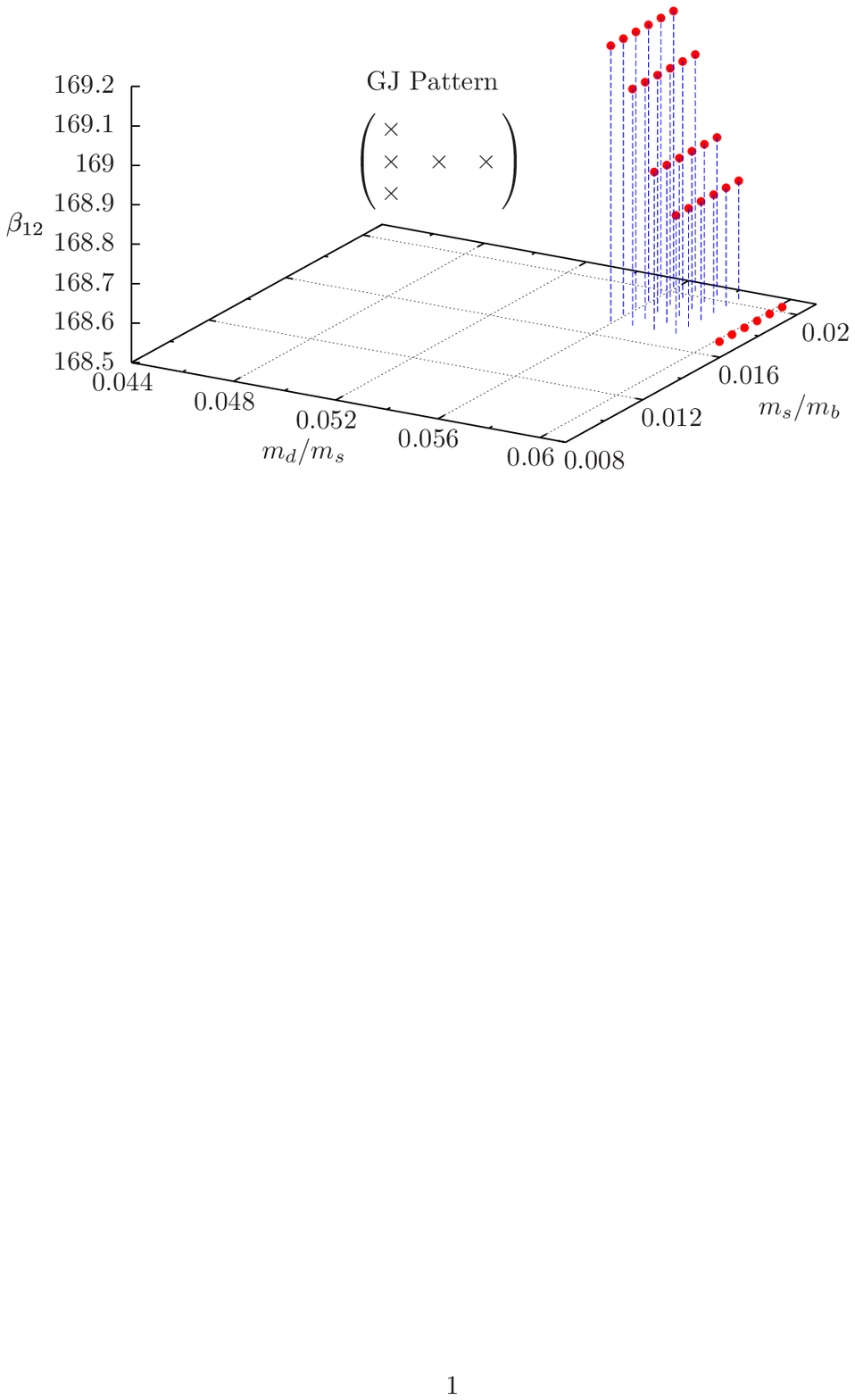}}
\scalebox{0.5}{\includegraphics*[140,470][500,670]{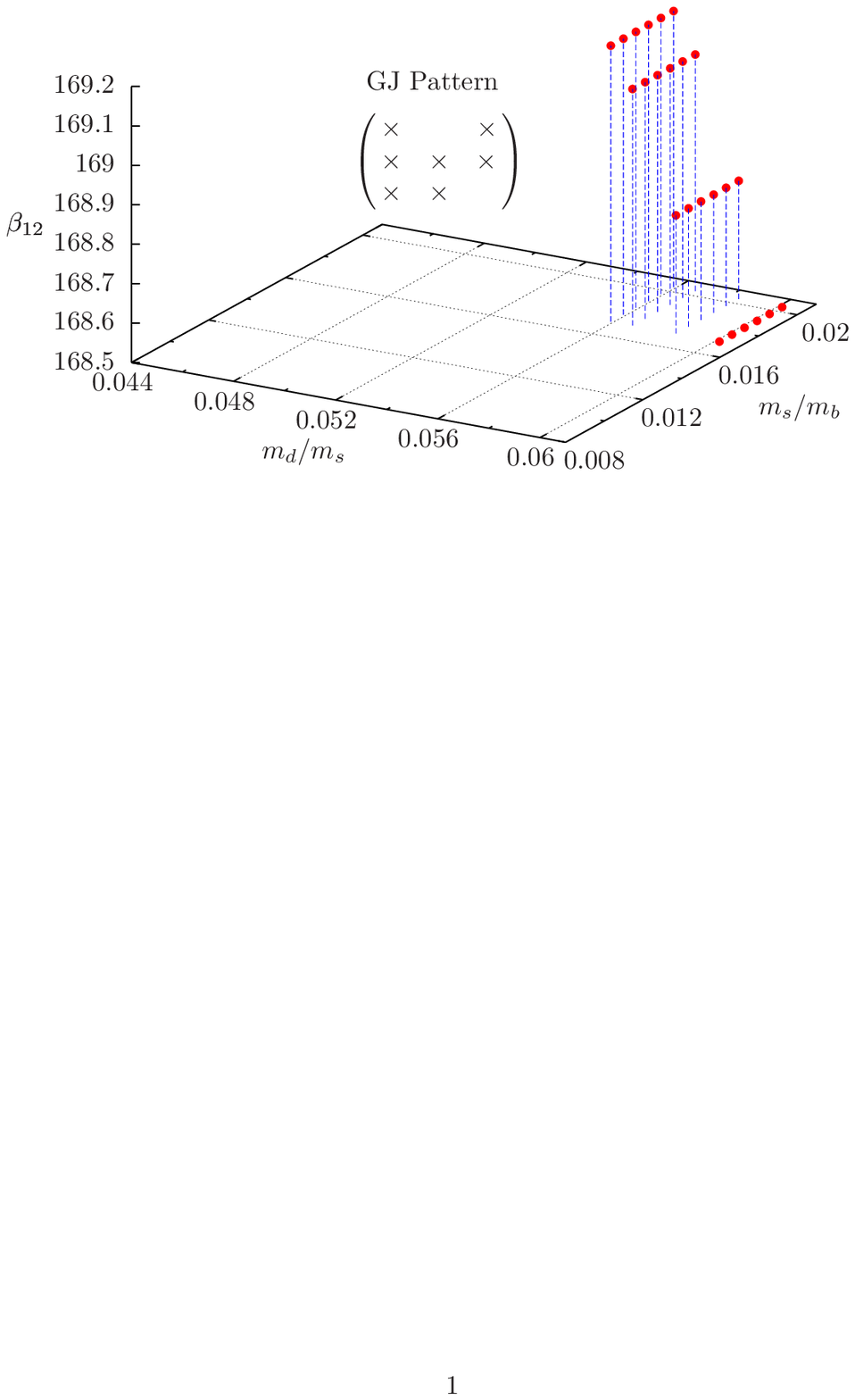}}
\caption{Allowed GJ patterns and parameter space for $m_d/m_s$, $m_s/m_b$ and $\beta_{12}$ in the one-angle Bimaximal mixing solutions. Each solution labelled by a red dot always pairs with another solution (not shown here) with $\beta_{12}$ shifted by $180^\circ$. No such pairing exists for solutions labelled by a green dot. Dashed vertical lines project to the corresponding values of $m_d/m_s$ and $m_s/m_b$ for each solution.}
\label{fg:1ABM}
\end{figure}

\begin{figure}[H]
\scalebox{0.5}{\includegraphics*[140,470][500,670]{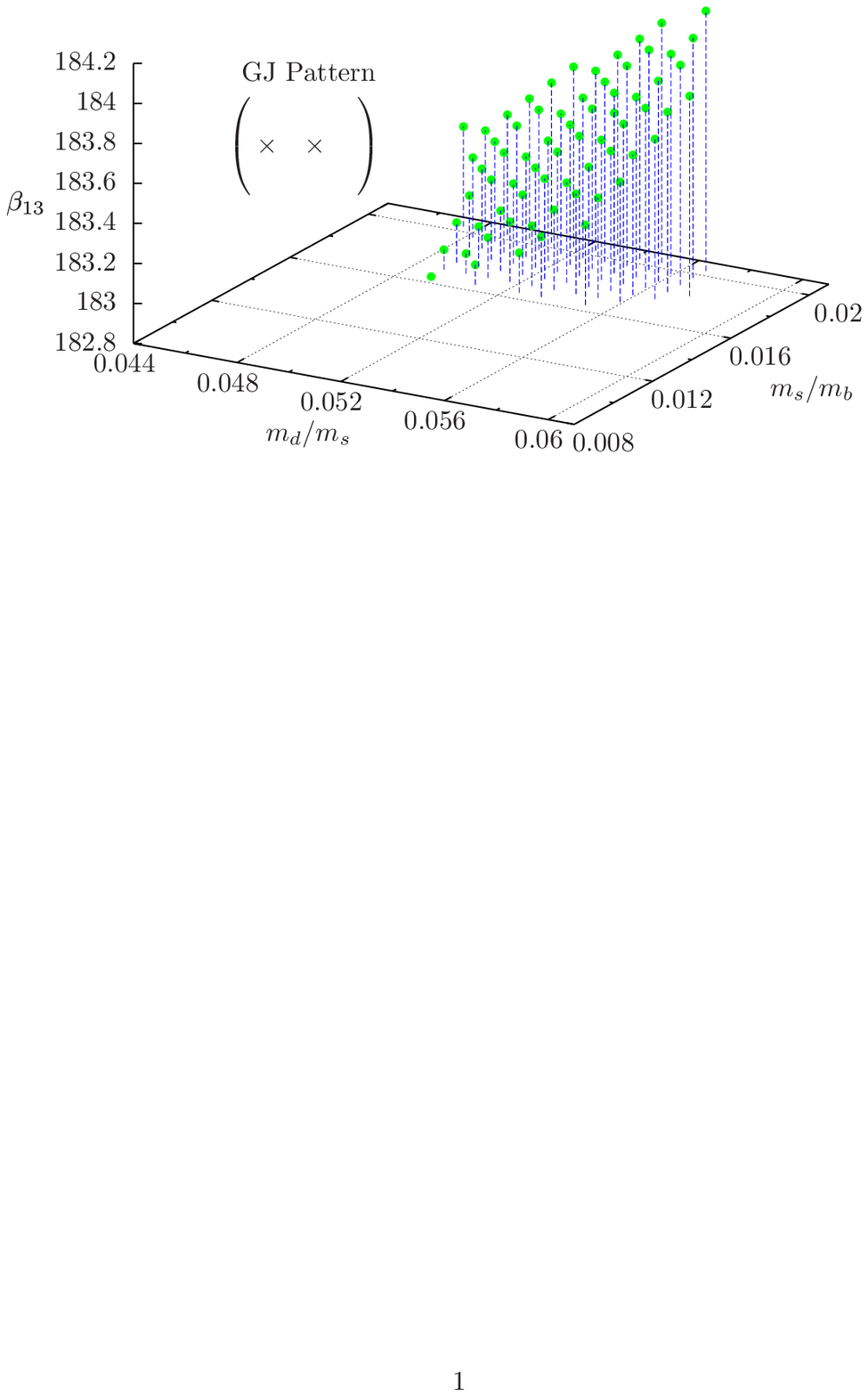}}
\scalebox{0.5}{\includegraphics*[140,470][500,670]{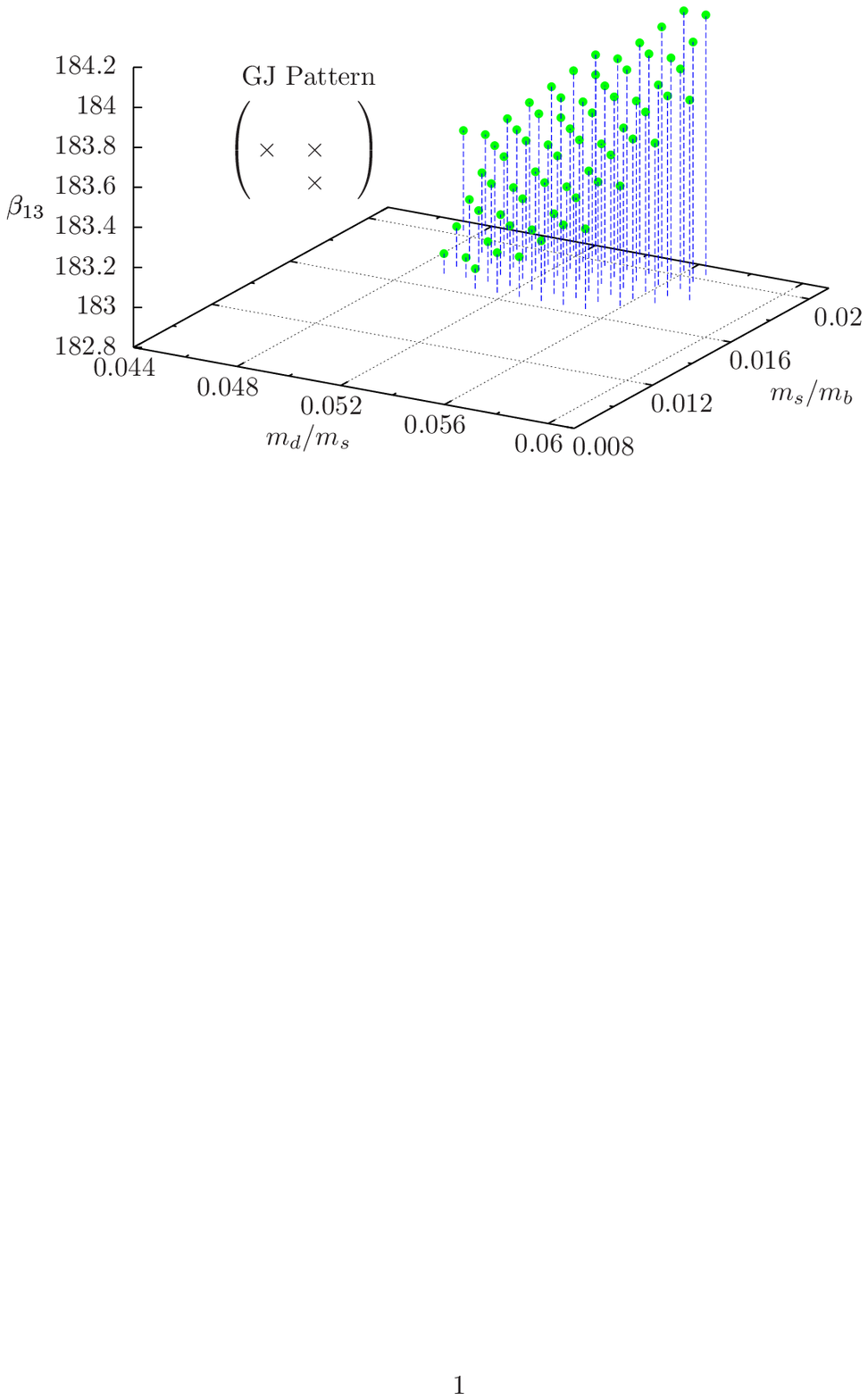}}
\scalebox{0.5}{\includegraphics*[140,470][500,670]{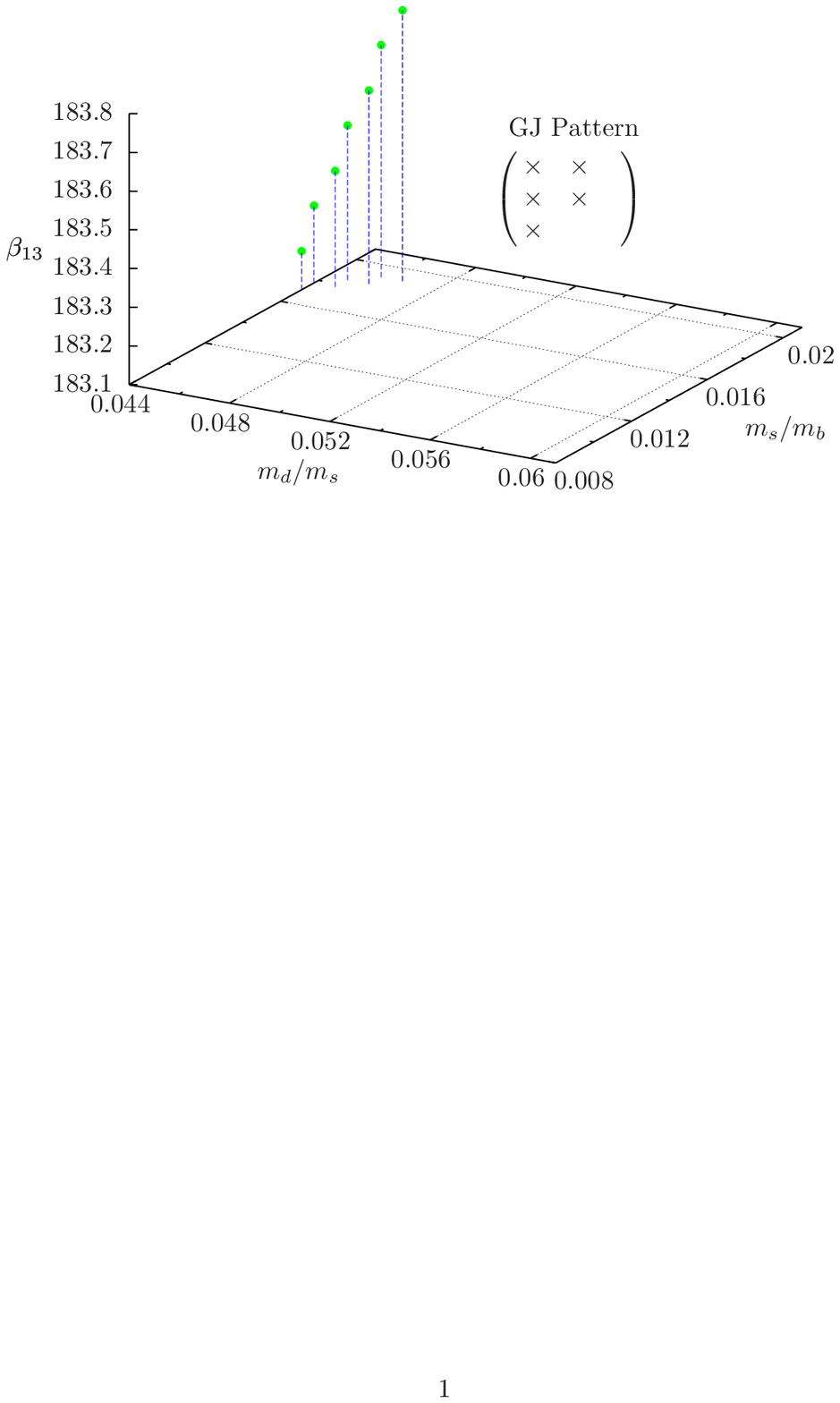}}
\scalebox{0.5}{\includegraphics*[140,470][500,670]{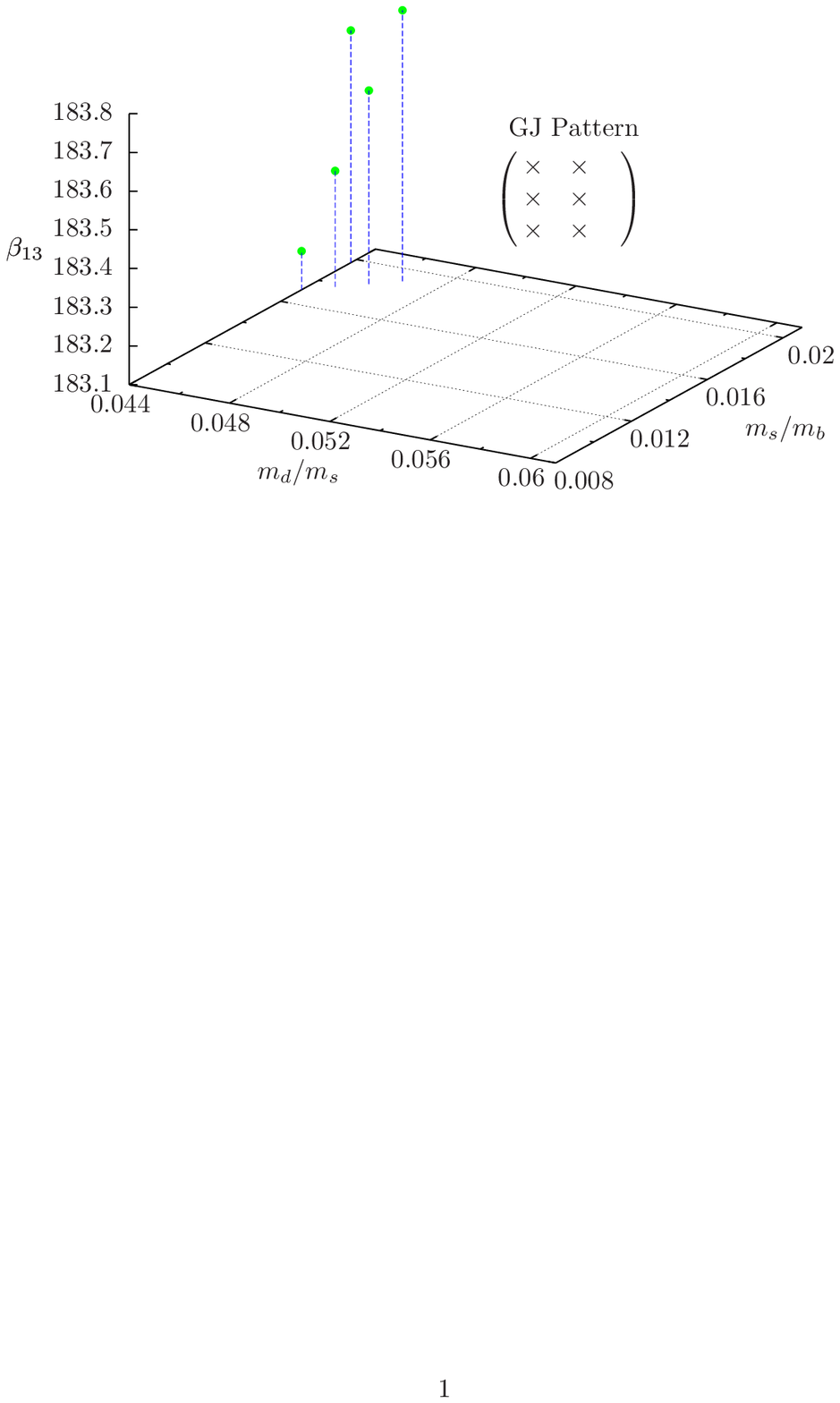}}
\caption{Allowed GJ patterns and parameter space for $m_d/m_s$, $m_s/m_b$ and $\beta_{13}$ in the one-angle Tri-bimaximal mixing solutions.}
\label{fg:1ATBM}
\end{figure}

\begin{figure}[H]
\scalebox{0.6}{\includegraphics*{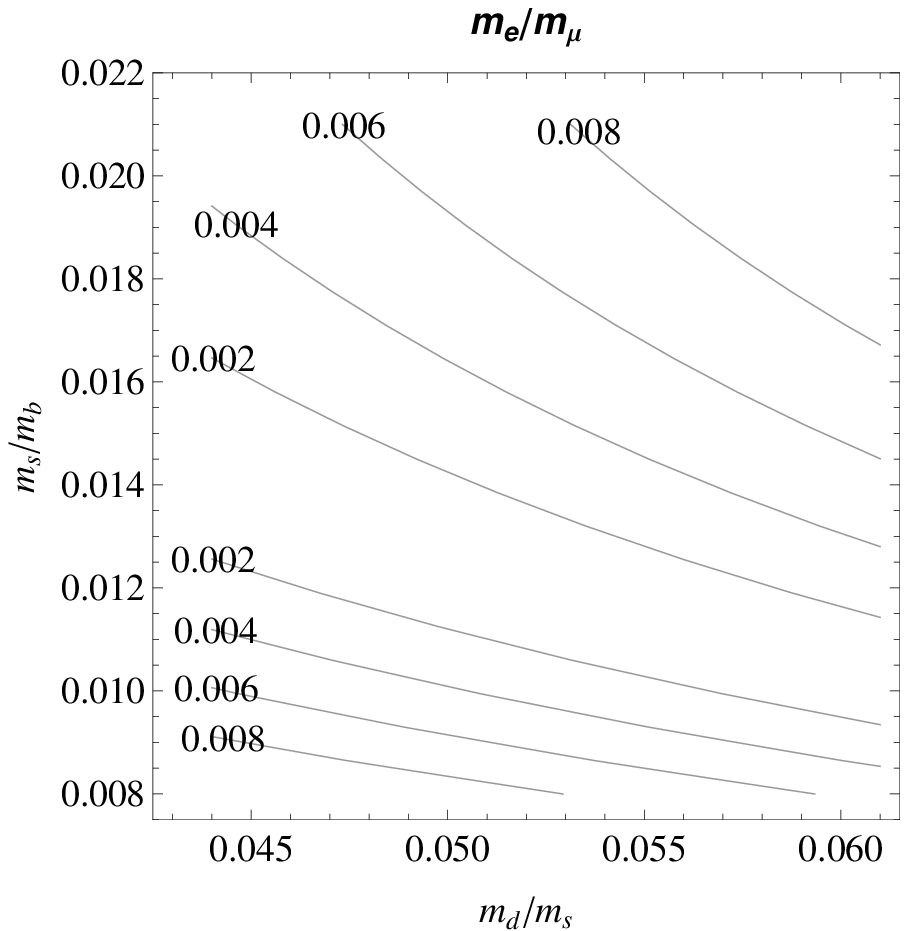}}
\scalebox{0.6}{\includegraphics*{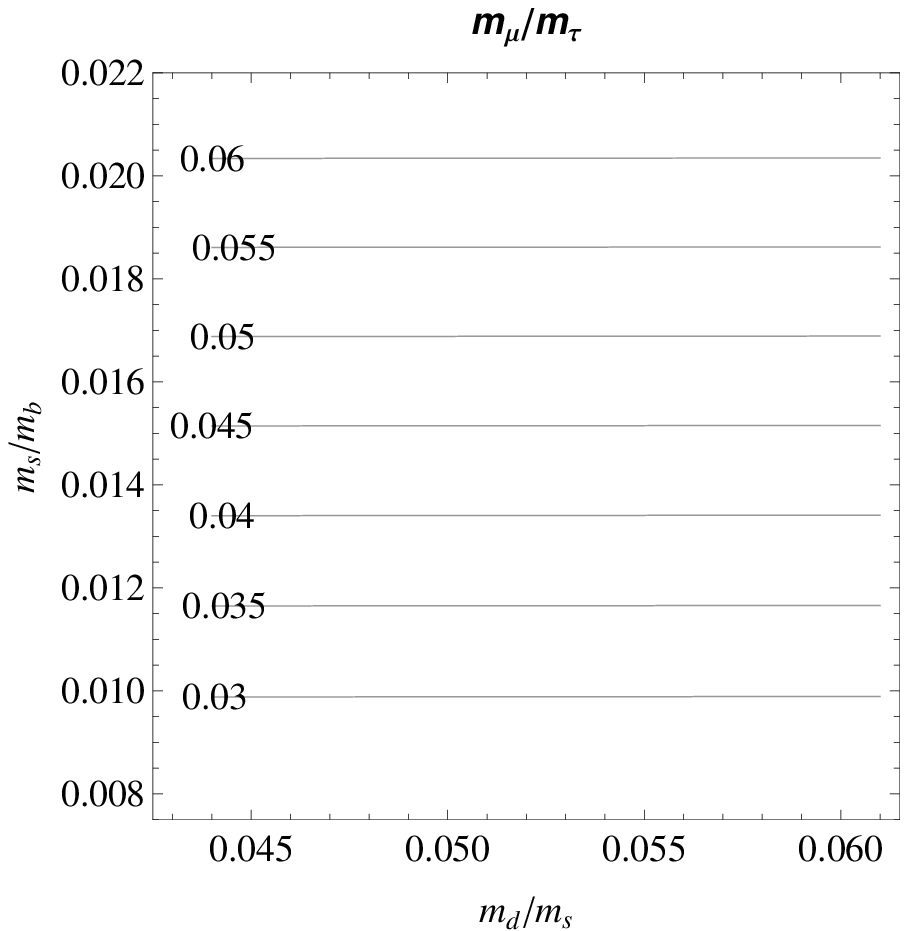}}
\scalebox{0.6}{\includegraphics*{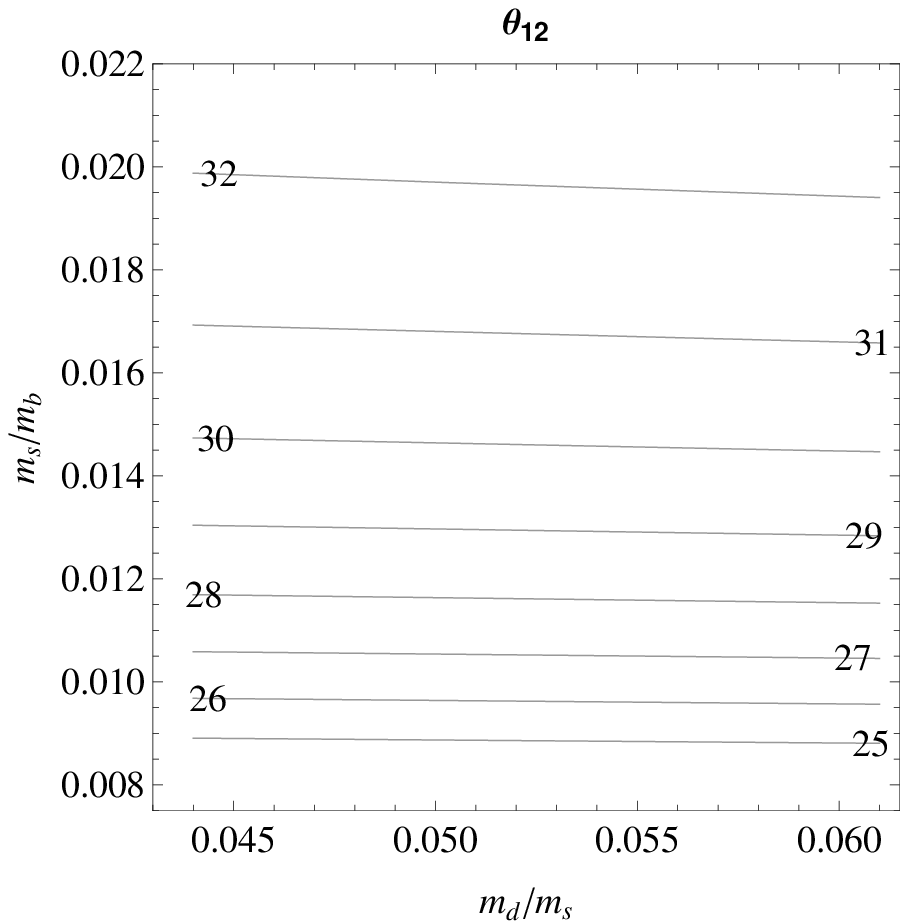}}
\scalebox{0.6}{\includegraphics*{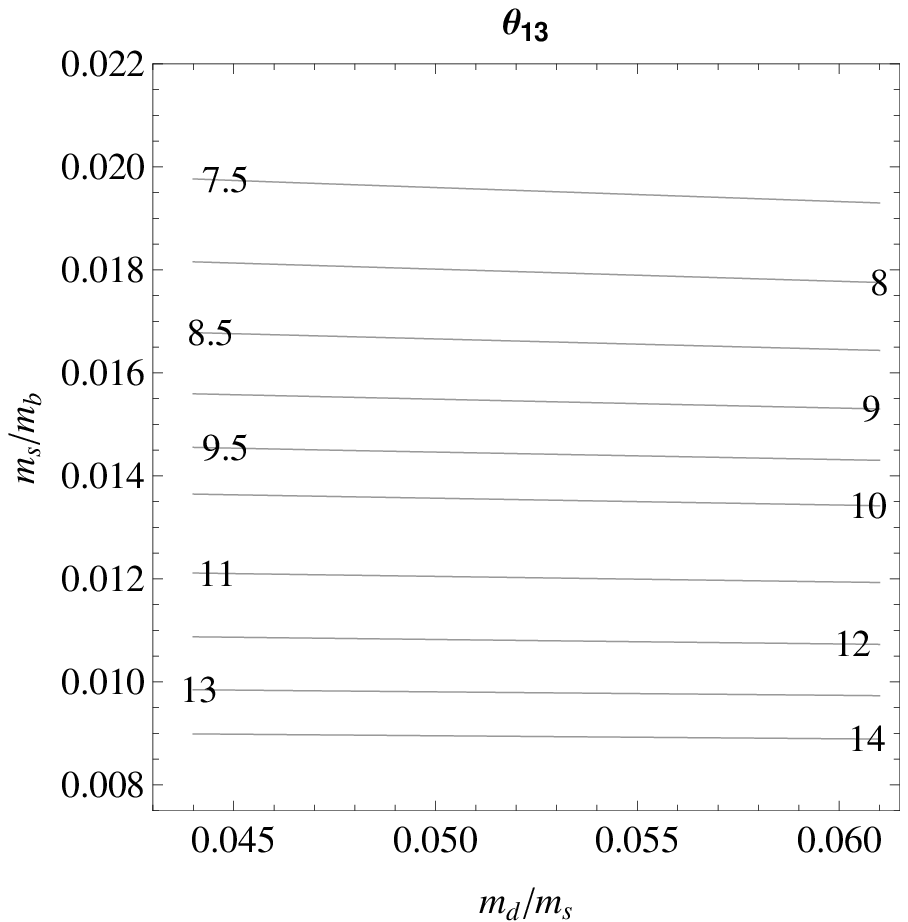}}
\scalebox{0.6}{\includegraphics*{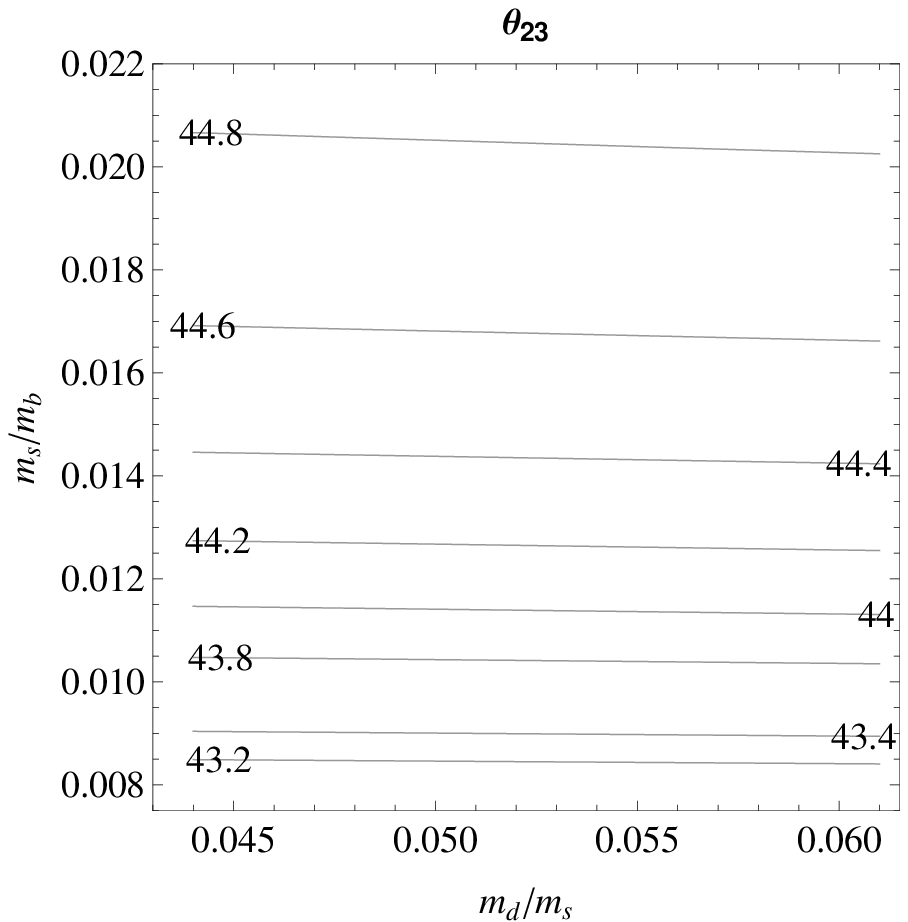}}~~~~~~~~~~
\scalebox{0.6}{\includegraphics*{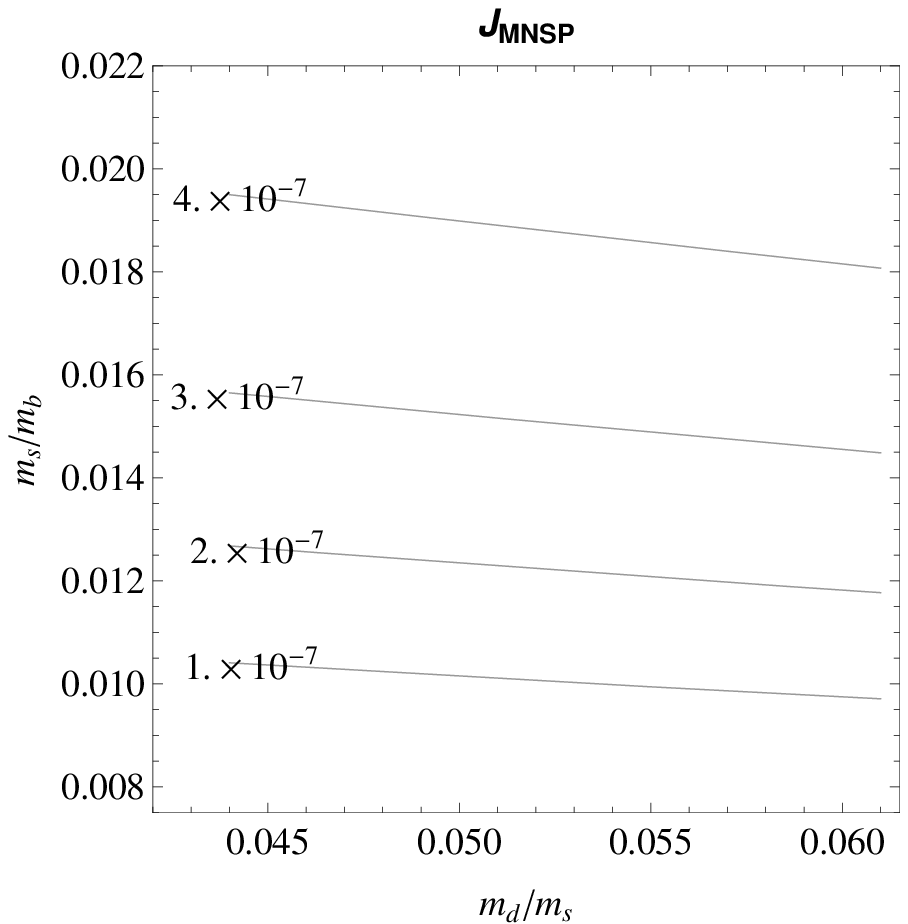}}
\caption{Contour plots of the output parameters as functions of $m_d/m_s$ and $m_s/m_b$. We have used the same input parameters (except for $m_d/m_s$ and $m_s/m_b$) and GJ pattern as those in Example I.}
\label{fg:contours}
\end{figure}



\section{Almost Symmetric $Y^{(-1/3)}_{}$} 
\label{sec:symmetric}
In the previous sections, we chose to introduce the asymmetry in $Y^{(-1/3)}$ by expanding $\m V$  away from unity. One could have equally well introduced the asymmetry by starting from a symmetric  $Y^{(-1/3)}_{}$ by taking ,

\be 
\m V =  \m U_{\rm CKM}^{T}\begin{pmatrix}
1 & 0 & 0 \\ 0 & \hat c_{23} &\hat s_{23} \\ 0 & -\hat s_{23} & \hat c_{23}  
\end{pmatrix} \begin{pmatrix}
\hat c_{13} & 0 & \hat s_{13}\\ 0 & 1 & 0 \\ -\hat s_{13} & 0 & \hat c_{13}
\end{pmatrix} \begin{pmatrix}
\hat c_{12} & \hat s_{12} & 0 \\ -\hat s_{12} & \hat c_{12} & 0 \\ 0 & 0 & 1
\end{pmatrix}.
\ee
where $\hat s_{ij}=\sin\gamma_{ij}$, and $\hat c_{ij}=\cos\gamma_{ij}$.
The  angles $\gamma_{ij}$ indicate directly the deviation from symmetry. In this case, the search for one-angle solutions is akin to the generic three-angle solutions where the angles and phases obey several relations among themselves. This form may be more convenient for model building purposes. 

Results are tabulated in Appendix B.


\section{Summary and Conclusions}
\label{sec:conclusion}

The Flavor Ring demonstrates how the simultaneous consideration of constraints on flavor observables and ideas inspired by grand unification can significantly restrict the flavor sector of the SM.  While these constraints do not uniquely determine the flavor structure of the SM fermions, they reduce the parameter space sufficiently to allow one to begin to do numerical searches for solutions compatible with all the fermion GUT-scale mass ratios and quark and lepton mixing angles.  The solutions found by these searches can then potentially be compared against the predictions of specific flavor models. 

In this work, we have performed a search for such solutions.  Inspired by a $\bs{\m Z_7 \rtimes \m Z_3}$ flavor symmetry, we take a diagonal $Y^{(2/3)}$, which then allows us to write $Y^{(-1/3)}$ in terms of the CKM matrix, a diagonal matrix, and a single right-handed unitary mixing matrix, $\m V$.  We first consider a symmetric $Y^{(-1/3)}$, setting $\m V_{}^{\dagger}=\m U_{\rm CKM}^{T}$, and then expand the search to include forms of $\m V$ containing one, two, and three nonzero mixing angles.  For each of these cases, we obtain the corresponding $Y^{(-1)}$, including up to eight Georgi-Jarlskog insertions.  We then retain those forms of $Y^{(-1)}$ which produce phenomenologically acceptable values for the neutrino mixing angles and the charged-lepton mass ratios, assuming either a BM or TBM form of the neutrino seesaw mass matrix.

We find several interesting features of the search results.  A first point is that we find no symmetric forms of $Y^{(-1/3)}$ compatible with the measured value of $\theta_{13}\sim 9^\circ$, even if we include phases in $\m U_{\rm seesaw}$.  However, many solutions exist for $\theta_{13}\sim 3^\circ$.  This highlights two important points.  First, our results would place significant constraints on flavor models which predict a symmetric $Y^{(-1/3)}$.\footnote{For specific models, however, this will depend upon our assumption of a diagonal $Y^{(2/3)}$.}  Second, this underscores the importance of the recent measurement of $\theta_{13}$ in constraining models of flavor.  

Next, we examine solutions with asymmetric $Y^{(-1/3)}$.  Our solutions are discussed in Secs. \ref{sec:search} and \ref{sec:symmetric}, listed in Appendices A and B, and shown in Figs. \ref{fg:TBM2angle}-\ref{fg:TBM_vs_BM_y}.  We find that the values of the angles $\beta_{ij}$ cluster around the values $0$, $180^\circ$, and $\pm 90^\circ$.  This general feature is not surprising, as the form of the TBM and BM seesaw mixing matrices is chosen to approximately coincide with the MNSP matrix; details can be found in Section \ref{sec:analysis} and Appendix C.  Additionally, we find that the values of $m_e/m_{\mu}$ and $\theta_{13}$ to be particularly effective at constraining $\m V$ and the placement of GJ entries. 

Also, we should emphasize that the general framework of combining flavor observables and ideas from grand unification can be applied somewhat more widely than was done here.  For example, the diagonal form of $Y^{(2/3)}$, chosen here due to its relevance to $\bs{\m Z_7 \rtimes \m Z_3}$ flavor models, could be replaced with other forms on a model-specific basis.  The neutrino seesaw matrix could similarly take forms other than the BM and TBM models studied here.  Thus, similar, dedicated searches could be performed to constrain or possibly rule out particular models.

In summary, the consideration of flavor ideas under the umbrella of grand unification allows one to relate flavor observables from the neutrino, charged-lepton, and quark sectors in ways that can be compared to currently existing data.  We have performed such a comparison under the assumption of specific up-quark Yukawa and neutrino seesaw matrices, and other, similar comparisons are ripe for investigation.


\section{Acknowledgements}

We thank Graham Ross and Christoph Luhn for helpful discussions at the early stage of this work, and Lisa Everett for useful suggestions. MP would like to thank the McKnight Doctoral Fellowship Program for their continued support. PR wishes to thank the Aspen Center for Physics, where part of this work was performed. JZ is grateful for support from the CLAS Dissertation Fellowship and the Institute for Fundamental Theory. This research is also partially supported by the Department of Energy Grant No. DE-FG02-97ER41029.


\newpage

\appendix
\numberwithin{equation}{section}

\section{Search Results: Asymmetric $Y^{(-1/3)}$}

\subsection*{Solutions with two angles in $\m V$}

All search results are presented in the form of tables. Tables \ref{tb:2ATBM23} and \ref{tb:2ATBM12} give the results for the TBM case, while those for the BM case can be found in tables \ref{tb:2ABM23}, \ref{tb:2ABM12} and \ref{tb:2ABM13}. 

 The first three entries of each table list the angles $\beta_{ij}$ in $\m V$ for which there are solutions.
The second column, labelled `` GJ Entries", indicates the location of the GJ couplings in the down-quark Yukawa matrix, and the magnitudes of entries in powers of $\lambda$.

In Tables \ref{tb:2ATBM23} and \ref{tb:2ABM23} we only list the magnitudes in $\lambda$ for entries in the first column of $Y^{(-1/3)}$ (transposed to save space). The entries of the last two columns do not vary much, because the angles that yield solutions  are at most 20 degrees away from the x-axis. In terms of the Cabibbo angle, the range of  the absolute values of the $Y^{(-1/3)}_{}$  matrix elements are given by,

\begin{eqnarray}
|Y^{(-1/3)}_{}|\sim \begin{pmatrix}
\lambda^4 -\lambda^6 & (1.0-1.5)\lambda^4 & (1.0-1.56)\lambda^{4} \\
\lambda^{3}-\lambda^6 & (0.31-0.32)\lambda^2 & (0.70-0.75)\lambda^{2} \\
\lambda^{1}-\lambda^4 & \lambda^{5} & 1
\end{pmatrix}.
\end{eqnarray}
For example, the first solution in Table \ref{tb:2ATBM23} has the magnitude of the entries in $Y^{(-1/3)}$ given by

\begin{eqnarray}
|Y^{(-1/3)}_{}| = \begin{pmatrix}
\lambda^{5.0} & \lambda^{3.7} & \lambda^{3.8} \\
\lambda^{4.0} & \lambda^{2.8} & \lambda^{2.2} \\
\lambda^{1.3} & \lambda^{4.9} & 1
\end{pmatrix}.
\end{eqnarray}

In Tables \ref{tb:2ATBM12}, \ref{tb:2ABM12} and \ref{tb:2ABM13} we instead list the magnitudes in $\lambda$ for all entries in $Y^{(-1/3)}$, as a similar pattern to that above does not occur.

The numerical outputs are given in the remaining columns of the tables: the mass ratio $m_e/m_\mu$ and the mixing angles in $\m U_{-1}^{}$, including some phases, explained in Section \ref{sec:analysis}. The MNSP mixing angles and the predicted Jarlskog invariant in $\m U^{}_{\rm MNSP}$ appear in the last four columns. 

All the angles and phases are in degrees and rounded to the tenth place if necessary. Further information about each table can be found in its caption.

\begin{landscape}

\begin{table}
\centering
Table \ref{tb:2ATBM23}: TBM Two-angle Results ($\beta_{23} = 0$)
\vskip .3cm
\footnotesize
\begin{tabular*}{1.53\textwidth}{ c c c| c | c | c c c c c c || c c c c }
\hline
\hline

&  &  &  &  &  &  &  &  &  &  &  &  &  &  \\
$\beta_{12}$ & $\beta_{13}$ & $\beta_{23}$ &  GJ Entries & $m_e/m_\mu$ & $\theta_{12}^e$ & $\theta_{13}^e$ & $\theta_{23}^e$ & $\delta_{12}^e$ & $\delta_{13}^e$ & $\delta_{23}^e$ & $\theta_{12}$ & $\theta_{13}$ & $\theta_{23}$ & $J^{}_{MNSP}$ \\
&  &  &  &  &  &  &  &  &  &  &  &  &  &  \\
\hline
 &  &  &  [22] &  &  &  &  &  &  &  &  &  &  &  \\
168.3 & 351.0 & 0 & (5.0, 4.0, 1.3) & 0.0048 & 4.2 & 9.0 & 0.2 & 0.0 & 180.0 & -0.3 & 31.8 & 9.3 & 45.1 & $2.1\times 10^{-8}$ \\
348.3 & 351.0 & 0 & (5.1, 3.1, 1.3) & 0.0048 & 4.2 & 9.0 & 0.2 & 0.0 & -180.0 & 179.8 & 31.9 & 9.3 & 44.8 & $4.5\times 10^{-7}$ \\
\hline
 &  &  &  [21],[22] &  &  &  &  &  &  &  &  &  &  &  \\
0.0 & 183.1 & 0 & (4.7, 4.2, 2.0) & 0.0047 & 9.1 & 3.1 & 0.2 & 180.0 & 0.0 & -0.3 & 31.0 & 8.6 & 44.6 & $3.7\times 10^{-7}$ \\
170.5 & 354.8 & 0 & (5.2, 5.0, 1.6) & 0.0048 & 6.4 & 5.2 & 0.2 & 0.0 & 180.0 & -0.3 & 36.1 & 8.2 & 44.8 & $1.1\times 10^{-7}$ \\
350.5 & 185.2 & 0 & (5.2, 5.0, 1.6) & 0.0048 & 6.4 & 5.2 & 0.2 & 180.0 & 0.0 & -0.3 & 34.4 & 8.2 & 44.8 & $-5.9\times 10^{-8}$ \\
\hline
 &  &  &  [22],[23] &  &  &  &  &  &  &  &  &  &  &  \\
180.0 & 182.8 & 0 & (4.8, 4.1, 2.0) & 0.0048 & 9.0 & 2.7 & 0.3 & -180.0 & 0.0 & 0.1 & 30.8 & 8.3 & 44.8 & $4.1\times 10^{-7}$ \\
\hline
 &  &  & [(22],[23],[32] &  &  &  &  &  &  &  &  &  &  &  \\
10.5 & 174.8 & 0 & (4.4, 3.8, 3.8) & 0.0048 & 6.8 & 5.2 & 0.2 & 0.0 & -180.0 & 0.7 & 36.4 & 8.5 & 44.8 & $2.1\times 10^{-6}$ \\
190.5 & 5.2 & 0 & (4.4, 3.8, 3.8) & 0.0048 & 6.8 & 5.2 & 0.2 & -180.0 & 0.0 & 0.7 & 34.1 & 8.5 & 44.8 & $-2.2\times 10^{-6}$ \\
159.0 & 352.2 & 0 & (5.1, 4.9, 1.3) & 0.0047 & 3.9 & 7.8 & 0.1 & 0.0 & -180.0 & 0.8 & 32.5 & 8.3 & 45.1 & $3.4\times 10^{-6}$ \\
\hline
 &  &  &  [11],[13],[22] &  &  &  &  &  &  &  &  &  &  &  \\
164.6 & 186.0 & 0 & (5.1, 3.2, 1.5) & 0.0047 & 5.5 & 6.0 & 0.1 & 180.0 & 0.0 & -179.2 & 35.6 & 8.1 & 44.6 & $1.7\times 10^{-6}$ \\
164.6 & 354.0 & 0 & (5.4, 5.2, 1.5) & 0.0047 & 5.5 & 6.0 & 0.1 & 0.0 & -180.0 & 0.8 & 34.4 & 8.2 & 44.9 & $4.4\times 10^{-6}$ \\
344.6 & 186.0 & 0 & (5.4, 5.2, 1.5) & 0.0047 & 5.5 & 6.0 & 0.1 & -180.0 & 0.0 & 0.8 & 35.6 & 8.2 & 44.9 & $-4.5\times 10^{-6}$ \\
344.6 & 354.0 & 0 & (5.1, 3.2, 1.5) & 0.0047 & 5.5 & 6.0 & 0.1 & 0.0 & -180.0 & -179.2 & 34.9 & 8.2 & 44.6 & $-2.1\times 10^{-6}$ \\
\hline
 &  &  &  [22],[23],[31] &  &  &  &  &  &  &  &  &  &  &  \\
174.7 & 178.8 & 0 & (5.1, 4.7, 2.6) & 0.0048 & 9.6 & 3.6 & 0.3 & 179.9 & 0.0 & 0.1 & 31.0 & 9.3 & 44.7 & $-7.3\times 10^{-6}$ \\
\hline
 &  &  &  [11],[22],[31] &  &  &  &  &  &  &  &  &  &  &  \\
343.2 & 178.9 & 0 & (5.4, 3.5, 2.7) & 0.0048 & 9.6 & 3.2 & 0.1 & 180.0 & 0.0 & -0.3 & 30.8 & 9.0 & 44.5 & $2.6\times 10^{-6}$ \\
\hline
 &  &  & [11],[13],[21],[22] &  &  &  &  &  &  &  &  &  &  &  \\
164.3 & 353.0 & 0 & (5.3, 5.9, 1.4) & 0.0048 & 6.3 & 7.0 & 0.1 & 0.0 & -180.0 & -0.8 & 34.8 & 9.4 & 44.8 & $4.7\times 10^{-6}$ \\
\hline
 &  &  & [11],[21],[22],[31] &  &  &  &  &  &  &  &  &  &  &  \\
171.7 & 358.3 & 0 & (5.3, 4.4, 2.4) & 0.0047 & 9.0 & 5.1 & 0.1 & 180.0 & 0.0 & -0.3 & 32.5 & 10.0 & 44.5 & $-6.6\times 10^{-7}$ \\
351.7 & 358.3 & 0 & (5.6, 3.8, 2.4) & 0.0047 & 9.0 & 5.1 & 0.1 & -180.0 & 0.0 & 179.8 & 32.4 & 9.9 & 44.2 & $-4.9\times 10^{-7}$ \\
\hline
 &  &  & [11],[12],[22],[31] &  &  &  &  &  &  &  &  &  &  &  \\
175.5 & 1.6 & 0 & (5.1, 4.0, 2.4) & 0.0047 & 6.8 & 4.8 & 0.2 & 0.0 & -180.0 & 0.8 & 36.7 & 8.2 & 44.8 & $1.1\times 10^{-7}$ \\
342.3 & 183.1 & 0 & (5.5, 3.9, 2.0) & 0.0048 & 4.1 & 9.3 & 0.1 & 0.0 & -180.0 & 0.8 & 31.6 & 9.4 & 45.1 & $6.1\times 10^{-6}$ \\
355.5 & 178.4 & 0 & (5.1, 4.0, 2.4) & 0.0047 & 6.8 & 4.8 & 0.2 & -180.0 & 0.0 & 0.8 & 33.8 & 8.2 & 44.8 & $-1.2\times 10^{-7}$ \\
\hline
\hline
\end{tabular*}\\
\vskip .2cm
(Table continued on the next page)
\end{table}
\end{landscape}

\begin{landscape}
\begin{table}
\centering
Two-angle Results, TBM ($\beta_{23} = 0$) cont'd
\vskip .3cm
\footnotesize
\begin{tabular*}{1.53\textwidth}{ c c c| c | c | c c c c c c || c c c c }
\hline
\hline
&  &  &  &  &  &  &  &  &  &  &  &  &  &  \\
$\beta_{12}$ & $\beta_{13}$ & $\beta_{23}$ & GJ Entries & $m_e/m_\mu$ & $\theta_{12}^e$ & $\theta_{13}^e$ & $\theta_{23}^e$ & $\delta_{12}^e$ & $\delta_{13}^e$ & $\delta_{23}^e$ & $\theta_{12}$ & $\theta_{13}$ & $\theta_{23}$ & $J_{MNSP}$ \\
&  &  &  &  &  &  &  &  &  &  &  &  &  &  \\
\hline
 &  &  & [12],[21],[22],[23] &  &  &  &  &  &  &  &  &  &  &  \\
176.5 & 171.0 & 0 & (4.6, 3.6, 1.3) & 0.0046 & 4.2 & 9.0 & 0.3 & 0.0 & 180.0 & -0.4 & 31.8 & 9.3 & 45.3 & $-9.3\times 10^{-7}$ \\
356.5 & 171.0 & 0 & (5.1, 3.3, 1.3) & 0.0046 & 4.1 & 9.0 & 0.3 & 0.0 & -180.0 & 179.6 & 31.8 & 9.3 & 44.7 & $2.8\times 10^{-6}$ \\
\hline
 &  &  & [11],[13],[22],[23] &  &  &  &  &  &  &  &  &  &  &  \\
349.4 & 4.0 & 0 & (5.4, 5.1, 1.8) & 0.0048 & 9.0 & 4.0 & 0.3 & 180.0 & 0.0 & -0.4 & 31.7 & 9.2 & 44.7 & $2.5\times 10^{-8}$ \\
\hline
 &  &  & [11],[22],[23],[31] &  &  &  &  &  &  &  &  &  &  &  \\
351.0 & 358.4 & 0 & (5.7, 3.7, 2.4) & 0.0048 & 6.9 & 4.7 & 0.3 & -180.0 & 0.0 & 0.1 & 33.7 & 8.2 & 44.9 & $3.6\times 10^{-6}$ \\
\hline
 &  &  & [11],[12],[22],[23],[31] &  &  &  &  &  &  &  &  &  &  &  \\
162.0 & 182.0 & 0 & (5.2, 3.4, 2.3) & 0.0048 & 6.6 & 5.9 & 0.3 & 0.0 & 180.0 & -0.4 & 35.8 & 8.8 & 44.9 & $-4.5\times 10^{-6}$ \\
342.0 & 358.0 & 0 & (5.2, 3.4, 2.3) & 0.0048 & 6.6 & 5.9 & 0.3 & 180.0 & 0.0 & -0.4 & 34.8 & 8.8 & 44.9 & $4.4\times 10^{-6}$ \\
\hline
 &  &  & [11],[21],[22],[23],[31] &  &  &  &  &  &  &  &  &  &  &  \\
166.6 & 177.9 & 0 & (6.2, 4.0, 2.2) & 0.0049 & 6.2 & 6.3 & 0.3 & -180.0 & 0.0 & 0.1 & 35.3 & 8.8 & 44.9 & $-5.2\times 10^{-6}$ \\
346.6 & 2.1 & 0 & (6.2, 4.0, 2.2) & 0.0049 & 6.2 & 6.3 & 0.3 & 0.0 & -180.0 & 0.1 & 35.2 & 8.8 & 44.9 & $4.8\times 10^{-6}$ \\
\hline
 &  &  & [11],[12],[13],[21],[22] &  &  &  &  &  &  &  &  &  &  &  \\
339.4 & 3.8 & 0 & (5.2, 3.8, 1.8) & 0.0046 & 9.2 & 3.8 & 0.1 & -180.0 & 0.0 & 177.6 & 31.4 & 9.2 & 44.2 & $-8.6\times 10^{-6}$ \\
356.8 & 183.8 & 0 & (4.8, 4.4, 1.8) & 0.0046 & 8.2 & 3.8 & 0.1 & 180.0 & 0.0 & -2.3 & 32.2 & 8.4 & 44.7 & $7.8\times 10^{-6}$ \\
\hline
 &  &  & [11],[13],[21],[22],[23] &  &  &  &  &  &  &  &  &  &  &  \\
171.7 & 4.5 & 0 & (5.5, 3.5, 1.7) & 0.0047 & 9.0 & 4.5 & 0.3 & -180.0 & 0.0 & 179.6 & 32.0 & 9.5 & 44.1 & $-2.8\times 10^{-6}$ \\
351.7 & 4.5 & 0 & (5.1, 5.1, 1.7) & 0.0047 & 9.0 & 4.5 & 0.3 & 180.0 & 0.0 & -0.4 & 32.0 & 9.5 & 44.7 & $2.2\times 10^{-6}$ \\
\hline
 &  &  & [11],[12],[13],[22],[23] &  &  &  &  &  &  &  &  &  &  &  \\
339.4 & 4.9 & 0 & (5.2, 4.0, 1.7) & 0.0047 & 9.2 & 4.7 & 0.3 & -180.0 & 0.0 & 1.2 & 32.2 & 9.9 & 44.6 & $-9.4\times 10^{-6}$ \\
\hline
\hline
\end{tabular*}
\caption{Two-angle search results starting from an asymmetric $Y^{(-1/3)}$ with $\beta_{23}=0$, TBM, $A=0.77$, (++) and $3 \sigma$ range for neutrino mixing angles assumed. For each row there exists a corresponding solution with an additional GJ factor in the (32) entry.}
\label{tb:2ATBM23}
\end{table}

\end{landscape}

\begin{landscape}
\begin{table}
\vskip .3cm
\centering
Table \ref{tb:2ATBM12}: Two-angle Results, TBM ($\beta_{12} = 0$)
\vskip .3cm
\footnotesize
\begin{tabular*}{1.53\textwidth}{ c c c| c | c | c c c c c c || c c c c }
\hline
\hline
&  &  &  &  &  &  &  &  &  &  &  &  &  &  \\
$\beta_{12}$ & $\beta_{13}$ & $\beta_{23}$ & GJ Entries & $m_e/m_\mu$ & $\theta_{12}^e$ & $\theta_{13}^e$ & $\theta_{23}^e$ & $\delta_{12}^e$ & $\delta_{13}^e$ & $\delta_{23}^e$ & $\theta_{12}$ & $\theta_{13}$ & $\theta_{23}$ & $J_{MNSP}$ \\
&  &  &  &  &  &  &  &  &  &  &  &  &  &  \\
\hline
 &  &  & [21],[23] &  &  &  &  &  &  &  &  &  &  &  \\
$\star$ 0.0 & 3.1 & 90.0 & $\begin{pmatrix}
4.7 & 3.8 & 3.7 \\ 4.2 & 2.2 & 2.8 \\ 2.0 & 0.0 & 4.9
\end{pmatrix}$ & 0.0047 & 9.1 & 3.1 & 89.9 & 359.7 & -0.3 & -359.7 & 31.0 & 8.6 & 44.6 & $3.7\times 10^{-7}$ \\
\hline
 &  &  & [22],[23] &  &  &  &  &  &  &  &  &  &  &  \\
0.0 & 177.2 & 269.0 & $\begin{pmatrix}
4.8 & 3.8 & 3.7 \\ 4.1 & 2.2 & 2.7 \\ 2.0 & 0.0 & 2.8
\end{pmatrix}$ & 0.0048 & 9.0 & 2.7 & 88.7 & 0.0 & 0.0 & 0.0 & 31.0 & 8.4 & 45.8 & $4.1\times 10^{-7}$ \\
0.0 & 357.2 & 181.0 & $\begin{pmatrix}
4.8 & 3.7 & 3.8 \\ 4.1 & 2.7 & 2.2 \\ 2.0 & 2.8 & 0.0
\end{pmatrix}$  & 0.0048 & 9.0 & 2.7 & 1.3 & -180.0 & 0.0 & 0.0 & 31.0 & 8.4 & 45.8 & $4.1\times 10^{-7}$ \\
\hline
 &  &  & [21],[22] &  &  &  &  &  &  &  &  &  &  &  \\
$\star$ 0.0 & 183.1 & 0.0 & $\begin{pmatrix}
4.8 & 3.7 & 3.8 \\ 4.2 & 2.8 & 2.2 \\ 2.0 & 4.9 & 0.0
\end{pmatrix}$ & 0.0047 & 9.1 & 3.1 & 0.2 & 180.0 & 0.0 & -0.3 & 31.0 & 8.6 & 44.6 & $3.7\times 10^{-7}$ \\
\hline
 &  &  & [22],[23],[33] &  &  &  &  &  &  &  &  &  &  &  \\
0.0 & 177.2 & 270.0 & $\left(
\begin{array}{ccc}
 4.8 & 3.8 & 3.7 \\
 4.1 & 2.2 & 2.8 \\
 2.0 & 0.0 & 4.9
\end{array}
\right)$ & 0.0048 & 9.1 & 2.7 & 89.6 & 0.1 & 0.1 & 0.0 & 30.8 & 8.4 & 44.9 & $-5.6\times 10^{-8}$ \\
\hline
 &  &  & [22],[23],[32] &  &  &  &  &  &  &  &  &  &  &  \\
0.0 & 357.2 & 180.0 & $\left(
\begin{array}{ccc}
 4.8 & 3.7 & 3.8 \\
 4.1 & 2.8 & 2.2 \\
 2.0 & 4.9 & 0.0
\end{array}
\right)$  & 0.0048 & 9.1 & 2.7 & 0.5 & -180.0 & 0.0 & 0.1 & 30.8 & 8.4 & 44.9 & $-5.6\times 10^{-8}$ \\
\hline
\hline
\end{tabular*}
\caption{Two-angle search results starting from an asymmetric $Y^{(-1/3)}$ with $\beta_{12}=0$, TBM, $A=0.77$, (++) and $3 \sigma$ range for neutrino mixing angles assumed. The row labelled by a ``$\star$''' also has a corresponding solution with an additional GJ factor in the entry whose order of magnitude is 4.9.}
\label{tb:2ATBM12}
\end{table}
\end{landscape}

\begin{landscape}
\begin{table}
\centering
Table \ref{tb:2ABM23}: Two-angle Results, BM ($\beta_{23} = 0$)
\vskip .3cm
\footnotesize
\begin{tabular*}{1.58\textwidth}{ c c c| c | c | c c c c c c || c c c c }
\hline
\hline
&  &  &  &  &  &  &  &  &  &  &  &  &  &  \\
$\beta_{12}$ & $\beta_{13}$ & $\beta_{23}$ & GJ Entries & $m_e/m_\mu$ & $\theta_{12}^e$ & $\theta_{13}^e$ & $\theta_{23}^e$ & $\delta_{12}^e$ & $\delta_{13}^e$ & $\delta_{23}^e$ & $\theta_{12}$ & $\theta_{13}$ & $\theta_{23}$ & $J_{MNSP}$ \\
&  &  &  &  &  &  &  &  &  &  &  &  &  &  \\
\hline
 &  &  & [22] &  &  &  &  &  &  &  &  &  &  &  \\
173.4 & 164.0 & 0 & (4.5, 3.2, 0.9) & 0.0048 & 2.5 & 16.0 & 0.2 & -180.0 & 180.0 & 179.8 & 31.8 & 9.5 & 46.3 & $1.0\times 10^{-6}$ \\
353.4 & 164.0 & 0 & (4.8, 2.9, 0.9) & 0.0048 & 2.4 & 16.0 & 0.2 & 180.0 & -180.0 & -0.2 & 31.8 & 9.5 & 46.6 & $-1.2\times 10^{-6}$ \\
\hline
 &  &  & [21],[22] &  &  &  &  &  &  &  &  &  &  &  \\
187.0 & 176.9 & 0 & (4.4, 3.7, 2.0) & 0.0050 & 15.8 & 3.1 & 0.1 & -180.0 & 180.0 & 179.7 & 31.6 & 8.9 & 44.2 & $-6.6\times 10^{-7}$ \\
\hline
 &  &  & [12],[22],[23] &  &  &  &  &  &  &  &  &  &  &  \\
201.0 & 357.9 & 0 & (4.1, 3.3, 2.2) & 0.0048 & 13.7 & 2.0 & 0.3 & -180.0 & 180.0 & 179.7 & 33.8 & 8.2 & 44.2 & $-3.4\times 10^{-7}$ \\
\hline
 &  &  & [11],[13],[21],[22] &  &  &  &  &  &  &  &  &  &  &  \\
170.6 & 0.8 & 0 & (5.8, 3.8, 2.9) & 0.0047 & 12.5 & 0.8 & 0.2 & -180.0 & 0.0 & 0.8 & 36.6 & 9.4 & 44.4 & $-1.3\times 10^{-6}$ \\
350.6 & 359.2 & 0 & (5.8, 3.8, 2.9) & 0.0047 & 12.5 & 0.8 & 0.2 & 180.0 & -180.0 & -179.2 & 35.5 & 8.3 & 44.3 & $2.4\times 10^{-6}$ \\
353.8 & 357.4 & 0 & (5.3, 3.8, 2.1) & 0.0048 & 14.9 & 2.6 & 0.2 & 180.0 & -180.0 & -179.2 & 32.6 & 8.6 & 44.2 & $2.1\times 10^{-6}$ \\
\hline
 &  &  & [11],[13],[22],[23] &  &  &  &  &  &  &  &  &  &  &  \\
341.4 & 177.5 & 0 & (5.1, 3.3, 2.1) & 0.0049 & 14.7 & 2.4 & 0.3 & -180.0 & 180.0 & 179.6 & 32.8 & 8.6 & 44.1 & $-2.4\times 10^{-6}$ \\
\hline
 &  &  & [12],[21],[22],[23] &  &  &  &  &  &  &  &  &  &  &  \\
178.3 & 345.0 & 0 & (4.4, 3.1, 0.9) & 0.0048 & 2.5 & 15.0 & 0.3 & -180.0 & 180.0 & 179.6 & 32.6 & 8.8 & 46.0 & $2.1\times 10^{-6}$ \\
358.3 & 345.0 & 0 & (4.8, 3.0, 0.9) & 0.0048 & 2.4 & 15.0 & 0.3 & 180.0 & -180.0 & -0.4 & 32.6 & 8.8 & 46.6 & $-3.3\times 10^{-6}$ \\
\hline
 &  &  & [11],[13],[21],[22],[31] &  &  &  &  &  &  &  &  &  &  &  \\
171.6 & 0.7 & 0 & (5.6, 3.9, 3.0) & 0.0047 & 15.4 & 2.0 & 0.3 & 180.0 & -180.0 & -179.9 & 32.6 & 9.4 & 44.0 & $3.2\times 10^{-6}$ \\
350.4 & 359.8 & 0 & (5.8, 3.9, 3.8) & 0.0050 & 12.0 & 0.5 & 0.3 & -180.0 & 0.0 & 0.1 & 36.8 &8.9 & 44.6 & $4.3\times 10^{-6}$ \\
\hline
 &  &  & [11],[12],[13],[21],[22] &  &  &  &  &  &  &  &  &  &  &  \\
184.5 & 176.1 & 0 & (4.5, 3.8, 1.8) & 0.0048 & 15.8 & 3.9 & 0.1 & -180.0 & 180.0 & 177.7 & 31.0 & 8.4 & 44.4 & $-5.7\times 10^{-6}$ \\
\hline
 &  &  & [11],[12],[21],[22],[23],[31] &  &  &  &  &  &  &  &  &  &  &  \\
5.0 & 181.3 & 0 & (4.5, 4.1, 2.6) & 0.0050 & 16.2 & 3.8 & 0.3 & -180.0 & 180.0 & 179.6 & 30.8 & 8.7 & 44.1 & $-4.5\times 10^{-6}$ \\
\hline
\hline
\end{tabular*}
\caption{Two-angle search results starting from an asymmetric $Y^{(-1/3)}$, with $\beta_{23}=0$, BM, $A=0.77$, (++) and $3 \sigma$ range for neutrino mixing angles assumed. Each row, except for the last one, also has a corresponding solution with an additional GJ factor in the (32) entry. }
\label{tb:2ABM23}
\end{table}
\end{landscape}

\begin{landscape}
\begin{table}
\centering
Table \ref{tb:2ABM12}: Two-angle Results, BM ($\beta_{12} = 0$)
\vskip .3cm
\footnotesize
\begin{tabular*}{1.6\textwidth}{ c c c| c | c | c c c c c c || c c c c }
\hline
\hline
&  &  &  &  &  &  &  &  &  &  &  &  &  &  \\
$\beta_{12}$ & $\beta_{13}$ & $\beta_{23}$ & GJ Entries & $m_e/m_\mu$ & $\theta_{12}^e$ & $\theta_{13}^e$ & $\theta_{23}^e$ & $\delta_{12}^e$ & $\delta_{13}^e$ & $\delta_{23}^e$ & $\theta_{12}$ & $\theta_{13}$ & $\theta_{23}$ & $J_{MNSP}$ \\
&  &  &  &  &  &  &  &  &  &  &  &  &  &  \\
\hline
 &  &  & [22],[23] &  &  &  &  &  &  &  &  &  &  &  \\
0 & 174.8 & 5.0 & $\begin{pmatrix}
4.9 & 3.8 & 3.8 \\ 3.8 & 2.9 & 2.2 \\ 1.6 & 1.6 & 0.0
\end{pmatrix}$ & 0.0049 & 16.1 & 5.1 & 4.7 & 180.0 & -180.0 & 0.0 & 30.7 & 9.0 & 49.3 & $1.9\times 10^{-6}$ \\
0 & 354.8 & 85.0 & $\begin{pmatrix}
4.9 & 3.8 & 3.8 \\ 3.8 & 2.2 & 2.9 \\ 1.6 & 0.0 & 1.6
\end{pmatrix}$ & 0.0049 & 16.1 & 5.1 & 85.3 & 0.0 & 180.0 & 0.0 & 30.7 & 9.0 & 49.3 & $1.9\times 10^{-6}$  \\
\hline
 &  &  & [21],[22],[32] &  &  &  &  &  &  &  &  &  &  &  \\
0 & 5.6 & 178.0 & $\begin{pmatrix}
4.6 & 3.8 & 3.8 \\ 3.8 & 2.8 & 2.2 \\ 1.6 & 2.3 & 0.0
\end{pmatrix}$ & 0.0046 & 16.0 & 5.5 & 6.0 & 180.0 & -180.0 & 0.0 & 30.6 & 9.0 & 50.7 & $-2.4\times 10^{-6}$ \\
\hline
 &  &  & [13],[21],[22],[23] &  &  &  &  &  &  &  &  &  &  &  \\
0 & 15.7 & 263.0 & $\begin{pmatrix}
4.3 & 3.8 & 3.8 \\ 3.1 & 2.2 & 3.0 \\ 0.9 & 0.0 & 1.4
\end{pmatrix}$ & 0.0049 & 0.8 & 15.7 & 83.3 & -1.5 & 180.0 & 0.0 & 32.1 & 9.0 & 52.8 & $5.6\times 10^{-5}$ \\
0 & 164.3 & 84.0 & $\begin{pmatrix}
4.3 & 3.8 & 3.7 \\ 3.0 & 2.2 & 2.6 \\ 0.9 & 0.0 & 1.6
\end{pmatrix}$ & 0.0048 & 0.7 & 15.7 & 83.7 & 1.6 & 180.0 & 0.0 & 32.2 & 9.2 & 52.4 & $4.6\times 10^{-5}$ \\
\hline
 &  &  & [12],[21],[22],[23] &  &  &  &  &  &  &  &  &  &  &  \\
0 & 195.7 & 187.0 & $\begin{pmatrix}
4.3 & 3.8 & 3.8 \\ 3.1 & 3.0 & 2.2 \\ 0.9 & 1.4 & 0.0
\end{pmatrix}$ & 0.0050 & 0.8 & 15.7 & 6.7 & 178.5 & -180.0 & 0.0 & 32.1 & 9.0 & 52.8 & $-5.6\times 10^{-5}$ \\
0 & 344.3 & 6.0 & $\begin{pmatrix}
4.3 & 3.7 & 3.8 \\ 3.0 & 2.6 & 2.2 \\ 0.9 & 1.6 & 0.0
\end{pmatrix}$ & 0.0048 & 0.7 & 15.7 & 83.7 & 1.6 & 180.0 & 0.0 & 32.2 & 9.2 & 52.4 & $4.6\times 10^{-5}$ \\
\hline
\hline
\end{tabular*}
\caption{Two-angle search results starting from an asymmetric $Y^{(-1/3)}$, with $\beta_{12}=0$, BM, $A=0.77$, (++) and $3 \sigma$ range for neutrino mixing angles assumed. In this case there are no additional solutions found by the addition of GJ factors into small entries. }
\label{tb:2ABM12}
\end{table}
\end{landscape}

\begin{landscape}
\begin{table}
\centering
Table \ref{tb:2ABM13}: Two-angle Results, BM ($\beta_{13} = 0$)
\vskip .3cm
\footnotesize
\begin{tabular*}{1.53\textwidth}{ c c c| c | c | c c c c c c || c c c c }
\hline
\hline
&  &  &  &  &  &  &  &  &  &  &  &  &  &  \\
$\beta_{12}$ & $\beta_{13}$ & $\beta_{23}$ & GJ Entries & $m_e/m_\mu$ & $\theta_{12}^e$ & $\theta_{13}^e$ & $\theta_{23}^e$ & $\delta_{12}^e$ & $\delta_{13}^e$ & $\delta_{23}^e$ & $\theta_{12}$ & $\theta_{13}$ & $\theta_{23}$ & $J_{MNSP}$ \\
&  &  &  &  &  &  &  &  &  &  &  &  &  &  \\
\hline
 &  &  & [12],[13],[22] &  &  &  &  &  &  &  &  &  &  &  \\
123.2 & 0 & 172.1 & $\begin{pmatrix}
4.0 & 3.9 & 3.8 \\ 2.9 & 2.9 & 2.2 \\ 5.0 & 1.3 & 0.0
\end{pmatrix}$ & 0.0048 & 16.5 & 0.0 & 7.8 & 180.0 & -9.5 & -180.0 & 31.7 & 9.9 & 36.1 & $9.2\times 10^{-6}$ \\
316.0 & 0 & 185.8 & $\begin{pmatrix}
4.2 & 3.8 & 3.8 \\ 3.0 & 2.8 & 2.2 \\ 5.2 & 1.5 & 0.0
\end{pmatrix}$ & 0.0048 & 12.9 & 0.0 & 5.7 & -180.0 & 170.2 & 0.1 & 36.7 & 10.0 & 50.0 & $-3.8\times 10^{-6}$ \\
\hline
 &  &  & [12],[13],[23] &  &  &  &  &  &  &  &  &  &  &  \\
136.0 & 0 & 84.2 & $\begin{pmatrix}
4.2 & 3.8 & 3.8 \\ 3.0 & 2.2 & 2.8 \\ 5.2 & 0.0 & 1.5
\end{pmatrix}$ & 0.0048 & 12.9 & 0.0 & 84.3 & 0.1 & 170.3 & -0.1 & 36.7 & 10.0 & 50.0 & $-3.8\times 10^{-6}$ \\
303.2 & 0 & 97.9 & $\begin{pmatrix}
4.0 & 3.8 & 3.9 \\ 2.9 & 2.2 & 2.9 \\ 5.0 & 0.0 & 1.3
\end{pmatrix}$ & 0.0048 & 16.5 & 0.0 & 82.2 & -180.0 & 170.5 & 180.0 & 31.7 & 9.9 & 36.1 & $9.2\times 10^{-6}$ \\
\hline
 &  &  & [13],[22],[23],[33] &  &  &  &  &  &  &  &  &  &  &  \\
134.3 & 0 & 88.2 & $\begin{pmatrix}
4.1 & 3.8 & 3.8 \\ 3.0 & 2.2 & 2.9 \\ 5.1 & 0.0 & 2.3
\end{pmatrix}$ & 0.0049 & 13.8 & 0.1 & 85.0 & -180.0 & 0.3 & 180.0 & 34.3 & 8.7 & 39.2 & $2.6\times 10^{-6}$ \\
316.5 & 0 & 91.9 & $\begin{pmatrix}
4.2 & 3.8 & 3.8 \\ 3.0 & 2.2 & 2.9 \\ 5.2 & 0.0 & 2.3
\end{pmatrix}$ & 0.0046 & 12.7 & 0.1 & 5.3 & 0.0 & 0.3 & 0.0 & 36.9 & 9.8 & 49.6 & $-1.8\times 10^{-6}$ \\
\hline
 &  &  & [12],[22],[23],[32] &  &  &  &  &  &  &  &  &  &  &  \\
314.3 & 0 & 181.8 & $\begin{pmatrix}
4.1 & 3.8 & 3.8 \\ 3.0 & 2.9 & 2.2 \\ 5.1 & 2.3 & 0.0
\end{pmatrix}$ & 0.0049 & 13.8 & 0.1 & 5.0 & 180.0 & -179.7 & -180.0 & 34.3 & 8.7 & 39.2 & $2.6\times 10^{-6}$ \\
136.5 & 0 & 178.1 & $\begin{pmatrix}
4.2 & 3.8 & 3.8 \\ 3.0 & 2.9 & 2.2 \\ 5.2 & 2.3 & 0.0
\end{pmatrix}$ & 0.0046 & 12.7 & 0.1 & 5.3 & -180.0 & 0.3 & 0.0 & 36.9 & 9.8 & 49.6 & $-1.8\times 10^{-6}$ \\
\hline
\hline
\end{tabular*}
\caption{Two-angle search results starting from an asymmetric $Y^{(-1/3)}$, with $\beta_{13}=0$, BM, $A=0.77$, (++) and $3 \sigma$ range for neutrino mixing angles assumed. Each row also has a corresponding solution with an additional GJ factor in the entry whose order of magnitude is around or above 5.0.}
\label{tb:2ABM13}
\end{table}
\end{landscape}

\subsection*{Solutions with three angles in $\m V$}

We organize the three-angle solutions according to the cases of TBM or BM, $\beta_{23}$ close to the x-axis or the y-axis, and normal hierarchy(NH) or inverted hierarchy(IH) solutions. $1\sigma$ ranges for neutrino mixing angles are used.

\begin{table}[h!]
\centering
\caption{Full three-angle NH solutions (also serve as part of the three-angle IH solutions) for TBM and $\beta_{23}$ close to the x-axis, with $A=0.77$, (+,+) and $1\sigma$ ranges for neutrino mixing angles assumed.}
\vskip 0.3cm
\footnotesize
\begin{tabular}{l l}
\begin{tabular}{|c | c  c  c|}
\hline
GJ Entries & $\beta_{12}$ & $\beta_{13}$ & $\beta_{23}$ \\
\hline
[22],[32] & 168.5 & 188.1 & 181.0 \\
	& 348.5 & 351.9 & 1.0 \\
	& 168.3 & 351.6 & 1.8 \\
	& 348.3 & 188.4 & 181.8 \\
\hline
[21],[22] & 169.2 & 353.9 & 175.8 \\
& 349.2 & 186.1 & 355.8 \\
\hline
[12],[21],[22] & 7.9 & 174.7 & 176.6 \\
	& 187.9 & 5.3 & 356.6 \\
	& 12.4 & 173.7 & 176.7\\
	& 192.4 & 6.3 & 356.2 \\
\hline
[22],[23],[32] & 154.6 & 174.3 & 181.9 \\
	& 334.6 & 5.7 & 1.8 \\
	& 163.4 & 175.2 & 181.5 \\
	& 343.4 & 4.8 & 1.5 \\
\hline
[11],[13],[22] & 163.8 & 351.5 & 355.3 \\
	& 343.8 & 188.5 & 175.3 \\
\hline
[11],[22],[31] & 170.1 & 357.3 & 176.7 \\
	& 350.1 & 182.7 & 356.7\\
\hline
[21],[22],[32] & 171.0 & 354.7 & 181.2 \\
	& 351.0 & 185.3 & 1.2\\
\hline
[12],[21],[22],[32] & 11.2 & 174.5 & 181.4 \\
	& 191.2 & 5.5 & 1.4\\
\hline
[11],[13],[22],[32] & 164.9 & 187.7 & 181.3 \\
	& 344.9 & 352.3 & 1.3 \\
	& 167.1 & 351.1 & 1.9 \\
	& 344.6 & 187.0 & 181.3\\
\hline
[11],[13],[21],[22] & 163.4 & 352.4 & 355.3\\
	& 343.4 & 187.6 & 175.3\\
\hline
\end{tabular}
&
\begin{tabular}{|c|c c c|}
\hline
[11],[21],[22],[31] & 171.4 & 358.0 & 356.0 \\
	& 351.4 & 181.6 & 176.5 \\
\hline
[11],[12],[22],[31] & 173.5 & 1.8 & 177.6 \\
	& 353.5 & 178.1 & 356.4 \\
\hline
[12],[21],[22],[23] & 177.0 & 171.6 & 355.0\\
	& 357.0 & 8.4 & 175.0 \\
\hline
[11],[22],[31],[32] & 347.9 & 182.7 & 1.0 \\
	& 167.9 & 357.3 & 181.0\\
\hline
[12],[21],[22],[23],[31]	& 11.0 & 358.2 & 355.0 \\
	& 191.0 & 181.8 & 175.0\\
\hline
[11],[12],[22],[23],[31] & 161.1 & 182.1 & 174.4 \\
	& 341.1 & 357.9 & 354.4\\
\hline
[11],[21],[22],[23],[31] & 166.6 & 177.9 & 354.3 \\
	& 346.6 & 2.1 & 174.3 \\
\hline
[11],[22],[23],[31],[32] & 168.6 & 181.9 & 180.9 \\
	& 348.6 & 358.0 & 1.4\\
	&171.5 & 181.9 & 181.3 \\
	& 351.5 & 358.1 & 1.3 \\
\hline
[11],[12],[13],[21],[22] & 173.3 & 354.8 & 176.5 \\
	& 353.3 & 185.1 & 356.8 \\
\hline
[11],[12],[22],[31],[32] & 174.4 & 1.9 & 181.2 \\
	& 354.4 & 178.1 & 1.2 \\
\hline
[12],[21],[22],[23],[32] & 175.6 & 171.9 & 1.1 \\
	& 355.6 & 8.1 & 181.1 \\
\hline 
[11],[12],[13],[22],[23],[32] & 157.4 & 174.5 & 181.7 \\
\hline
[11],[13],[21],[22],[23],[32] & 171.1 & 174.1 & 181.4 \\
	& 351.1 & 4.9 & 1.4 \\
\hline
\end{tabular}

\end{tabular}
\end{table}

\newpage
\begin{table}[h!]
\centering
\caption{Other three-angle IH solutions for TBM and $\beta_{23}$ close to the x-axis, with $A=0.77$, (+,+) and $1\sigma$ ranges for neutrino mixing angles assumed.}
\vskip 0.3 cm
\footnotesize
\begin{tabular}{l l}
\begin{tabular}{|c | c  c  c|}
\hline
GJ Entries & $\beta_{12}$ & $\beta_{13}$ & $\beta_{23}$ \\
\hline
[22] & 168.3 & 187.9 & 179.6 \\
	& 168.1 & 352.1 & 359.7 \\
	& 348.1 & 187.9 & 179.7 \\
	& 348.3 & 352.0 & 359.6 \\
\hline
[21],[22] & 155.3 & 175.1 & 185.4 \\
	& 335.3 & 4.9 & 5.4 \\
	& 351.0 & 185.2 & 0.0 \\
\hline
[22],[32] & 168.2 & 352.3 & 359.0 \\
	& 347.8 & 352.4 & 358.2 \\
\hline
[11],[13],[22] & 164.6 & 186.6 & 179.9 \\
	& 344.6 & 352.4 & 0.0 \\
\hline
[12],[21],[22] & 186.5 & 4.3 & 0.1 \\
\hline
[11],[22],[32] & 350.6 & 185.2 & 358.5 \\
\hline
[11],[22],[31],[32] & 167.0 & 352.1 & 358.3 \\
	& 344.0 & 353.2 & 357.8 \\
\hline
[11],[13],[22],[23] & 169.2 & 176.3 & 184.9 \\
	& 349.2 & 3.7 & 4.9 \\
\hline
[11],[21],[22],[31] & 171.6  & 181.3 &  183 \\
	& 351.6 & 358.3 & 4.1 \\
\hline
[12],[21],[22],[23] & 175.4 & 7.1 & 186.5 \\
	& 355.4 & 172.4 & 6.2 \\
	& 356.5 & 172.1 & 359.6 \\
\hline
[12],[21],[22],[32] & 185.5 & 3.9 & 357.8 \\
\hline
\end{tabular}
&
\begin{tabular}{| c | c c c|}
\hline
[11],[12],[13],[21],[22] & 157.5 & 175.9 & 184.0 \\
	& 337.5 & 4.1 & 4.0\\
\hline
[11],[12],[22],[23],[31] & 160.2 & 182.1 & 5.7 \\
	& 176.6 & 178.9 & 6.4 \\
	& 340.2 & 357.9 & 185.7 \\
	& 356.6 & 1.1 & 186.4 \\
\hline
[11],[13],[21],[22],[32] & 164.4 & 353.1 & 0.0 \\
	& 166.6 & 353.7 & 357.8 \\
	& 344.4 & 186.9 & 180 \\
	& 346.6 & 186.3 & 177.8 \\
\hline
[11],[21],[22],[23],[31] & 166.6 & 177.9 & 182.7 \\
	& 346.6 & 2.1 & 2.7 \\
\hline
[12],[21],[22],[23],[32] & 176.0 & 7.7 & 179.1 \\
	& 356.0 & 172.3 & 359.1 \\
\hline
[11],[12],[13],[21],[22],[32] & 159.0 & 175.7 & 178.8 \\
	& 174.0 & 353.5 & 179 \\
	& 338.7 & 4.3 & 358.8 \\
	& 354.0 & 184.4 & 358.0\\
\hline
[11],[12],[22],[23],[31],[32] & 161.7 & 182.1 & 180.2\\
	& 341.7 & 357.9 & 0.2\\
\hline
[11],[13],[21],[22],[23],[32] & 171.3 & 3.8 & 358.6 \\
	& 351.3 & 176.2 & 178.6\\
\hline
\end{tabular}

\end{tabular}
\end{table}

\newpage
\begin{table}[h!]
\centering
\caption{Full three-angle NH solutions (also serve as part of three-angle IH solutions) for TBM and $\beta_{23}$ close to the y-axis, with $A=0.77$, (+,+) and $1\sigma$ ranges for neutrino mixing angles assumed.}
\vskip 0.3 cm
\footnotesize
\begin{tabular}{l l}
\begin{tabular}{|c | c  c  c|}
\hline
GJ Entries & $\beta_{12}$ & $\beta_{13}$ & $\beta_{23}$ \\
\hline
[23],[33] & 168.5 & 8.1 & 269.0\\
	& 168.3 & 171.6 & 88.1 \\
\hline
[21],[23] & 169.2 & 173.9 & 274.1 \\
	& 349.2 & 6.1 & 94.1 \\
\hline
[13],[21],[23] & 6.7 & 355.3 & 272.4 \\
	& 186.7 & 184.7 & 92.4 \\
\hline
[22],[23],[33] & 154.6 & 354.3 & 268.1 \\
	& 334.6 & 185.7 & 88.1 \\
	& 343.0 & 184.6 & 89.1 \\
\hline
[11],[12],[23] & 163.8 & 171.5 & 94.4 \\
	& 343.8 & 8.0 & 274.6 \\
\hline
[11],[23],[31] & 170.1 & 177.3 & 273.2 \\
	& 350.1 & 2.7 & 93.2 \\
\hline
[21],[23],[33] & 171.0 & 174.7 & 268.8 \\
	& 351.0 & 5.3 & 88.8 \\
\hline
[13],[21],[23],[33] & 7.2 & 355.4 & 269.3 \\
	& 11.0 & 354.6 & 269.1 \\
	& 187.2 & 184.6 & 89.3 \\
\hline
[11],[12],[23],[33] & 164.8 & 6.8 & 269.2 \\
	& 164.5 & 173.1 & 88.9 \\
	& 344.5 & 6.7  & 269.2 \\
\hline
[11],[23],[31],[33] & 167.9 & 177.3 & 268.1 \\
	& 347.9 & 2.7 & 88.1 \\
\hline
[11],[21],[23],[31] & 171.4 & 178.0 & 93.7 \\
	& 351.4 & 1.6 & 273.2 \\
\hline
\end{tabular}
&
\begin{tabular}{| c | c c c |}
\hline
[13],[21],[22],[23] & 176.8 & 351.7 & 93.4 \\
	& 356.8 & 188.0 & 273.0 \\
\hline
[11],[12],[21],[23] & 343.4 & 7.6 & 274.2 \\
\hline
[11],[13],[23],[31] & 353.5 & 358.1 & 93.3 \\
\hline
[13],[21],[22],[23],[31] & 11.0 & 178.2 & 94.3 \\
	& 191.0 & 1.8 & 274.3 \\
\hline
[11],[13],[22],[23],[31] & 161.1 & 2.1 & 275.4 \\
	& 341.1 & 177.9 & 95.4 \\
\hline
[11],[21],[22],[23],[31] & 166.6 & 357.9 & 93.4 \\
	& 346.6 & 182.1 & 273.4 \\
\hline
[11],[22],[23],[31],[33] & 168.6 & 1.9 & 269.0 \\
	& 348.6 & 178.0 & 88.5 \\
\hline
[11],[12],[13],[21],[23] & 173.3 & 174.8 & 273.5 \\
	& 353.3 & 5.1 & 93.0 \\
\hline
[11],[13],[23],[31],[33] & 174.4 & 181.9 & 268.4 \\
	& 354.4 & 358.1 & 88.3 \\
\hline
[13],[21],[22],[23],[33] & 175.6 & 351.9 & 88.9 \\
	& 355.6 & 188.1 & 268.9 \\
\hline
[11],[12],[13],[22],[23],[33] & 157.4 & 354.5 & 268.3 \\
	& 337.4 & 185.5 & 88.3 \\
\hline
[11],[12],[21],[22],[23],[33] & 171.1 & 354.1 & 268.6 \\
	& 351.1 & 184.9 & 88.6 \\
\hline
\end{tabular}
\end{tabular}
\end{table}

\newpage
\begin{table}[h!]
\centering
\caption{Other three-angle IH solutions for TBM and $\beta_{23}$ close to the y-axis, with $A=0.77$, (+,+) and $1\sigma$ ranges for neutrino mixing angles assumed.}
\vskip 0.3 cm
\footnotesize
\begin{tabular}{l l}
\begin{tabular}{|c | c  c  c|}
\hline
GJ Entries & $\beta_{12}$ & $\beta_{13}$ & $\beta_{23}$ \\
\hline
[23] & 168.1 & 172.1 & 90.3 \\
	& 168.3 & 7.9 & 270.0 \\
	& 348.1 & 7.9 & 270.3 \\
	& 348.3 & 172.0 & 90.2 \\
\hline
[21],[23] & 155.3 & 355.1 & 264.5 \\
	& 170.0 & 174.3 & 272.2 \\
	& 335.3 & 184.9 & 84.5 \\
	& 350.0 & 5.7 & 92.2 \\
\hline
[23],[33] & 167.8 & 7.4 & 271.8 \\
	& 347.8 & 172.4 & 91.8 \\
\hline
[13],[21],[23] & 6.5 & 355.6 & 270.6 \\
	& 10.5 & 354.8 & 270.0 \\
	& 186.5 & 184.3 & 89.9 \\
	& 190.5 & 185.2 & 90.0 \\
\hline
[22],[23],[33] & 155.5 & 354.8 & 269.2 \\
	& 162.9 & 355.7 & 269.8 \\
	& 335.5 & 185.2 & 89.2 \\
	& 342.9 & 184.3 & 89.8 \\
\hline
[11],[12],[23] & 164.6 & 6.6 & 269.8 \\
	& 167.1 & 171.8 & 90.3 \\
	& 344.1 & 8.1 & 272.6 \\
	& 344.6 & 172.4 & 89.9 \\
	& 347.1 & 171.6 & 90.1 \\
\hline
[11],[23],[31] & 170.2 & 177.3 & 272.0 \\
	& 350.2 & 2.6 & 91.0 \\
\hline
[21],[23],[33] & 170.6 & 174.8 & 271.3 \\
	& 350.6 & 5.2 & 91.3 \\
\hline
[13],[21],[23],[33] & 5.5 & 356.1 & 272.1 \\
	& 185.5 & 183.9 & 92.1 \\
\hline
[11],[12],[23],[33] & 164.0 & 6.8 & 272.2 \\
	& 164.5 & 173.2 & 89.3\\
	& 344.5 & 6.7 & 269.3 \\
	& 344.1 & 173.0 & 91.7 \\
\hline
[11],[23],[31],[33] & 167.9 & 177.3 & 269.2 \\
\hline
[11],[12],[21],[23] & 163.8 & 172.7 & 92.4 \\
	& 171.6 & 1.3 & 266.4 \\
	& 343.8 & 7.3 & 272.4 \\
\hline
[11],[12],[22],[23] & 169.2 & 356.3 & 264.3 \\
\hline
[11],[21],[23],[31] & 171.5 & 178.3 & 92.0 \\
\hline
\end{tabular}
&
\begin{tabular}{|c | c  c  c|}
\hline
[13],[21],[22],[23] & 175.4 & 187.1 & 263.3 \\
	& 176.6 & 351.9 & 91.5 \\
	& 355.4 & 352.4 & 83.7 \\
	& 356.6 & 187.9 & 271.7 \\
\hline
[11],[13],[23],[31] & 353.6 & 358.2 & 92.0 \\
\hline
[11],[12],[13],[21],[23] & 157.5 & 355.9 & 265.8 \\
	& 173.5 & 175.1 & 271.9 \\
	& 337.5 & 184.1 & 85.8 \\
	& 353.5 & 4.9 & 91.9 \\
\hline
[11],[13],[22],[23],[31] & 160.2 & 2.1 & 84.2 \\
	& 168.7 & 1.8 & 269.6 \\
	& 341.9 & 178.0 & 91.6 \\
	& 340.2 & 177.9 & 264.2 \\
	& 356.6 & 181.1 & 263.5 \\
\hline
[11],[12],[21],[23],[33] & 164.4 & 173.1 & 89.7 \\
	& 344.4 & 6.9 & 269.7 \\
\hline
[11],[21],[22],[31],[23] & 166.6 & 357.9 & 263.7 \\
	& 346.6 & 182.1 & 83.7 \\
\hline
[11],[13],[23],[31],[33] & 176.0 & 181.8 & 269.3 \\
	& 356.0 & 358.2 & 89.3 \\
\hline
[13],[21],[22],[23],[33] & 176.0 & 187.7 & 270.9 \\
	& 355.7 & 188.1 & 269.0 \\
\hline
[11],[22],[23],[31],[33] & 348.7 & 178.1 & 89.2 \\
\hline
[11],[12],[13],[22],[23],[33] & 157.7 & 354.9 & 269.2 \\
	& 337.7 & 185.1 & 89.2 \\
\hline
[11],[12],[13],[21],[23],[33] & 159.0 & 355.7 & 271.1 \\
	& 174.0 & 175.5 & 270.8 \\
	& 339.0 & 184.3 & 91.1 \\
	& 161.7 & 2.1 & 269.8 \\
\hline
[11],[13],[22],[23],[31],[33] & 161.7 & 2.1 & 269.8 \\
	& 341.7 & 177.9 & 89.8 \\
\hline
[11],[12],[21],[22],[23],[33] & 171.3 & 183.8 & 91.4 \\
	& 171.4 & 354.8 & 269.2 \\
	& 351.4 & 184.9 & 89.2 \\
	& 351.3 & 356.2 & 271.4 \\
\hline
\end{tabular}
\end{tabular}
\end{table}

\newpage
\begin{table}[!h]
\centering
\caption{Full three-angle NH solutions (also serve as part of three-angle IH solutions) for BM and $\beta_{23}$ close to the x-axis, with $A=0.77$, (+,+) and $1\sigma$ ranges for neutrino mixing angles assumed. Note that solutions with ``$\star$'' 's belong to NH solutions only.}
\vskip 0.3 cm
\footnotesize
\begin{tabular}{|c | c  c  c|}
\hline
GJ Entries & $\beta_{12}$ & $\beta_{13}$ & $\beta_{23}$ \\
\hline
[22] & 174.7 & 14.1 & 175.8 \\
	& 354.7 & 165.9 & 355.8 \\
\hline
[22],[32] & 172.7 & 14.5 & 182.0 \\
	& 173.0 & 165.4 & 1.8 \\
	& 353.0 & 14.4 & 181.9 \\
	& 352.8 & 165.5 & 2.1\\
\hline
[12],[22],[23] & 24.4 & 182.0 & 177.5 \\
	& 204.6 & 358.0 & 357.3 \\
\hline
[12],[13],[22] & $\star$ 131.3 & 0.7 & 174.9 \\
	& $\star$ 311.3 & 179.3 & 354.9 \\
\hline
[12],[13],[22],[31] & 131.8 & 359.9 & 175.2 \\
	& 311.8 & 180.1 & 355.2 \\
\hline
[12],[22],[23],[32] & $\star$ 133.0 & 180.0 & 1.8 \\
	& $\star$ 313.0 & 0.0 & 181.8 \\
\hline
[12],[21],[22],[23] & 176.5 & 194.4 & 173.5 \\
	& 176.6 & 345.4 & 353.0 \\
	& 356.6 & 194.4 & 173.0 \\
	& 356.5 & 345.6 & 353.5 \\
\hline
[11],[13],[21],[22],[32] & 171.0 & 181.5 & 180.9 \\
	& 351.0 & 358.5 & 0.9\\
\hline
\end{tabular}
\end{table}

\begin{table}[h!]
\centering
\caption{Other three-angle IH solutions for BM and $\beta_{23}$ close to the x-axis, with $A=0.77$, (+,+) and $1\sigma$ ranges for neutrino mixing angles assumed.}
\vskip 0.3 cm
\footnotesize
\begin{tabular}{l l}
\begin{tabular}{|c | c  c  c|}
\hline
GJ Entries & $\beta_{12}$ & $\beta_{13}$ & $\beta_{23}$ \\
\hline
[22] & 172.4 & 164.7 & 4.3 \\
	& 352.4 & 15.3 & 184.3 \\
	& 353.2 & 165.2 & 0.5 \\
\hline
[22],[32] & 173.4 & 165.1 & 359.6 \\
	& 353.4 & 165.1 & 359.8 \\
\hline
[21],[22],[32] & 3.6 & 3.6 & 177.8 \\
	& 183.6 & 176.4 & 357.8 \\
\hline
[12],[22],[23] & 199.4 & 357.9 & 4.0 \\
\hline
[12],[21],[22],[23] & 359.5 & 194.9 & 185.2 \\
	& 178.4 & 194.6 & 180.3 \\
	& 178.5 & 345.1 & 1.0 \\
	& 358.4 & 345.1 & 0.3 \\
\hline
[12],[22],[23],[31] & 21.6 & 179.0 & 184.4 \\
	& 201.6 & 1.0 & 4.4 \\
\hline
[12],[22],[23],[32] & 21.4 & 183.3 & 178.6 \\
	& 201.4 & 356.7 & 358.6 \\
\hline
[11],[22],[23],[31] & 161.4 & 358.8 & 184.0 \\
	& 341.4 & 181.2 & 4.0 \\
\hline
[11],[13],[21],[22] & 170.8 & 180.9 & 179.0 \\
	& 350.8 & 358.9 & 359.9 \\
\hline
\end{tabular}
&
\begin{tabular}{| c | c c c|}
\hline
[13],[21],[22],[23],[31] & 3.0 & 181.3 & 5.7 \\
	& 183.0 & 358.7 & 185.7 \\
\hline
[12],[22],[23],[31],[32] & 28.5 & 179.0 & 179.0 \\
	& 208.5 & 1.0 & 359.0 \\
\hline
[11],[13],[22],[23],[32] & 161.4 & 2.5 & 180.0 \\
	& 341.4 & 177.3 & 359.8 \\
\hline
[11],[21],[22],[23],[31] & 173.4 & 0.7 & 2.7 \\
	& 353.4 & 179.3 & 182.7 \\
\hline
[11],[13],[21],[22],[32] & 173.5 & 182.0 & 178.7 \\
	& 350.9 & 358.8 & 0.0 \\
\hline
[12],[21],[22],[23],[32] & 178.3 & 194.7 & 179.8 \\
	& 177.8 & 344.7 & 358.7 \\
	& 357.8 & 194.5 & 178.7 \\
	& 358.3 & 345.1 & 359.7 \\
\hline
[11],[21],[22],[23],[31],[32] & 171.0 & 0.5 & 359.7 \\
	& 351.0 & 179.3 & 178.9 \\
\hline
\end{tabular}
\end{tabular}
\end{table}

\newpage

\begin{table}[h!]
\centering
\caption{Full three-angle NH solutions (also serve as part of three-angle IH solutions) for BM and $\beta_{23}$ close to the y-axis, with $A=0.77$, (+,+) and $1\sigma$ ranges for neutrino mixing angles assumed.}
\vskip 0.3 cm
\footnotesize
\begin{tabular}{|c | c  c  c|}
\hline
GJ Entries & $\beta_{12}$ & $\beta_{13}$ & $\beta_{23}$ \\
\hline
[23] & 174.5 & 194.0 & 274.0 \\
	& 354.5 & 346.0 & 94.0 \\
\hline
[23],[33] & 172.6 & 194.4 & 267.6 \\
	& 172.9 & 345.4 & 87.6 \\
	& 352.9 & 194.3 & 267.6 \\
	& 352.6 & 345.4 & 87.6 \\
\hline
[22],[13],[23] & 24.3 & 2.0 & 272.3 \\
	& 131.3 & 180.7 & 275.1 \\
	& 204.3 & 178.0 & 92.3 \\
	& 311.3 & 359.2 & 95.1 \\
\hline
[13],[12],[31],[23] & 131.8 & 179.9 & 274.8 \\
	& 311.8 & 0.1 & 94.8 \\
\hline
[22],[13],[23],[33] & 133.0 & 0.0 & 88.2 \\
	& 313.0 & 180.0 & 268.2 \\
\hline
[21],[22],[13],[23] & 176.5 & 14.4 & 276.5 \\
	& 356.5 & 165.6 & 96.5 \\
\hline
[13],[22],[31],[23],[33] & 133.0 & 0.1 & 88.4 \\
	& 313.0 & 179.9 & 268.4 \\
\hline
[11],[12],[21],[23],[33] & 170.9 & 1.4 & 269.3 \\
	& 350.9 & 178.6 & 89.3 \\
\hline
\end{tabular}
\end{table}

\newpage

\begin{table}[h!]
\centering
\caption{Other three-angle IH solutions for BM and $\beta_{23}$ close to the y-axis, with $A=0.77$, (+,+) and $1\sigma$ ranges for neutrino mixing angles assumed.}
\vskip 0.3 cm
\footnotesize
\begin{tabular}{l l}
\begin{tabular}{|c | c  c  c|}
\hline
GJ Entries & $\beta_{12}$ & $\beta_{13}$ & $\beta_{23}$ \\
\hline
[23] & 172.4 & 344.7 & 85.7 \\
	& 173.2 & 194.8 & 269.5 \\
	& 352.4 & 195.3 & 265.7 \\
	& 353.2 & 345.2 & 89.3 \\
\hline
[23],[33] & 172.9 & 194.5 & 268.7 \\
	& 173.1 & 345.3 & 88.7 \\
	& 353.1 & 194.5 & 268.7 \\
	& 352.9 & 345.4 & 88.7 \\
\hline
[21],[23],[33] & 3.6 & 183.6 & 272.2 \\
	& 183.6 & 356.4 & 92.2 \\
\hline
[13],[22],[23] & 19.4 & 2.1 & 265.9 \\
	& 199.4 & 177.9 & 85.9 \\
\hline
[13],[22],[23],[31] & 21.6 & 359.0 & 265.6 \\
	& 201.6 & 181.0 & 85.6 \\
\hline
[13],[22],[23],[33] & 23.0 & 2.1 & 270.0 \\
	& 201.4 & 176.7 & 91.4 \\
\hline
[11],[22],[23],[31] & 161.4 & 178.8 & 262.9 \\
	& 341.4 & 1.2 & 82.9 \\
\hline
[11],[12],[22],[23] & 162.0 & 182.6 & 267.1 \\
	& 342.0 & 357.4 & 87.1 \\
\hline
[11],[12],[21],[23] & 170.8 & 0.9 & 271.0 \\
	& 350.8 & 178.9 & 90.0 \\
\hline
\end{tabular}
&
\begin{tabular}{|c | c c c|}
\hline
[13],[21],[22],[23] & 177.2 & 14.5 & 274.0 \\
	& 177.4 & 165.3 & 93.6 \\
	& 357.2 & 165.5 & 94.0 \\
	& 357.4 & 14.7 & 273.6 \\
\hline
[12],[21],[22],[23],[31] & 3.0 & 1.3 & 84.1 \\
	& 183.0 & 178.7 & 264.1 \\
\hline
[13],[22],[23],[31],[33] & 28.5 & 359.0 & 271.0 \\
	& 208.5 & 181.0 & 91.0 \\
\hline
[11],[12],[22],[23],[33] & 170.8 & 1.1 & 269.9 \\
	& 350.8 & 178.8 & 89.6 \\
\hline
[11],[21],[22],[23],[31] & 173.4 & 180.7 & 86.0 \\
	& 353.4 & 359.3 & 266.0 \\
\hline
[13],[21],[22],[23],[33] & 178.3 & 14.7 & 270.1 \\
	& 177.8 & 164.7 & 91.3 \\
	& 357.8 & 14.5 & 271.3 \\
	& 358.3 & 165.1 & 90.1 \\
\hline
[11],[21],[22],[23],[31],[33] & 171.0 & 180.5 & 90.2 \\
	& 351.0 & 359.3 & 271.0 \\
\hline
\end{tabular}
\end{tabular}
\end{table}

\newpage

\section{Search Results: Almost Symmetric $Y^{(-1/3)}$}

As in Appendix A, we list the results of our search in table form. Table \ref{tb:sym1ATBM13} gives our result for the TBM case, while Table \ref{tb:sym1ABM12} shows the result for BM mixing. 

The layout of the tables is similar to that used in Appendix A, where we now study the deviation $\gamma_{ij}$ from the symmetric matrix. In addition, we have also varied the Wolfenstein parameter $A$, and these two parameters may be found in the second column. In the third column we list the corresponding $\beta_{ij}$ for a given solution matrix, to allow comparison with the three-angle results starting from an asymmetric matrix. 

As for these examples all of the entries of $Y^{(-1/3)}$ tend to vary, we list their orders of magnitude as done for the BM case in column 4. The rest of the columns give the various outputs defined earlier. 

\begin{landscape}

\begin{table}
\begin{center}
Table \ref{tb:sym1ATBM13}: One-angle Almost-Symmetric Results TBM
\vskip .3cm
\scriptsize 

\begin{tabular*}{1.7\textwidth}{ c | c | c| c c c| c | c | c c c c c c || c c c c }
\hline
\hline
& & & & &  &  &  &  &  &  &  &  &  &  &  &  &  \\
& A & $\gamma_{13}$ & $\beta_{12}$ & $\beta_{13}$ & $\beta_{23}$ & GJ Entries & $m_e/m_\mu$ & $\theta_{12}^e$ & $\theta_{13}^e$ & $\theta_{23}^e$ & $\delta_{12}^e$ & $\delta_{13}^e$ & $\delta_{23}^e$ & $\theta_{12}$ & $\theta_{13}$ & $\theta_{23}$ & $J_{MNSP}$ \\
& & & & &  &  &  &  &  &  &  &  &  &  &  &  &  \\
\hline
\multirow{13}{*}{(++)} & & & & & & [12],[21],[22] & & & & & & & & & & &  \\
& 0.76 & 173.2 & 13.4 & 6.7 & 2.3 & $\left(
\begin{array}{ccc}
 4.4 & 3.8 & 3.8 \\
 4.7 & 2.7 & 2.2 \\
 1.4 & 2.2 & 0.0
\end{array}
\right)$ & 0.0047 & 7.3 & 6.7 & 2.1 & -356.4 & 178.6 & 180.0 & 36.1 & 9.8 & 42.4 & $-1.6\times10^{-3}$\\
& & & & & & [12],[22],[23],[31] & & & & & & & & & & &  \\
& 0.71 & 359.0 & 13.2 & 0.9 & 2.1 & $\left(
\begin{array}{ccc}
 4.3 & 3.8 & 3.9 \\
 3.9 & 2.7 & 2.2 \\
 2.8 & 2.2 & 0.0
\end{array}
\right)$ & 0.0049 & 9.5 & 2.7 & 2.4 & -174.3 & 350.0 & 0.0 & 30.7 & 8.8 & 46.8 & $3.9\times10^{-3}$\\
& & & & & & [12],[21],[22],[23],[31] & & & & & & & & & & &  \\
& 0.76 & 357.9 & 13.2 & 2.0 & 2.2 & $\left(
\begin{array}{ccc}
 4.3 & 3.8 & 3.8 \\
 4.1 & 2.7 & 2.2 \\
 2.3 & 2.2 & 0.0
\end{array}
\right)$ & 0.0048 & 6.7 & 6.0 & 2.5 & -165.8 & -4.8 & 0.0 & 35.3 & 8.9 & 47.2 & $6.4\times10^{-3}$\\
\hline
\multirow{17}{*}{(+-)} & & & & & & [11],[13],[22],[32] & & & & & & & & & & &  \\
& 0.60 & 7.1 & 12.9 & 7.2 & 1.8 & $\left(
\begin{array}{ccc}
 5.4 & 3.7 & 4.0 \\
 6.4 & 2.8 & 2.3 \\
 1.4 & 2.3 & 0.0
\end{array}
\right)$ & 0.0047 & 4.0 & 7.2 & 5.4 & -179.8 & 1.1 & 180.0 & 36.7 & 8.1 & 39.5 & $-3.6\times10^{-4}$\\
& 0.60 & 170.9 & 13.4 & 9.0 & 1.8 & $\left(
\begin{array}{ccc}
 5.2 & 3.7 & 4.0 \\
 3.2 & 2.8 & 2.4 \\
 1.3 & 2.3 & 0.0
\end{array}
\right)$ & 0.0047 & 5.5 & 9.0 & 5.4 & 0.2 & 179.1 & 0.0 & 31.9 & 10.0 & 50.2 & $-4.2\times10^{-4}$\\
& & & & & & [11],[13],[21],[22],[32] & & & & & & & & & & &  \\
& 0.60 & 7.2 & 12.9 & 7.3 & 1.8 & $\left(
\begin{array}{ccc}
 5.4 & 3.7 & 4.0 \\
 6.4 & 2.8 & 2.3 \\
 1.4 & 2.3 & 0.0
\end{array}
\right)$ & 0.0049 & 4.1 & 7.3 & 5.4 & -184.4 & 1.1 & 180.0 & 36.7 & 8.2 & 39.5 & $-1.2\times10^{-3}$\\
& & & & & & [11],[21],[22],[23],[31] & & & & & & & & & & &  \\
& 0.67 & 182.1 & 13.0 & 2.2 & 2.0 & $\left(
\begin{array}{ccc}
 6.2 & 3.7 & 3.9 \\
 4.0 & 2.8 & 2.3 \\
 2.2 & 2.3 & 0.0
\end{array}
\right)$ & 0.0046 & 4.9 & 6.6 & 1.7 & -14.7 & -176.2 & 180.0 & 34.2 & 8.1 & 43.0 & $4.3\times10^{-3}$\\
\hline
\hline
\end{tabular*}
\vskip .3cm
(Table continued on the next page)
\end{center}
\end{table}
\end{landscape}

\begin{landscape}

\begin{table}
\begin{center}
One-angle Almost-Symmetric Case TBM, cont'd
\vskip .3cm
\scriptsize 

\begin{tabular*}{1.7\textwidth}{ c | c | c| c c c| c | c | c c c c c c || c c c c }
\hline
\hline
& & & & &  &  &  &  &  &  &  &  &  &  &  &  &  \\
& A & $\gamma_{13}$ & $\beta_{12}$ & $\beta_{13}$ & $\beta_{23}$ & GJ Entries & $m_e/m_\mu$ & $\theta_{12}^e$ & $\theta_{13}^e$ & $\theta_{23}^e$ & $\delta_{12}^e$ & $\delta_{13}^e$ & $\delta_{23}^e$ & $\theta_{12}$ & $\theta_{13}$ & $\theta_{23}$ & $J_{MNSP}$ \\
& & & & &  &  &  &  &  &  &  &  &  &  &  &  &  \\
\hline
\multirow{17}{*}{(-+)} & & & & & & [11],[13],[21],[22] & & & & & & & & & & &  \\
& 0.61 & 172.1 & 13.4 & 7.8 & 1.8 & $\left(
\begin{array}{ccc}
 5.4 & 3.7 & 4.0 \\
 5.6 & 2.7 & 2.3 \\
 1.3 & 2.4 & 0.0
\end{array}
\right)$ & 0.0048 & 5.1 & 7.8 & 1.7 & 3.5 & 1.0 & 180.0 & 33.7 & 9.2 & 43.1 & $-1.1\times10^{-3}$\\
& & & & & & [11],[13],[21],[22],[32] & & & & & & & & & & &  \\
& 0.61 & 172.1 & 13.4 & 7.8 & 1.8 & $\left(
\begin{array}{ccc}
 5.4 & 3.7 & 4.0 \\
 5.6 & 2.7 & 2.3 \\
 1.3 & 2.4 & 0.0
\end{array}
\right)$ & 0.0048 & 6.5 & 7.8 & 5.4 & 3.1 & -181.0 & 0.0 & 33.5 & 10.0 & 50.2 & $-1.2\times10^{-3}$\\
& & & & & & [11],[21],[22],[23],[31] & & & & & & & & & & &  \\
& 0.64 & 357.5 & 13.2 & 2.4 & 1.9 & $\left(
\begin{array}{ccc}
 6.2 & 3.7 & 3.9 \\
 4.0 & 2.7 & 2.3 \\
 2.1 & 2.3 & 0.0
\end{array}
\right)$ & 0.0049 & 5.5 & 7.2 & 2.1 & -168.2 & -3.3 & 0.0 & 36.8 & 8.9 & 46.9 & $4.6\times10^{-3}$\\
& & & & & & [11],[21],[22],[23],[31],[32] & & & & & & & & & & &  \\
& 0.6 & 357.1 & 13.2 & 2.8 & 1.8 & $\left(
\begin{array}{ccc}
 6.1 & 3.7 & 4.0 \\
 4.0 & 2.7 & 2.4 \\
 2.0 & 2.4 & 0.0
\end{array}
\right)$ & 0.0048 & 5.0 & 8.4 & 4.9 & -171.2 & -2.7 & 180.0 & 36.8 & 9.6 & 39.8 & $3.2\times10^{-3}$\\
\hline
\multirow{4}{*}{(--)} & & & & & & [12],[21],[22],[32] & & & & & & & & & & &  \\
& 0.8 & 5.9 & 12.9 & 6.0 & 2.4 & $\left(
\begin{array}{ccc}
 4.4 & 3.8 & 3.7 \\
 4.8 & 2.8 & 2.1 \\
 1.5 & 2.1 & 0.0
\end{array}
\right)$ & 0.0048 & 7.3 & 6.0 & 7.1 & -184.3 & 1.7 & 180.0 & 33.1 & 9.2 & 37.3 & $-1.9\times10^{-3}$\\
\hline
\hline
\end{tabular*}
\end{center}
\caption{One-angle almost-symmetric case with TBM, non-zero $\gamma_{13}$ and $3\sigma$ range for neutrino mixing angles.}
\label{tb:sym1ATBM13}
\end{table}

\begin{table}
\begin{center}
\caption{One-angle almost-symmetric case with BM, non-zero $\gamma_{12}$ and $3\sigma$ range for neutrino mixing angles.}
\vskip .3cm
\scriptsize 

\begin{tabular*}{1.7\textwidth}{ c | c | c| c c c| c | c | c c c c c c || c c c c }
\hline
\hline
& & & & &  &  &  &  &  &  &  &  &  &  &  &  &  \\
& A & $\gamma_{12}$ & $\beta_{12}$ & $\beta_{13}$ & $\beta_{23}$ & GJ Entries & $m_e/m_\mu$ & $\theta_{12}^e$ & $\theta_{13}^e$ & $\theta_{23}^e$ & $\delta_{12}^e$ & $\delta_{13}^e$ & $\delta_{23}^e$ & $\theta_{12}$ & $\theta_{13}$ & $\theta_{23}$ & $J_{MNSP}$ \\
& & & & &  &  &  &  &  &  &  &  &  &  &  &  &  \\
\hline
\multirow{9}{*}{(++)} & & & & & & [11],[21],[22],[23],[31] & & & & & & & & & & &  \\
& 0.7 & 158.3 & 8.6 & 0.6 & 1.7 & $\left(
\begin{array}{ccc}
 5.6 & 3.7 & 4.0 \\
 3.9 & 2.7 & 2.3 \\
 3.1 & 2.4 & 0.0
\end{array}
\right)$ & 0.0048 & 14.8 & 2.0 & 2.2 & -175.3 & 167.4 & 178.5 & 32.8 & 8.6 & 42.1 & $2.2\times10^{-3}$\\
& & & & & & [11],[21],[22],[23],[31],[32] & & & & & & & & & & &  \\
& 0.61 & 337.3 & 9.6 & 0.6 & 1.7 & $\left(
\begin{array}{ccc}
 5.8 & 3.7 & 4.0 \\
 3.8 & 2.7 & 2.3 \\
 3.1 & 2.4 & 0.0
\end{array}
\right)$ & 0.0047 & 12.0 & 1.8 & 4.7 & -176.2 & -11.9 & 178.1 & 36.9 & 9.1 & 39.5 & $3.2\times10^{-3}$\\
\hline
\hline
\end{tabular*}
\label{tb:sym1ABM12}
\end{center}
\end{table}

\end{landscape}

\section{Subtleties}

In this appendix we address several issues concerning the quadrants of the angles in $\m U_{\rm MNSP}$ and $\m U_{\rm seesaw}$. Adopting the parametrization of Eq.(\ref{eq:V}) for $\m V$ to decompose $\m U_{\rm MNSP}$ and $\m U_{\rm seesaw}$, we find that all of the TBM angles lie in the fourth quadrant, up to an ambiguity for $\theta^\nu_{13}$, which is zero. As experiments only measure the $\sin^2$ or $\cos^2$ of the MNSP mixing angles, the quadrants in which they lie are not determined. 

The alert reader may wonder whether this opens a back door for solutions with large mixing angles in $\m U_{-1}$; having large mixing angles in $\m U_{-1}$ would contradict the conventional wisdom that the seesaw matrix accounts for large mixing angles in $\m U_{\rm MNSP}$. In a simple numerical study (see later), we indeed find that large mixing angles in $\m U_{-1}$ can occur if one allows for such an ambiguity in $\m U_{\rm MNSP}$. Fortunately, because in the search we restrict some of the input angles in $\m V$ to be close to the x-axis, we find the cases with large angles in $\m U_{-1}$ do not occur.

Another quadrant issue that one may worry about in the search is whether the quadrants of $\m U_{\rm MNSP}$ and $\m U_{\rm seesaw}$ match. This would be expected if the seesaw matrix indeed accounts for the large mixing angles and $\m U_{-1}$ provides only small corrections. Although in the search we do not require such a match, we will show that such an issue actually does not exist, due to the restriction to small mixing angles in $\m U_{-1}$ and an ambiguity in relocating phases when diagonalizing $Y^{(-1)}$.

Finally, we highlight the freedom available in choosing the quadrants of the angles in $\m U_{\rm seesaw}$, and show that transferring to a different set of quadrants does not affect the search results, up to a change in the quadrants of the input angles. 

\subsection*{Large mixing angles in $\m U_{-1}$?}


We begin by discussing the issue of large mixing angles in $\m U_{-1}$ though a numerical example given in Table \ref{tb:langles}. Fixing the seesaw mixing matrix $\m U_{\rm seesaw}$ to be the TBM matrix in Eq.(\ref{eq:Useesaw}), we allow for the mixing angles in $\m U_{\rm MNSP}$ to be in any quadrant. 

The mixing angles in $\m U_{-1}$ can then be obtained through 

\begin{equation}
\m U_{-1}=\m U_{\rm seesaw} \m U_{\rm MNSP}^\dagger,
\end{equation} 
assuming the best fit values for $\m U_{\rm MNSP}$, and fall into two categories according to the choice for the quadrants of $\m U_{\rm MNSP}$. Half lead to three large angles, while the other half give small angles.

\begin{table}[!h]
\centering
\begin{tabular}{c | c c c}
\hline
\hline
Quadrants of $(\theta_{12}, \theta_{13}, \theta_{23})$ in $\m U_{\rm MNSP}$ & $\theta_{12}^e$ & $\theta_{13}^e$ & $\theta_{23}^e$ \\
\hline
111, 113, 132, 134, 321, 323, 342, 344 & $64.5^\circ$ & $34.5^\circ$ & $69.6^\circ$ \\
112, 114, 131, 133, 322, 324, 341, 343 & $62.2^\circ$ & $40.6^\circ$ & $35.5^\circ$ \\
121, 123, 142, 144, 311, 313, 332, 334 & $64.5^\circ$ & $34.5^\circ$ & $20.4^\circ$ \\
122, 124, 141, 143, 312, 314, 331, 333 & $62.2^\circ$ & $40.6^\circ$ & $54.5^\circ$ \\
\hline
211, 213, 232, 234, 421, 423, 442, 444 & $7.0^\circ$ & $5.6^\circ$ & $3.8^\circ$ \\
212, 214, 231, 233, 422, 424, 441, 443 & $4.6^\circ$ & $7.7^\circ$ & $85.9^\circ$ \\
221, 223, 242, 244, 411, 413, 432, 434 & $7.0^\circ$ & $5.6^\circ$ & $86.2^\circ$ \\
222, 224, 241, 243, 412, 414, 431, 433 & $4.6^\circ$ & $7.7^\circ$ & $4.1^\circ$ \\
\hline
\hline
\end{tabular}
\caption{Mixing angles in $\m U_{-1}$ for a given choice of quadrants of the mixing angles in $\m U_{\rm MNSP}$. Best fit values in Table \ref{tb:fit} are used for $\theta_{ij}$, $\m U_{\rm seesaw}=\m U_{\rm TBM}$, and $\theta_{ij}^e$ are interpreted as deviations from the x-axis.}
\label{tb:langles}
\end{table}

Since this occurrence of large angles in $\m U_{-1}$ would destroy the idea of using the seesaw matrix to explain large angles in $\m U_{\rm MNSP}$, we therefore restrict ourselves only to the cases where the mixing angles in $\m U_{-1}$ are small, or close to $90^\circ$.\footnote{The reason that we also study the latter case is because we think having $\theta_{23}^e$ close to $90^\circ$ seems special and may have implications for model building.}

In Table \ref{tb:langles}, one also notices that when the quadrants of $\m U_{\rm MNSP}$, say $(444)$ or $(414)$, match with those of $\m U_{\rm seesaw}$, the mixing angles in $\m U_{-1}$ are small. However, there also exist other choices that give rise to small angles in $\m U_{-1}$ but are not matched to the quadrants of $\m U_{\rm seesaw}$. Since in the search we have included these choices as well, one may wonder whether we still have such a matching requirement. Next we will try to argue that that requirement is still present, although we do not impose it explicitly.


\subsection*{Matching the quadrants of $\m U_{\rm MNSP}$ and $\m U_{\rm seesasw}$}


We begin by invoking the Iwasawa decompositions of $\m U_{-1}$ and $\m U_{\rm seesaw}$ in Eq.(\ref{eq:parametrization}). Although in our current search phases are not included in $\m V$ and $\m U_{\rm seesaw}$, the phase matrices $P_L$ and $P_R$ can still occur if we have fixed a particular convention for the quadrants of angles.
In other words, these phase matrices contain the information about the quadrants of the angles.

We then obtain the MNSP matrix through

\begin{eqnarray} \label{eq:MNSP_quadrant}
\m U_{\rm MNSP} &=& \m U_{-1}^\dagger \m U_{\rm seesaw} \nonumber \\
&=& P_R^{e\dagger} R_{12}^{e\dagger}U_{13}^{e\prime\dagger}R_{23}^{e\dagger} P_L^{e\dagger}P_L^\nu R_{23}^\nu U_{13}^{\nu\prime} R_{12}^\nu P_R^\nu \nonumber \\
&=& (P_R^{e\dagger} P_L^{e\dagger}P_L^\nu)~ U_{12}^{e\dagger}U_{13}^{e\dagger}U_{23}^{e\dagger}  R_{23}^\nu U_{13}^{\nu\prime} R_{12}^\nu ~P_R^\nu,
\end{eqnarray}
where instead of commuting all the phase matrices to the left as we did in Eq.(\ref{eq:MNSP_Product}), we now only commute $P_L^{e\dagger}P_L^\nu$ to the left, with the phases in $U_{ij}^e$ given by
\begin{eqnarray}
\delta_{12}^e &=& (\varphi_2^\nu-\varphi_2^e) - (\varphi_1^\nu  -\varphi_1^e), \nonumber \\
\delta_{13}^e &=& \delta_{13}^{e\prime} - (\varphi_3^\nu -\varphi_3^e)+ (\varphi_1^\nu-\varphi_1^e), \nonumber \\
\delta_{23}^e &=& (\varphi_3^\nu-\varphi_3^e) -(\varphi_2^\nu-\varphi_2^e). \nonumber
\end{eqnarray}

Matching the quadrants of the angles in $\m U_{\rm seesaw}$ and $\m U_{\rm MNSP}$ can then be discussed based on the last line of Eq.(\ref{eq:MNSP_quadrant}). First, because in the search we restrict ourselves to the case where the mixing angles in $\m U_{-1}$ are small, multiplying $U_{12}^{e\dagger}U_{13}^{e\dagger}U_{23}^{e\dagger}$ by $R_{23}^\nu U_{13}^{\nu\prime} R_{12}^\nu$ will not change the quadrants of the mixing angles in $R_{23}^\nu U_{13}^{\nu\prime} R_{12}^\nu$. We are then left to match the left and right diagonal phase matrices between $\m U_{\rm seesaw}$ and $\m U_{\rm MNSP}$. From Eq.(\ref{eq:parametrization}), one can see that the right phase matrix $P_R^\nu$ has already been matched, while for the left phase matrix we must require

\begin{eqnarray}
P_R^{e\dagger} P_L^{e\dagger}P_L^\nu \approx P_L^\nu,
\end{eqnarray}
which implies that in the decomposition of $\m U_{-1}$, the phases in $P_R^e$ have to cancel those in $P_L^e$. For a generic $\m U_{-1}$, this certainly may not hold. However, in the diagonalizaion of $Y^{(-1)}$,

\begin{eqnarray}
Y^{(-1)} &=& \m U_{-1} \m D_{-1} \m V_{-1}^\dagger, \nonumber \\
&=& (P_L^e R_{23}^e U_{13}^e R_{12}^e P_R^e)~ \m D_{-1} \m V_{-1}^\dagger,
\end{eqnarray}
the ambiguity of relocating phases among $P_R^e$, $\m D_{-1}$, and $\m V_{-1}$  allows one to \emph{choose} the phases in  $P_R^e$ to cancel those in $P_L^e$, eventually resulting in a match of the quadrants between $\m U_{\rm seesaw}$ and $\m U_{\rm MNSP}$. We therefore do not have to impose such an extra matching criterion in our current search, although one may have to include it if one allows for large mixing angles in $\m U_{-1}$.

\subsection*{Quadrants of the angles in $\m U_{\rm seesaw}$}

As a final point, we discuss the freedom available in choosing the quadrants of the angles in $\m U_{\rm seesaw}$. In the main text a particular form of $\m U_{\rm seesaw}$ is chosen, where the mixing angles $\theta_{12}^\nu$ and $\theta_{13}^\nu$ are both in the fourth quadrant according to the parametrization used in Eq.(\ref{eq:V}). Although we made this choice in order to be consistent with our previous publication \cite{Kile:2013gla}, one can certainly choose a different form of $\m U_{\rm seesaw}$ by setting its angles in different quadrants. The search results will be altered by doing so, however, we find that the only change is in the quadrants of the input angles $\beta_{ij}$; their actual deviations from the x-axis stay the same as before. Next we will give an example to illustrate this.

Suppose that we were to instead set both $\theta_{12}^\nu$ and $\theta_{13}^\nu$ in $\m U_{\rm seesaw}$ to be in the first quadrant. A transformation can then be found between this new seesaw matrix and the previous one,

\begin{eqnarray} \label{eq:quadtrans}
\m U_{\rm seesaw}^\prime = \begin{pmatrix}
1 & & \\ & -1 & \\ & & 1
\end{pmatrix} \m U_{\rm seesaw} \begin{pmatrix}
1 & & \\ & -1 & \\ & & 1
\end{pmatrix},
\end{eqnarray}
where a superscript ``$\prime$'' is used to distinguish these two cases.

With this new $\m U_{\rm seesaw}^\prime$, the MNSP matrix now becomes

\begin{eqnarray}
\m U_{\rm MNSP}^\prime = \m U_{-1}^{\prime\dagger}\begin{pmatrix}
1 & & \\ & -1 & \\ & & 1
\end{pmatrix} \m U_{\rm seesaw} \begin{pmatrix}
1 & & \\ & -1 & \\ & & 1
\end{pmatrix},
\end{eqnarray}
which implies that if one could also relate the above new $\m U_{-1}^\prime$ with $\m U_{-1}$ through

\begin{eqnarray}
\m U_{-1}^\prime = \begin{pmatrix}
1 & & \\ & -1 & \\ & & 1
\end{pmatrix} \m U_{-1},
\end{eqnarray}
the new MNSP matrix would have the same mixing angles as the previous case, up to some unknown Majorana phases. Since in our search $Y^{(-1)}$ is related to $Y^{(-1/3)}$ through $SU(5)$, such a relation between $\m U_{-1}^\prime$ and $\m U_{-1}$ can indeed be realized by performing a transformation on the input matrix $\m V$

\begin{eqnarray}
\m V^\prime = \begin{pmatrix}
1 & & \\ & -1 & \\ & & 1
\end{pmatrix} \m V,
\end{eqnarray}
which gives
\begin{eqnarray}
\beta_{23}^\prime &=& -\beta_{23}, \nonumber \\
\beta_{13}^\prime &=& \pi + \beta_{13}, \nonumber \\
\beta_{12}^\prime &=& \beta_{12}.
\end{eqnarray}

Other cases with different choices of quadrants can be studied similarly, and the same conclusion holds. Therefore in the search one can make an arbitrary choice for the quadrants of the angles in $\m U_{\rm seesaw}$, as a particular choice may be related to any other by a change of the quadrants of the input angles.

\newpage

{}
\end{document}